\documentclass[12pt,twoside,dvips]{article}
\usepackage{epsfig}
\usepackage{amsmath}
\usepackage{amssymb}
\usepackage{float}
\newcommand{\dissum}[2]{\displaystyle \sum_{#1}^{#2}}
\newcommand{\fnd}[2]{\frac{\textstyle #1}{\textstyle #2}}

\newcommand{\ket}[1]{\mbox{$\left| #1\right\rangle$}}
\newcommand{\bra}[1]{\mbox{$\left\langle #1\right|$}}
\newcommand{\boldtau}{\mbox{\boldmath $\tau$}}
\newcommand{\boldpi}{\mbox{\boldmath $\pi$}}
\newcommand{\be}{\begin{equation}}
\newcommand{\ee}{\end{equation}}
\newcommand{\ba}{\begin{array}{rcl}}
\newcommand{\ea}{\end{array}}
\newcommand{\bma}{\begin{displaymath}}
\newcommand{\ema}{\end{displaymath}}
\newcommand{\bfi}{\begin{figure}}
\newcommand{\efi}{\end{figure}}
\newcommand{\bc}{\begin{center}}
\newcommand{\ec}{\end{center}}
\newcommand{\kkb}          {\mbox{$|$K$\bar{\mbox{K}}>$}}

\begin{document}

\title{Miniproceedings for the International Workshop\\
{\it Hadron Physics at COSY}\\
{7.-10.7.2003}}

\author{Edited by: A. Gillitzer, C. Hanhart, V. Kleber,
S. Krewald,\\ H.P. Morsch, F. Rathmann, A. Sibirtsev}

\maketitle

\tableofcontents

\newpage

\phantom{.}
\begin{figure}
\vspace{7cm}
\includegraphics{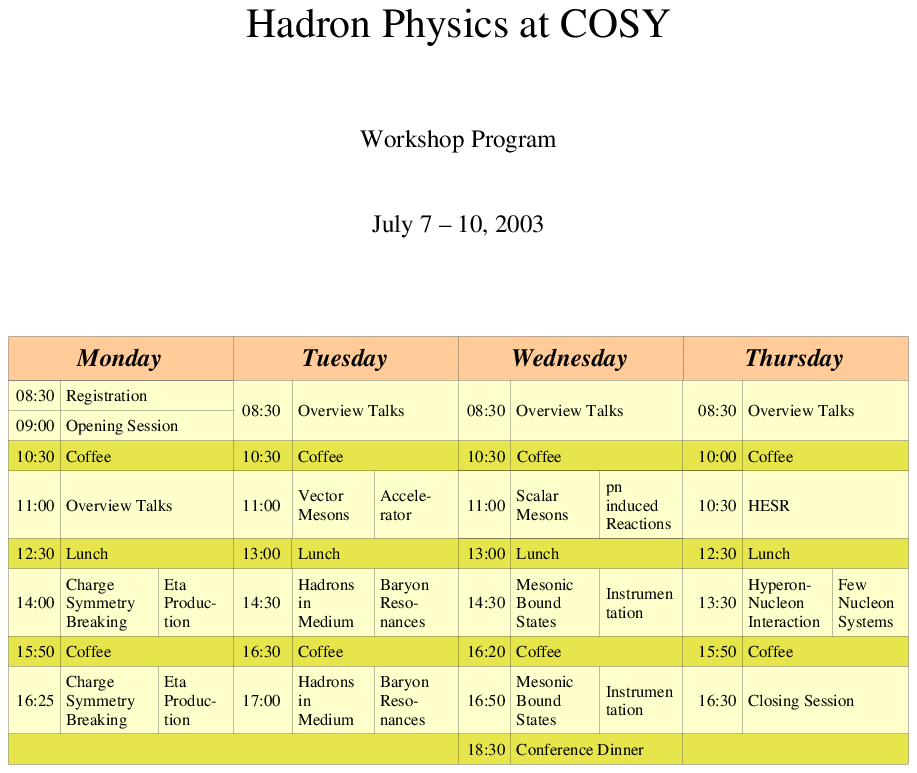}
\end{figure}

\newpage

\section{Foreword}

COSY allows the study of reactions with (un)polarized proton and deuteron
beams up to momenta of about 3.88 GeV/c. The goal of the workshop
{\it Hadron Physics at COSY} was to bring
together experimentalists and theoreticians from various fields of hadron
physics to identify the key physics questions, which can be addressed with
proton and deuteron induced reactions at COSY. Recent reviews on the subject
can be found in Refs. \cite{rev1,rev2}. The possibilities of approaching
particular problems with hadronic probes were discussed.

These Miniproceedings are meant to present a useful collection of key
references to the field as well as short summaries of the
presentations\footnote{The speakers were asked for a one/two page summary for
  a parallel/overview talk.}. For
most of the talks the corresponding transparencies can be found on our home
page under program/talks (http://www.fz-juelich.de/ikp/theorie/hpc2003/).

\newpage

\section{List of Participants}

\small{\begin{enumerate}
\item Aichelin, J\"org;  Universite de Nantes, France
\\   e-mail: {\tt aichelin@subatech.in2p3.fr}
 
\item Bacelar, Jose; Kernfysisch Versneller Instituut Groningen, The Netherlands
\\   e-mail: {\tt bacelar@kvi.nl}
 
\item Balewski, Jan; Indiana University Cyclotron Facility, USA
\\   e-mail: {\tt balewski@iucf.indiana.edu},
 
\item Bashkanov, Mikhail; Universit\"at T\"ubingen, Germany
\\   e-mail: {\tt bashkanov@pit.physik.uni-tuebingen.de}
 
\item Beck, Reinhard; UniversitŽät Mainz, Germany
\\   e-mail: {\tt rbeck@kph.uni-mainz.de}
 
\item Bisplinghoff, Jens; Universit\"at Bonn, Germany
\\   e-mail: {\tt jens@iskp.uni-bonn.de}
 
\item Bongardt, Klaus; Forschungszentrum J\"ulich, Germany
\\   e-mail: {\tt k.Bongardt@fz-juelich.de}
 
\item Brinkmann, Kai-Thomas; Technische Universit\"at Dresden, Germany
\\   e-mail: {\tt brinkman@pktw09.phy.tu-dresden.de}
 
\item Buballa, Michael; Technische Universit\"at Darmstadt, Germany
\\   e-mail: {\tt buballa@dirac.ikp.physik.tu-darmstadt.de}
 
\item B\"uscher, Markus; Forschungszentrum J\"ulich, Germany
\\   e-mail: {\tt M.Buescher@fz-juelich.de}
 
\item Calen, Hans; The Svedberg Laboratory Uppsala, Sweden
\\   e-mail: {\tt hans.calen@tsl.uu.se}
 
\item Cassing, Wolfgang; Universit\"at Giessen, Germany
\\   e-mail: {\tt wolfgang.cassing@theo.physik.uni-giessen.de}
 
\item Clement, Heinz; Universit\"at T\"ubingen, Germany
\\   e-mail: {\tt clement@pit.physik.uni-tuebingen.de}
 
\item Crede, Volker; Universit\"at Bonn, Germany
\\   e-mail: {\tt crede@iskp.uni-bonn.de}
 
\item Demiroers, Levent; DESY Hamburg, Germany
\\   e-mail: {\tt levent@kaa.desy.de}
 
\item Dillig, Manfred; Universit\"at Erlangen, Germany
\\   e-mail: {\tt mdillig@theorie3.physik.uni-erlangen.de}
 
\item Dorochkevitch, Evgueni; Universit\"at T\"ubingen, Germany
\\   e-mail: {\tt evd@pit.physik.uni-tuebingen.de}
 
\item Dshemuchadse, Solomon; Technische Universit\"at Dresden, Germany
\\   e-mail: {\tt solomon@pktw09.phy.tu-dresden.de}
 
\item Durso, John W.; Forschungszentrum J\"ulich, Germany
\\   e-mail: {\tt j.durso@fz-juelich.de}
 
\item Elster, Charlotte; Ohio University, USA
\\   e-mail: {\tt c.elster@fz-juelich.de}
 
\item Erhardt, Arthur; Universit\"at T\"ubingen, Germany
\\   e-mail: {\tt erhardt@pit.physik.uni-tuebingen.de}
 
\item Eyrich, Wolfgang; Universit\"at Erlangen, Germany
\\   e-mail: {\tt Eyrich@physik.uni-erlangen.de}
 
\item Eyser, K. Oleg; Universit\"at Hamburg, Germany
\\   e-mail: {\tt oleg@kaa.desy.de}
 
\item Fedorets, Pawel; Forschungszentrum J\"ulich, Germany
\\   e-mail: {\tt p.fedorets@fz-juelich.de}

\item Fix, Alexander; Universit\"at Mainz, Germany
\\   e-mail: {\tt fix@kph.uni-mainz.de}
 
\item Franzke, Bernhard; Gesellschaft f\"ur Schwerionenforschung, Germany
\\   e-mail: {\tt b.franzke@gsi.de}
 
\item Fr\"ohlich, Ingo; Universit\"at Giessen, Germany
\\   e-mail: {\tt ingo.froehlich@exp2.physik.uni-giessen.de}
 
\item Garcilazo, Humberto; Instituto Politecnico Nacional, Mexico
\\   e-mail: {\tt humberto@esfm.ipn.mx}
 
\item Gasparyan, Ashot; Forschungszentrum J\"ulich, Germany
\\   e-mail: {\tt g.achot@fz-juelich.de}
 
\item Gibson, Ben; Los Alamos National Laboratory, USA
\\   e-mail: {\tt bfgibson@lanl.gov}
 
\item Gillitzer, Albrecht; Forschungszentrum J\"ulich, Germany
\\   e-mail: {\tt a.gillitzer@fz-juelich.de}
 
\item Goldenbaum, Frank; Forschungszentrum J\"ulich, Germany
\\   e-mail: {\tt f.goldenbaum@fz-juelich.de}
 
\item Gotta, Detlev; Forschungszentrum J\"ulich, Germany
\\   e-mail: {\tt D.Gotta@fz-juelich.de}
 
\item Grzonka, Dieter; Forschungszentrum J\"ulich, Germany
\\   e-mail: {\tt D.Grzonka@fz-juelich.de}
 
\item Gutbrod, Hans; Gesellschaft f\"ur Schwerionenforschung, Germany
\\   e-mail: {\tt h.gutbrod@gsi.de}
 
\item Haidenbauer, Johann; Forschungszentrum J\"ulich, Germany
\\   e-mail: {\tt j.haidenbauer@fz-juelich.de}
 
\item Hanhart, Christoph; Forschungszentrum J\"ulich, Germany
\\   e-mail: {\tt c.hanhart@fz-juelich.de}
 
\item Hartmann, Michael; Forschungszentrum J\"ulich, Germany
\\   e-mail: {\tt M.Hartmann@fz-juelich.de}
 
\item Hejny, Volker; Forschungszentrum J\"ulich, Germany
\\   e-mail: {\tt V.Hejny@fz-juelich.de}
 
\item Hinterberger, Frank; Universit\"at Bonn, Germany
\\   e-mail: {\tt fh@iskp.uni-bonn.de}
 
\item Jaekel, Rene; Technische Universit\"at Dresden, Germany
\\   e-mail: {\tt jaekel@physik.phy.tu-dresden.de}
 
\item Janusz, Michal; Jagiellonian University Cracow, Poland
\\   e-mail: {\tt michal\_janusz@yahoo.com}
 
\item Kacharava, Andro; Universit\"at Erlangen, Germany
\\   e-mail: {\tt A.Kacharava@fz-juelich.de}
 
\item Kaiser, Norbert; Technische Universit\"at M\"unchen, Germany
\\   e-mail: {\tt nkaiser@physik.tu-muenchen.de}
 
\item Kaskulov, Murat; Universit\"at T\"ubingen, Germany
\\   e-mail: {\tt kaskulov@pit.physik.uni-tuebingen.de}
 
\item Kilian, Kurt; Forschungszentrum J\"ulich, Germany
\\   e-mail: {\tt k.kilian@fz-juelich.de}
 
\item Kleber, Vera; Universit\"at zu K\"oln, Germany
\\   e-mail: {\tt v.kleberfz-juelich.de}
 
\item Klempt, Eberhard; Universit\"at Bonn, Germany
\\   e-mail: {\tt klempt@iskp.uni-bonn.de}
 
\item Kobushkin, Alexander; Bogolyubov Institute for Theoretical Physics, Kiev, Ukraine
\\   e-mail: {\tt akob@ap3.bitp.kiev.ua}
 
\item Koch, Helmut; Universit\"at Bochum, Germany
\\   e-mail: {\tt hkoch@ep1.ruhr-uni-bochum.de}
 
\item Krewald, Siegfried; Forschungszentrum J\"ulich, Germany
\\   e-mail: {\tt s.krewald@fz-juelich.de}
 
\item Kroll, Peter; Universit\"at Wuppertal, Germany
\\   e-mail: {\tt kroll@physik.uni-wuppertal.de}
 
\item Kuhlmann, Eberhart; Forschungszentrum J\"ulich, Germany
\\   e-mail: {\tt e.kuhlmann@fz-juelich.de}
 
\item Kulessa, Pawel; Forschungszentrum J\"ulich, Germany
\\   e-mail: {\tt p.kulessa@fz-juelich.de}
 
\item Lang, Norbert; Universit\"at M\"unster, Germany
\\   e-mail: {\tt langn@ikp.uni-muenster.de}
 
\item Lehmann, Inti; Forschungszentrum J\"ulich, Germany
\\   e-mail: {\tt i.lehmann@fz-juelich.de}
 
\item Lehrach, Andreas; Forschungszentrum J\"ulich, Germany
\\   e-mail: {\tt a.lehrach@fz-juelich.de}
 
\item Lenske, Horst; Universit\"at Giessen, Germany
\\   e-mail: {\tt horst.lenske@theo.physik.uni-giessen.de}
 
\item Lorentz, Bernd; Forschungszentrum J\"ulich, Germany
\\   e-mail: {\tt b.lorentz@fz-juelich.de}
 
\item Machner, Hartmut; Forschungszentrum J\"ulich, Germany
\\   e-mail: {\tt H.Machner@fz-juelich.de}
 
\item Maeda, Yoshikazu; Forschungszentrum J\"ulich, Germany
\\   e-mail: {\tt Y.Maeda@fz-juelich.de}
 
\item Magiera, Andrzej; Jagiellonian University Cracow, Poland
\\   e-mail: {\tt magiera@if.uj.edu.pl}
 
\item Maier, Rudolf; Forschungszentrum J\"ulich, Germany
\\   e-mail: {\tt R.Maier@fz-juelich.de}
 
\item Martin, Sig; Forschungszentrum J\"ulich, Germany
\\   e-mail: {\tt S.Martin@fz-juelich.de}
 
\item Meier, Rudolf; Universit\"at T\"ubingen, Germany
\\   e-mail: {\tt r.meier@pit.physik.uni-tuebingen.de}
 
\item Metag, Volker; Universit\"at Giessen, Germany
\\   e-mail: {\tt Volker.Metag@exp2.physik.uni-giessen.de}
 
\item Metsch, Bernhard; Universit\"at Bonn, Germany
\\   e-mail: {\tt metsch@itkp.uni-bonn.de}
 
\item Miller, Gerald A.; University of Washington, USA
\\   e-mail: {\tt miller@phys.washington.edu}
 
\item Morsch, Hans-Peter; Forschungszentrum J\"ulich, Germany
\\   e-mail: {\tt Morsch@fz-juelich.de}
 
\item Moskal, Pawel; Forschungszentrum J\"ulich, Germany
\\   e-mail: {\tt p.moskal@fz-juelich.de}
 
\item Nakayama, Kanzo; University of Georgia, USA
\\   e-mail: {\tt k.nakayama@fz-juelich.de}
 
\item Nekipelov, Mikhail; Forschungszentrum J\"ulich, Germany
\\   e-mail: {\tt M.Nekipelov@fz-juelich.de}
 
\item Nevelainen, Kirsi; University of Helsinki, Finland
\\   e-mail: {\tt kirsi.nevalainen@helsinki.fi}
 
\item Nikolaev, Nikolai N.; Forschungszentrum J\"ulich, Germany
\\   e-mail: {\tt n.nikolaev@fz-juelich.de}
 
\item Niskanen, Jouni; University of Helsinki, Finland
\\   e-mail: {\tt jouni.niskanen@helsinki.fi}
 
\item Oelert, Walter; Forschungszentrum J\"ulich, Germany
\\   e-mail: {\tt W.Oelert@fz-juelich.de}
 
\item Opper, Alena Ohio; University, USA
\\   e-mail: {\tt opper@ohiou.edu} 
 
\item Pavlov, Fedor; Forschungszentrum J\"ulich, Germany
\\   e-mail: {\tt f.pavlov@fz-juelich.de}
 
\item Petry, Herbert; Universit\"at Bonn, Germany
\\   e-mail: {\tt petry@itkp.uni-bonn.de}
 
\item Pfeiffer, Marco; Universit\"at Giessen, Germany
\\   e-mail: {\tt Marco.Pfeiffer@exp2.physik.uni-giessen.de}

\item Piskor-Ignatowicz, Cezary; Jagellonian University, Cracow, Poland
\\   e-mail: {\tt  hobit0@poczta.onet.pl}

\item Pizzolotto, Cecilia; Universit\"at Erlangen, Germany
\\   e-mail: {\tt pizzolotto@physik.uni-erlangen.de}
 
\item Polyakov, Maxim; Universit\"at Bochum, Germany
\\   e-mail: {\tt Maxim.Polyakov@tp2.ruhr-uni-bochum.de}
 
\item Prasuhn, Dieter; Forschungszentrum J\"ulich, Germany
\\   e-mail: {\tt D.Prasuhn@fz-juelich.de}
 
\item Prokofiev, Alexander; Petersburg Nuclear Physics Institute, Russia
\\   e-mail: {\tt prokan@hep486.pnpi.spb.ru}
 
\item Przerwa, Joanna Jagiellonian; University Cracow, Poland
\\   e-mail: {\tt j\_przerwa@hotmail.com}
 
\item Rathmann, Frank; Forschungszentrum J\"ulich, Germany
\\   e-mail: {\tt F.Rathmann@fz-juelich.de}
 
\item Ritman, James; Universit\"at Giessen, Germany
\\   e-mail: {\tt james.ritman@exp2.physik.uni-giessen.de}
 
\item Roderburg, Eduard; Forschungszentrum J\"ulich, Germany
\\   e-mail: {\tt E.Roderburg@fz-juelich.de}
 
\item Rohdjess, Heiko; Universit\"at Bonn, Germany
\\   e-mail: {\tt rohdjess@iskp.uni-bonn.de}
 
\item Rossen, Peter; Forschungszentrum J\"ulich, Germany
\\   e-mail: {\tt p.v.rossen@fz-juelich.de}
 
\item Roth, Thomas; Technische Universit\"at Darmstadt, Germany
\\   e-mail: {\tt thomas.roth@physik.tu-darmstadt.de}
 
\item Roy, Bidyut; Forschungszentrum J\"ulich, Germany
\\   e-mail: {\tt b.roy@fz-juelich.de}
 
\item Rozek, Thomasz; Forschungszentrum J\"ulich, Germany
\\   e-mail: {\tt t.rozek@fz-juelich.de}
 
\item Rudy, Zbigniew Jagiellonian University Cracow, Poland
\\   e-mail: {\tt  ufrudy@cyf-kr.edu.pl}
 
\item Sassen, Felix; Forschungszentrum J\"ulich, Germany
\\   e-mail: {\tt f.p.sassen@fz-juelich.de}
 
\item Schadmand, Susan; Universit\"at Giessen, Germany
\\   e-mail: {\tt susan.schadmand@exp2.physik.uni-giessen.de}
 
\item Schieck, Hans Paetz gen.; Universit\"at K\"oln, Germany
\\   e-mail: {\tt schieck@ikp.uni-koeln.de}
 
\item Schleichert, Ralf; Forschungszentrum J\"ulich, Germany
\\   e-mail: {\tt R.Schleichert@fz-juelich.de}
 
\item Schnase, Alexander; Forschungszentrum J\"ulich, Germany
\\   e-mail: {\tt a.schnase@fz-juelich.de}
 
\item Schneider, Sonja; Forschungszentrum J\"ulich, Germany
\\   e-mail: {\tt s.schneider@fz-juelich.de}
 
\item Schr\"oder, Wolfgang; Universit\"at Erlangen, Germany
\\   e-mail: {\tt wolfgang.schroeder@physik.uni-erlangen.de}
 
\item Schulte-Wissermann, Martin; Technische Universit\"at Dresden, Germany
\\   e-mail: {\tt schulte@pktw09.phy.tu-dresden.de}
 
\item Sefzick, Thomas; Forschungszentrum J\"ulich, Germany
\\   e-mail: {\tt t.sefzik@fz-juelich.de}
 
\item Shyam, Radhey; Universit\"at Giessen, Germany
\\   e-mail: {\tt Radhey.Shyam@theo.physik.uni-giessen.de}
 
\item Sibirtsev, Alexander; Forschungszentrum J\"ulich, Germany
\\   e-mail: {\tt a.sibirtsev@fz-juelich.de}
 
\item Skorodko, Tatiana; Universit\"at T\"ubingen, Germany
\\   e-mail: {\tt skorodko@pit.physik.uni-tuebingen.de}
 
\item Smyrski, Jerzy; Jagiellonian University Cracow, Poland
\\   e-mail: {\tt smyrski@sigma.if.uj.edu.pl}
 
\item Soyeur, Madeleine; DAPNIA, Saclay, France
\\   e-mail: {\tt msoyeur@cea.fr}
 
\item Speth, Josef; Forschungszentrum J\"ulich, Germany
\\   e-mail: {\tt j.speth@fz-juelich.de}
 
\item Stein, Jochen; Forschungszentrum J\"ulich, Germany
\\   e-mail: {\tt j.stein@fz-juelich.de}
 
\item Stockhorst, Hans; Forschungszentrum J\"ulich, Germany
\\   e-mail: {\tt h.stockhorst@fz-juelich.de}
 
\item Str\"oher, Hans; Forschungszentrum J\"ulich, Germany
\\   e-mail: {\tt H.Stroeher@fz-juelich.de}
 
\item Suzuki, Ken; Technische Universit\"at M\"unchen, Germany
\\   e-mail: {\tt ken.suzuki@physik.tu-muenchen.de}
 
\item Tegner, Per-Erik Stockholm University, Sweden
\\   e-mail: {\tt tegner@physto.se}
 
\item Thomas, Anthony; University of Adelaide, Australia
\\   e-mail: {\tt athomas@physics.adelaide.edu.au}
 
\item Tiator, Lothar; Universit\"at Mainz, Germany
\\   e-mail: {\tt tiator@kph.uni-mainz.de}
 
\item Uhlig, Florian; Technische Universit\"at Darmstadt, Germany
\\   e-mail: {\tt F.Uhlig@gsi.de}
 
\item Ucar, Aziz; Forschungszentrum J\"ulich, Germany
\\   e-mail: {\tt a.ucar@fz-juelich.de}
 
\item Uzikov, Yuri Joint Institute for Nuclear Research Dubna, Russia
\\   e-mail: {\tt uzikov@nusun.jinr.dubna.su}
 
\item van Beveren, Eef; Universidade de Coimbra, Portugal
\\   e-mail: {\tt eef@teor.fis.uc.pt}
 
\item Winter, Peter; Forschungszentrum J\"ulich, Germany
\\   e-mail: {\tt p.winter@fz-juelich.de}
 
\item Wintz, Peter; Forschungszentrum J\"ulich, Germany
\\   e-mail: {\tt p.wintz@fz-juelich.de}
 
\item Wirzba, Andreas; Universit\"at Bonn, Germany
\\   e-mail: {\tt a.wirzba@fz-juelich.de}

\item Zychor, Izabella; Andrzej Soltan Institute for Nuclear Studies Swierk, Poland
\\   e-mail: {\tt i.zychor@fz-juelich.de}
\end{enumerate}

}
\newpage 

\section{Opening session}

\subsection{Overview and Highlights of COSY}
\addtocontents{toc}{\hspace{2cm}(D. Grzonka)\par}

D. Grzonka

{\em Institut f\"ur Kernphysik, Forschungszentrum J\"ulich GmbH, Germany}

The Cooler Synchrotron COSY \cite{may97} delivers polarized and unpolarized
phase\-space cooled proton- and deuteron-beams in the momentum range
between 300 - 3650 MeV/c for internal as well as external experimental
installations. 
The physics studied at the different experiments includes a variety of
topics which are mostly covered by dedicated contributions to these
proceedings.

A few recent results from experiments at COSY 
ranging from elementary pp interactions up to reactions on nuclei
are briefly discussed, clearly demonstrating the
high-performance 
detection capabilities of the experimental installations which allow in
combination with the low emittance COSY beam for high-precision data.

The elastic pp scattering studies at EDDA \cite{roj03}
 resulted in accurate data for excitation functions \cite{alb97},
analyzing power\cite{alt00} and spin correlation
parameters\cite{bau03} which extend drastically the data base
for phase shift analysis and improve the understanding of the
elementary pp interaction. No hint was found for dibaryons which
couple to NN.

The study of deuteron breakup performed at ANKE \cite{kom03}
probes the short range
NN dynamics. The use of the CD Bonn potential including a rather soft
short range NN potential gives a much better description of the data
compared to RSC or Paris potentials \cite{hai03}.

High statistics $\eta$ production has been performed at both the COSY-11
\cite{mos03}  and the TOF \cite{rod03}
installation from which  background free differential distributions
could be extracted. The distribution of the proton-proton invariant
mass $m_{pp}$ shows a drastic enhancement over the expected s-wave 
phase space distribution with pp FSI. A simple incoherent additional 
p-$\eta$ FSI contribution
is much too weak to account for this effect.
Higher partial waves in the pp-system might explain the enhancement 
\cite{nak03} and additionally a coherent inclusion of pp and p-$\eta$
FSI seems to be necessary.
The measurement of spin correlation parameters would clarify this
point.
First investigations of  $\eta$ production with a polarized beam 
\cite{win02} result in a rough determination of 
the angular distribution of the analyzing power. The analysis of a
higher statistics data sample will allow to disentangle the dominant
contributing exchange mesons in the production ($\pi$, $\eta$
\cite{nak02} or $\rho$ \cite{fal01}).

The hyperon production performed at TOF\cite{eyr03} and
COSY-11 \cite{bal96} \cite{sew99}
 yield very  
informative and surprising results. 
The geometrical decay spectrometer at TOF allows
a nearly background free sample of $\Lambda$ events. In Dalitz
plots for beam momenta of 2.95 and 3.2 GeV/c the excitation of nucleon
resonances as well as the p-$\Lambda$ FSI are clearly visible. A
comparison of the data with the resonance model \cite{tsu97} 
including a coherent
addition of $N^*$ excitation and  p-$\Lambda$ FSI
shows a strong
dominance of the $N^*$(1650) at 2.95 GeV/c and equal contributions
from $N^*$(1650) and   $N^*$(1720) at 3.2 GeV/c.
The most actual result at COSY-11 \cite{sew99}, the
extreme high cross section
ratio between $\Lambda$ and $\Sigma$ production close to threshold with
a value of about 28 compared to the factor 2.5 at high energies, is
still not unambiuously explained. Several very different models with
only $\pi$ and K exchange, coherently or incoherently added,
additional exchange mesons with or without intermediate 
$N^*$ excitation are able
to at least describe the trend of the data within a factor of two
\cite{lsr03}. A
solid data base for Y-N data is really missing which would allow to
fix more parameters for the model descriptions.

The K$^+$ production in nuclei far below the free NN production threshold
performed at ANKE \cite{kop01}
addresses topics like medium modification, reaction
mechanism and collectivity. The K$^+$ potential in nuclei could be
determined from the comparison of the cross section ratio between
different nuclei (Au/C) with model calculations by varying the
potential depth. A value of 20 $\pm$3 MeV results whereby compared to
relativistic heavy ion collisions the potential is determined at
normal nuclear density. Another interesting aspect in the K$^+$
production studies are the K$^+$-d correlations \cite{kop03}
which yield informations
on the production mechanism. From a first test run at 1.2 GeV it is
estimated that about 30 \% of the productions is mediated by a two step
mechanism ($pN \rightarrow d \pi$, $\pi N \rightarrow K^+ \Lambda$).
 
Much more highlights and results are available
which can not be presented within this short overview. Not even all
experimental installations were covered by the selected topics.

The near future physics program at COSY will certainly
continue the present activities. 
Several results answered some but 
opened new questions as well which need more selective studies 
by a full control of the spin and isospin degrees of freedom. 
Furthermore new topics will be included which were in a large part 
presented during this workshop.

A main point in the future must be an additional detection capability for
photons which will drastically improve the performance for present
investigations but mainly will open new possibilities.
Concerning e.g. the $\eta$ production an additionally detection of photons
would allow to remove the unavoidable background in missing mass
measurements by a tag on the $\eta \rightarrow \gamma \gamma$ decay.
Only the precise measurement of both charged and neutral particles
will give 
the opportunity to separate overlapping structures like the 
important but not fully understood $\Lambda$(1405) and the $\Sigma$(1385).
Further, production studies with more than one neutral ejectile 
will only be possible with a 4 $\pi$ charged and neutral particle
detector, 
and will guarantee an active, topical and exciting future at COSY.

\subsection{Aspects of NN induced Reactions}
\addtocontents{toc}{\hspace{2cm}(C. Hanhart)\par}

C. Hanhart

{\em Institut f\"ur Kernphysik, Forschungszentrum J\"ulich GmbH, Germany}

The goal of this presentation is to introduce the audience to the opportunities
and challenges of studying $NN$ induced reactions. 

As an example, the steps necessary to extract in a controlled way the
$\Lambda$--nucleon scattering lengths from the invariant mass spectrum of the
reaction  $pp\to pK^+\Lambda$ were discussed in detail: first of all a Dalitz
plot analysis is necessary to prove, that the structure in the spectrum
actually stems for the hyperon--nucleon interaction. Although in a large
momentum transfer reaction the production operator itself should not introduce
any significant energy dependence, the occurrence of resonances and their
interference with final state interaction effects might well distort the
signal \cite{eyrich}. In the next step polarization
observables are to be used to disentangle  the possible spin states. Once this
is done the scattering lengths can be extracted from the data directly
\cite{achotprep}.

As an additional illustration of the potential of polarization observables,
the use of double polarization observables as spin filters was discussed. For
details we refer to Ref. \cite{report}.

\newpage 

\section{Charge symmetry breaking}

Convenors: H. Machner and J. Niskanen

\subsection{Charge Symmetry Breaking-2003}
\addtocontents{toc}{\hspace{2cm}(Gerald A. Miller)\par}

Gerald A. Miller

{\em  University of Washington
  Seattle, WA 98195-1560}

Charge symmetry is the invariance of the QCD Lagrangian under the interchange
of $u$ and $d$ quarks, if one makes the reasonable assumption that the masses
these two light quarks are the same\cite{Miller:iz}. In isospin space,  
 this is invariance under rotations of 90$^\circ$ about the $y$ axis. This is
to be contrasted with isospin symmetry which is invariance under arbitrary
rotations\cite{emh69}. 

Charge symmetry breaking CSB arises only from the mass difference between the
 $u$ and $d$ quarks and from electromagnetic effects\cite{Miller:iz}. The
quark origin of the effects offers the special opportunity to examine the
direct influence of quarks in hadronic physics.  The positive nature of
$m_d-m_u$ causes the neutron to be more massive than the proton which gives 
the hydrogen atom its stability and causes the sun to shine. 

 CSB in the  nucleon-nucleon system causes the scattering length $a_{nn}$ to
be more negative than  $a_{pp}$ and a positive value of the difference neutron
and proton
analyzing powers at the (non-zero)
angle at which the analyzing power goes through
zero\cite{Miller:1994zh}. This 
establishes the existence of class III and IV potentials\cite{hm79}. The use of
a NN potential that reproduces the scattering length difference leads to an
explanation of the Okamoto-Nolen-Schiffer 
anomaly\cite{Miller:iz,Machleidt:2000vh} (within 10-20\%) in heavy
nuclei.

A new era in studies of CSB was initiated by studies of effective chiral
Lagrangians in which QCD is mapped onto onto a chiral Lagrangian that 
respects chiral symmetry\cite{Weinberg:1994tu}. This leads
to terms of the form $N^\dagger\pi_3 \boldtau\cdot\boldpi N$ and 
$N^\dagger\tau_3 \pi^2 N$ that cause charge symmetry to be severely violated
in $\pi^0$ interactions with nucleons.
Thus it is worthwhile to examine
 CSB in pion production processes\cite{vanKolck:2000ip}.
 Hence the new striking observations of
CSB in the reactions $np\to d \pi^0$\cite{Opper:2003sb}
 and $dd \to \alpha\pi^0$\cite{Stephenson:2003dv} are particularly exciting.
A consortion of theorists including: A. Fonseca, A. Gardestig, C. Hanhart,
C.~J.~Horowitz, G.~A.~Miller, A. Nogga and U. van Kolck has been formed
to compute the cross section for the $dd \to \alpha\pi^0$ reaction. 
C. Hanhart and J.~Niskanen are recomputing the forward-backward asymmetry
for the reaction $np\to d\pi^0$. These new experiments and the 
related studies planned at COSY will  provide new insights into the role
of the light quark mass difference in hadronic physics and into the origin of
mass.

I thank the USDOE for partial support of this work.

\subsection{Isospin Symmetry breaking studies at GEM }
\addtocontents{toc}{\hspace{2cm}(H. Machner)\par}

H. Machner for the GEM collaboration

{\it IKP, Forschungszentrum, J\"{u}lich, Germany}

The concept of isospin symmetry was introduced already in the
thirties into subatomic physics. Heisenberg \cite{Hei32} assumed
the neutron soon after its discovery and the proton to be two
different states of the otherwise identical nucleon. He introduced
a variable $\tau$, the isotopic spin in order to distinguish
neutrons and protons. Breit et al. and Cassen et al. \cite{BCP36}
claimed the strong part of nucleon-nucleon s-wave interactions to
be almost independent from isotopic spin. The latest summary with
these physics input is Ref. \cite{Henley_Miller}. Isospin symmetry
breaking was assumed to occur only via the Coulomb force. The next
summary \cite{MNS90} took differences between up and quark masses
into account. On the level of quarks isospin symmetry is broken by
\begin{equation}
\frac{{m_u  - m_d }}{{m_u  + m_d }} \approx  - 0.3
\end{equation}
Here we have used the current masses of the quarks. However, in
experiments we are dealing with real mesons and the isospin
symmetry violation effects can be estimated from their masses.
Quarks mass differences can be estimated from
\begin{equation}
\frac{m(K^0)-m(K^+)}{m(\omega)}=0.0051
\end{equation} with the $\omega$ mass as a typical hadron mass.
The mass difference between pions is believed to be almost only
due to Coulombian forces thus
\begin{equation}
\frac{m(\pi^+)-m(K^0)}{m(\omega)}=0.0059.
\end{equation}
Both effects seem to be of the same order of magnitude. The first
effect leads to $\pi^0-\eta$ mixing. In principle also the mixing
with $\eta'$ has to be included but is believed to be small.
Following a conjecture by C. Y. Yang we compare cross sections for
pion production in $pp$ and $np$ reaction \cite{Dro87,Hut90}.
Isospin violation is found in close to threshold values for the
total spin and at $\eta\approx 0.5$ for the anisotropy. In pion
production in $pd\to (A=3)+\pi$ we found angular distributions to
consist of two components: one coherent part which violates
isospin symmetry and an isotropic incoherent part which obeys
isospin symmetry to a high degree \cite{Betigeri,Abdel-Samad}. The
latter shows deviation from isospin symmetry in the vicinity of
the $\eta$-threshold \cite{Abdel-Bary}.

\subsection{The isospin forbidden $dd\to {^4}He\pi^0$ reaction
  near
threshold}
\addtocontents{toc}{\hspace{2cm}(E. Stephenson)\par}
  
E. Stephenson\footnote{E. Stephenson could not come to Bad Honnef. The essence
  of the talk was presented by A. Opper in front of her presentation.}

{\it Indiana University Cyclotron Facility, Bloomington, IN 47408.}

The results on the total cross section for the $d + d \to ^4He\pi^0$
reaction have been accepted for publication by Physical Review
Letters.  A preprint may be found in Ref. \cite{ed}.
   A review of the chiral Lagrangians is contained in Ref. \cite{KNM}.
   The cross sections are normalized against the work on $dp$
elastic scattering from the KVI that is now available in Ref.
\cite{KVI}.

\subsection{CSB in forward--backward Asymmetry in $pn\to
  d\pi^0$}
\addtocontents{toc}{\hspace{2cm}(A. Opper)\par}

A. Opper

{\em Department of Physics and Astronomy, Ohio University, Athens, OH 45701}

\subsection{Search for $\pi^0-\eta$ mixing}
\addtocontents{toc}{\hspace{2cm}(A. Magiera)\par}

  A. Magiera for the GEM Collaboration

  {\em Institute of Physics, Jagellonian University, Cracow}

  On the quark level the isospin symmetry is broken due to the $u$ and $d$
  quarks current mass difference and their electroweak interaction.  In some
  part this quark mass difference is responsible for the hadron mass
  difference. This effect can be partially described in the chiral
  perturbation theory or in the lattice calculations which lead to an
  estimation of the $u$ and $d$ quark mass ratio or their average mass (see
  Ref. \cite{PDG02} for review). Beside this the dynamical effects in the
  strong hadron interaction are induced by the quark mass difference and one
  of the most important is the light meson mixing.  Such effects can be
  calculated perturbatively and may be directly related to the quark mass
  difference. On the other hand the mixing may be also easily studied
  experimentally via the isospin or charge symmetry breaking processes.  The
  strength of the isospin or charge symmetry breaking depends on the mixing
  angle $\theta_m$.  Magnitude of this angle plays an important role in
  various processes where pions appear in the intermediate or final states
  (e.g. in the analysis of CP violation sources \cite{ECK00,GAR99a} or in the
  analysis of $B\rightarrow\pi\pi$ decays \cite{GAR99b}).  Various QCD based
  models predict the $\pi^0-\eta$ mixing angle in the range of 0.014-0.015 rad
  (see \cite{MAG00} for references). Experimental evidence comes from isospin
  or charge symmetry forbidden meson decays (see \cite{NEF92} for references).
  This mixing was also observed in hadronic reactions
  $\pi^+$d$\rightarrow$pp$\eta$ and $\pi^-$d$\rightarrow$nn$\eta$, where the
  $\pi^0-\eta$ mixing angle of 0.026$\pm$0.007 rad \cite{TIP01} was extracted.

The importance of the $\pi^0-\eta$ meson mixing was predicted for the
pd$\to^3$H$\pi^+$/$^3$He$\pi^0$ reactions \cite{MAG00}.  The isospin symmetry
predicts the ratio of the cross section for these reactions to be equal 2.
Due to meson mixing the real $\pi^0$ meson have a small admixture of isospin
0, which leads to some deviation of this ratio. A large effect of isospin
symmetry breaking due to meson mixing should appear for beam momenta close to
the $\eta$ meson production threshold and for outgoing products angles
corresponding to large relative proton-pion angles.  In order to extract the
mixing angle, the beam momentum dependence of the ratio was measured for five
beam momenta over a narrow region close to the $\eta$ production threshold. In
order to reduce the systematic uncertainty a simultaneous measurement was
performed.  Details on the experimental procedure and on data evaluation can
be found in Refs. \cite{BOJ02}.  The ratio of the cross sections as well as
the absolute values of the cross sections were obtained.  The measured values
of the cross sections ratio indicate the isospin symmetry breaking effects.
The present data lead to the ratio values larger than 2 at beam momenta far
above the $\eta$ threshold. This may be due to systematical uncertainty of the
data, however, it might be also attributed to isospin symmetry breaking
effects not related to the meson mixing. The largest effect may originate from
the differences in the wave functions of $^3$H and $^3$He nuclei \cite{KOH60}.
The simple model of Ref. \cite{MAG00} does not contain all the effects that
may lead to a deviation of the cross sections ratio from the isospin symmetry
predicted value of 2.  When comparing the model predictions to the
experimental results this effect as well as systematic uncertainty of the data
were combined together and treated as an overall normalization factor N.  In
the fitting procedure the global minimum $\chi^2/n$ was found for N=1.15 and
$\theta_m=0.006\pm0.005$ rad.  More details on the results and model analysis
may be found in \cite{ABD03}. The analysis with more advanced model based on
K-matrix formalism \cite{GRE02} might deliver more exact information on mixing
angle.

Much better for the interpretation in terms of meson mixing will be the
dd$\to^4$He$\pi^0$ reaction. When measured at the $\eta$ production threshold,
the cross section for this reaction is directly proportional to the
$\theta_m^2$ and the known cross section for dd$\to^4$He$\eta$ reaction. Such
a measurement was proposed at the COSY accelerator \cite{MAG97}. The cross
section for charge symmetry forbidden dd$\to^4$He$\pi^0$ reaction is expected
to be very small \cite{MAG00}. Therefore the detection of all ejectiles is
necessary in order to reduce background substantially. It would necessitate in
the photon detector for measurements of the $\pi^0$ decay products. The
construction of such detector is planned at COSY \cite{HEJ02}.

The charge symmetry breaking part of the nucleon-nucleon interaction induced
by $\pi^0-\eta$ mixing may be studied in pn$\to$d$\pi^0$ reaction. The center
of mass forward-backward asymmetry of the cross section have to be measured.
It is proposed to study this reaction as a quasi-free reaction
pd$\to$d$\pi^0$p on the neutron in the deuteron. The advantage of such
measurement is the simultaneous detection of pd$\to$d$\pi^+$n reaction, what
allows for continuous control of the systematic uncertainties. Even better
will be the investigation of the asymmetry of the analyzing power when the
polarized COSY proton beam is used. With the proton beam momentum of 750 MeV/c
the measurement may be performed in the region where the analyzing power
reaches maximum of -0.55 and the predicted asymmetry is largest. The advantage
of the asymmetry investigation is that the measurement is relative and no
acceptance corrections influence the final result. In order to unambiguously
identify both quasi-free reactions it is necessary to measure angle and energy
of deuteron and proton (for pd$\to$d$\pi^0$p) or deuteron and $\pi^+$ (for
pd$\to$d$\pi^+$n) and identify the deuteron. Such a measurement is possible
with the Germanium Wall detector \cite{BET99}. This detector with large
angular acceptance, very good angular and energy resolution and the axial
symmetry is ideally suited for such investigations with polarized beam.


\subsection{Charge symmetry breaking meson production}
\addtocontents{toc}{\hspace{2cm}(J. A. Niskanen)\par}

J. A. Niskanen 

{\em University of Helsinki}

Breaking of the isospin symmetry, in particular charge
symmetry breaking, has been studied for a long time as the
difference between the $pp$ and $nn$ systems and mirror
nuclei. During the past decade the study has increasingly
concentrated on the $np$ interaction, where actual mixing
of isospin values zero and one can take place. This makes it
possible, for example, to produce pions also from the initial
nuclear
isospin zero component in $np \rightarrow d\pi^0$, which in
turn can be observed as an asymmetry about 90$^\circ$ in the
differential cross section \cite{opper}. Without an interference of the
two isospin states the cross section would be symmetric.

Charge symmetry breaking is intimately related to the up-
and down-quark mass difference in QCD (and also to the
electromagnetic interaction). This effect has recently been
explicitly
incorporated as an isospin breaking effective pion-nucleon
interaction in a calculation of $np \rightarrow d\pi^0$ close
to threshold, and presently it appears to be the largest contribution
so far discussed \cite{kolck}.
Another important contribution previously
investigated is $\eta\pi$ meson mixing, which enters with the
 opposite sign \cite{fbs}, and so
the two are qualitatively different. The TRIUMF experiment
\cite{opper} gives the sign favouring the dominance of isospin
breaking pion rescattering and will be discussed elsewhere in
these proceedings.

This talk discusses theoretical consequences of the principal
mechanisms mainly in mesonic inelasticities along with their
uncertainties. Some connections to the more complex isospin
forbidden reaction $dd\rightarrow ^4$He$\pi^0$ are presented.


\subsection{$a_0-f_0$ mixing}
\addtocontents{toc}{\hspace{2cm}(C. Hanhart)\par}

C. Hanhart

{\em Institut f\"ur Kernphysik, Forschungszentrum J\"ulich GmbH, Germany}

Regardless the efforts of recent years to resolve the nature of the light
scalar resonances, no consensus is found up to date if they are predemonantly
of $\bar qq$, $\bar q\bar q qq$ or molecular nature. It is expected, that the
magnitude of the charge symmetry breaking (CSB) $a_0-f_0$ mixing amplitude, predicted to be unusually
large \cite{achasov}, will help to shed additional light on this important question.

In complete analogy to what was discussed e.g. by G.A. Miller at this
conference, also the forward backward asymmetry in $pn\to d\pi\eta$ as well as
$dd\to \alpha \pi\eta$ vanish in the absence of CSB and are thus clean
experiments to quantify the $a_0-f_0$ mixing \cite{sascha}. The fact that
$a_0$ and $f_0$ are rather narrow and overlapping (their nominal masses
according to the particle data group differ by a few MeV only) in addition
guarantees  that the CSB are to be completely dominated by the mixing effects
\cite{utica} and thus can be extracted even without detailed knowledge about
the production operator. This is in vast contrast to the situation for pion
production \cite{jouni}.

\subsection{Mixing of the pseudoscalar mesons}
\addtocontents{toc}{\hspace{2cm}(Peter Kroll)\par}

Peter Kroll

{\it Fachbereich Physik, Universit\"at Wuppertal,\\
D-42097 Wuppertal, Germany}

The mixing parameters of the pseudoscalar meson states and their decay
constants are calculated within QCD to first order of flavour symmetry
breaking on exploiting the divergencies of the axial vector currents
which embody the U(1)$_A$ anomaly. Starting point of this analysis is
the quark flavour basis and the assumption that the decay constants in
that basis follow the pattern of particle state mixing. The proposed
mixing scheme is tested against experiment and corrections to the
first order mixing parameters are determined
phenomenologically. It also allows for an estimate of isospin
violations for pseudoscalar mesons. It is found for instance that the $\eta$
admixture to the $\pi^0$ amounts to $1.4\%$.
The talk is based on Refs.\ \cite{F1,F2,F3}.


\newpage \section{$\eta$ Production}

Convenors: P. Moskal and C. Hanhart

\subsection{Eta production}
\addtocontents{toc}{\hspace{2cm}(P.~Moskal)\par}

{P.~Moskal$^{1,2}$ for the COSY-11 collaboration}

{\it $^1$ Institut f{\"u}r Kernphysik, Forschungszentrum J\"{u}lich, D-52425 J\"ulich, Germany}
            
{\it $^2$ Institute of Physics, Jagellonian University, PL-30-059 Cracow, Poland}

  Results of the study of the $\eta$pp dynamics presented in the talk 
  are avaiable via the e-Print Archive server~\cite{ppeta}.
  Here we abridged the text to  few remarks 
  --~from the introduction of this article~--
  concerning investigations
  of the interaction between the $\eta$ meson and the proton.

  Due to the short live time of the flavour-neutral 
  mesons~(eg. $\eta$ or $\eta^{\prime}$), the study of their 
  interaction with nucleons or with other mesons is at present not feasible
  in direct scattering experiments. One of the methods permitting such 
  investigations is the production of a meson in the nucleon-nucleon
  interaction close to the kinematical threshold or in kinematics regions
  where the outgoing particles possess small relative velocities.
  In the last decade  major
  experimental~\cite{etap_data,hibou,eta_data}
  and theoretical~\cite{eta_theo,nakayamaa} efforts
  were concentrated on the study of the creation of
  $\eta$ and  $\eta^{\prime}$ mesons
  via the hadronic interactions~\cite{review}.
   Measurements 
  have been performed
  in the vicinity of the kinematical threshold where only 
  a few partial waves in both
  initial and final state
  are expected to contribute to the production process.
  
  The determined energy dependences of the total cross section
  for $\eta^{\prime}$~\cite{etap_data,hibou} and 
  $\eta$~\cite{hibou,eta_data} 
  mesons in  proton-proton collisions
  reveals that the proton-proton FSI enhances the total cross section by more than an order
  of magnitude for low excess energies.
  Interestingly, in the case of the $\eta$  meson the increase of the
  total cross section for very low and very high energies is much larger than expected
  from the  $^1S_0$ final state interaction between protons.
  The excess at higher energies
  can be assigned to the significant onset of higher partial waves, and
  the influence of the attractive
  interaction between the $\eta$ meson and the proton 
  could be a plausible explanation for the enhancement at threshold.\\
The interaction between particles depends on their relative momenta.
Therefore it should show up as  modification
of the phase-space abundance in the kinematical regions where the outgoing
particles possess small relative velocities.
Only two  invariant masses of the three subsystems are independent
and therefore the entire  accessible information about the final
state interaction of the three-particle system can be presented in the
form of the Dalitz plot.  To some extent this information is still 
available from the projections of the phase space population onto the 
invariant mass distributions.
   These have been recently
    determined at Q~=~15~MeV by the COSY-TOF collaboration~\cite{TOFeta} and
    at Q~=~15.5~MeV by the COSY-11 group~\cite{ppeta}.
  The structure of the observed spectra may indicate  a non-negligible contribution from the P-waves in the 
  outgoing proton-proton subsystem~\cite{nakayamaa}.
  The amount of the P-wave admixture  derived from the proton-proton invariant mass
  distribution  leads to a good description of the excitation function 
  at higher excess energies
  while at the same time  it spoils significanly 
  the agreement with  the data at low values of Q~\cite{nakayamaa}.
  In contrast to the P-wave contribution
  the three-body treatment~\cite{Fixprivate} of the $pp\eta$
  system
  leads to an even larger enhancement of the cross section near threshold 
  than that based on 
  the Ansatz of the factorization of the proton-proton and proton-$\eta$ interactions.
  For the complete understanding of the low energy $pp\eta$ dynamics,
  a rigorous three-body approach to the $pp\eta$ system is required.
  A herald of such calculations have been already reported~\cite{Fixprivate,deloff}.

\subsection{Potential of the COSY-11 facility}
\addtocontents{toc}{\hspace{2cm}(Joanna Przerwa)\par}

Joanna Przerwa for the COSY-11 collaboration

{\em Institute of Physics, Jagellonian University, Cracow, PL-30-059 Poland}

One of the main goal of the COSY-11 collaboration is to study the eta meson
interaction with nucleons and the mechanism of the $\eta$ production in
different isospin channels. Over the last few years, we have performed several
experimental investigations on the close to treshold $\eta$ production in
nucleon-nucleon collisions using unpolarized as well as polarized beam.
Experiments of the $ p p \to p p $ $\eta$,
$ p d \to p d $ $\eta$  and $ p n \to p n $ $\eta$  are based on the
four-momentum registration of outgoing nucleons and nuclei. $\eta$ meson is
identified via the missing mass technique. 

The internal facility COSY-11 [1] installed at the Cooler Synchrotron COSY [2]
is shown in the figure below.

Results of the performed measurements and the details of the experimental
method can be found in references~[3].


\begin{figure}[H]
  \parbox{0.5\textwidth}{\epsfig{file=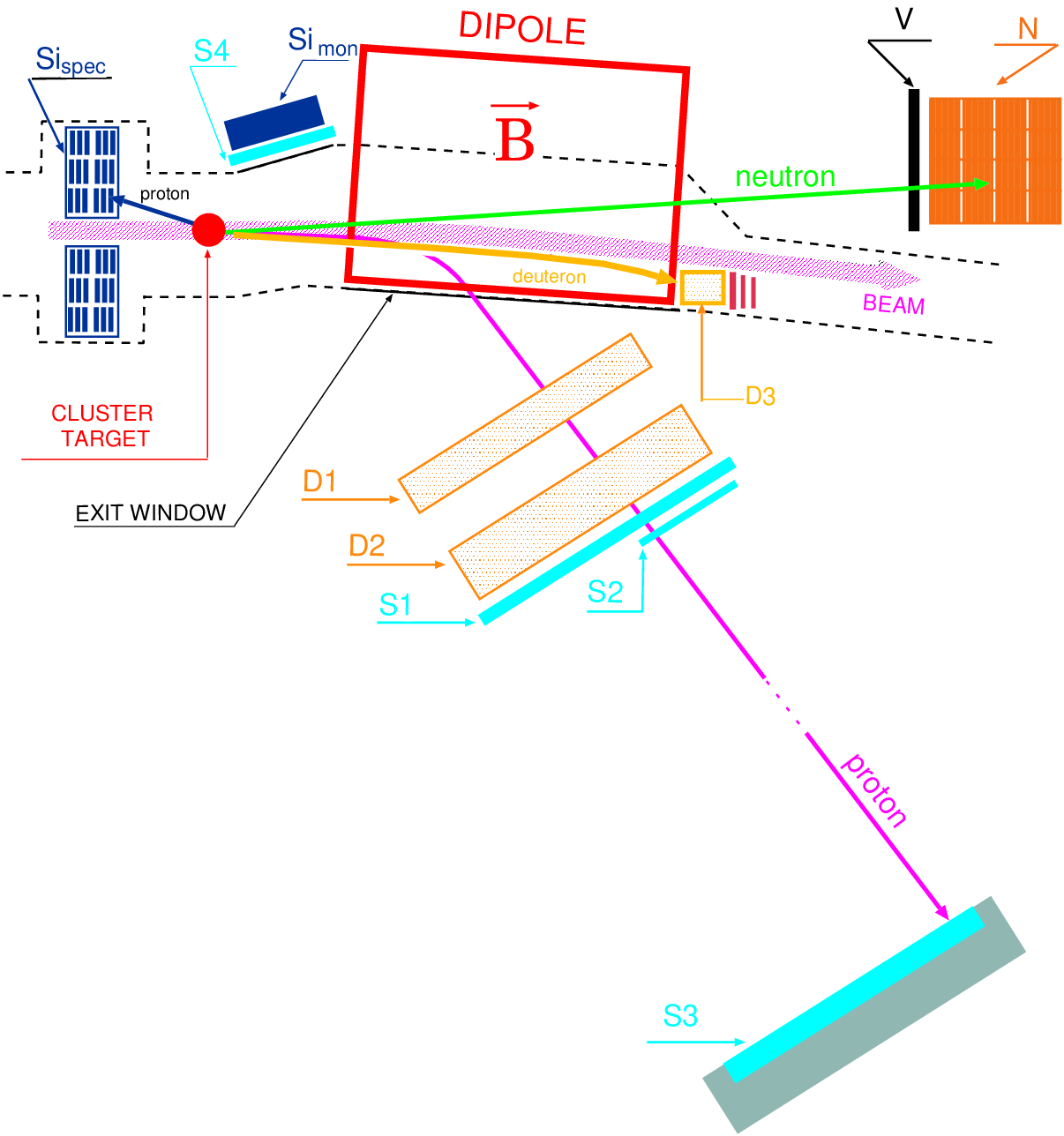,width=0.5\textwidth,angle=0.}}
  \parbox{0.55\textwidth} {\caption{\em Scheme of the COSY-11 detection
      system. Protons are registered in two drift chambers D1, D2 and in the
      scintillator hadoscopes S1, S2, S3. An array of silicon pad detectors $
      ( Si_{spec})$ is used for the registration of the spectator protons.
      Neutrons are registered in the neutron modular detector(N). In order to
      distinguish neutrons from charged particles veto detector is used.
      Deuterons are registered in deuteron chamber D3.  }}

\end{figure}

\vspace{-1cm}

\vspace{0.5cm}
{\large \bf References}

\begin{flushleft}

[1] S.Brauksiepe et al., Nucl. Instr. $\&$ Meth. \textbf{A 376} $397-410$ ($1996$) \\

[2] R.Maier, Nucl. Instr. $\&$ Meth. \textbf{A 390} $1-8$ ($1997$)\\

[3] P.Winter et al., Phys.Lett. \textbf{B 544} ($2002$) $251-258$\\

\hspace{0.45cm} P.Moskal et al., Phys.Lett. \textbf{B 517} ($2001$) $295-298$\\

\hspace{0.45cm} P.Moskal et al., Phys.Lett. \textbf{B 482} ($2000$) $356-362$\\

\hspace{0.45cm} P.Moskal et al., Phys.Lett. \textbf{B 474} ($2000$) $416-422$\\

\hspace{0.45cm} J.Smyrski, Phys.Lett. \textbf{B 474} ($2000$) $182-187$\\

\hspace{0.45cm} P.Moskal et al., Phys.Lett. \textbf{B 420} ($1998$) $211-216$\\

\hspace{0.45cm} C.Quentmaier et al., Phys.Lett. \textbf{B 515} ($2001$) $276-282$\\

\hspace{0.45cm} M.Wolke et al., Nucl.Phys.News \textbf{9} ($1999$) $27-32$\\

\hspace{0.45cm} S.Sewerin , Phys.Rev.Lett. \textbf{83} ($1999$) $2682-685$\\

\hspace{0.45cm} J.Balewski et al., Eur.Phys.Jou. \textbf{A 2} ($1998$) 1, $99-104$\\

\hspace{0.45cm} J.Balewski et al., Phys.Lett. \textbf{B 420} ($1998$) $211-216$

\end{flushleft}

\subsection{Recent Results from TOF}
\addtocontents{toc}{\hspace{2cm}(E. Roderburg)\par}

E. Roderburg

{\it IKP, Forschungszentrum, J\"{u}lich, Germany}

\subsection{Higher Partial Waves in $pp\to pp\eta$}
\addtocontents{toc}{\hspace{2cm}(K. Nakayama)\par}

{K. Nakayama$^{a,b}$, J. Haidenbauer$^b$, C. Hanhart$^b$, and J. Speth$^b$}

{\em $^a$Department of Physics and Astronomy, University of Georgia,
Athens, GA 30602, USA \\
$^b$Institut f\"ur Kernphysik, Forschungszentrum J\"ulich,
D-52425, J\"ulich, Germany }

It is shown that most of the available data on the $pp\rightarrow pp\eta$ reaction
\cite{EXP}, including the invariant mass distributions in the 
$pp\rightarrow pp\eta$ reaction recently measured at COSY \cite{Roderburg,Moskal},
can be understood in terms of the partial-wave amplitudes involving final $pp$ $S$
and $P$ states and the $\eta$ meson $s$-wave. This finding, together with the fact
that results within a meson--exchange model are especially sensitive to the details 
of the excitation mechanism of the $S_{11}(1535)$ resonance, demonstrates the
possibility of investigating the properties of this resonance in $NN$ collisions. The
spin correlation function $C_{xx}$ is shown to disentangle the $S$- and $P$-wave
contributions. It is also argued that spin correlations may be used to help constrain
the contributions of the amplitudes corresponding to the final $pp$ $^3P_0$ and $^3P_2$
states. Details of the present results may be found in Ref.\cite{Nak}.

\subsection{Three-body treatment of $\eta$ production in $pp$-collision}
\addtocontents{toc}{\hspace{2cm}(A.\ Fix)\par}

{A.\ Fix and H.\ Arenh\"ovel}

{\em Institut f\"ur Kernphysik, Johannes Gutenberg-Universit\"at Mainz,
D-55099 Mainz, Germany}

Effect of interaction between the final particles in the reaction $pp\to
pp\eta$ is studied. The calculation is based on the three-body approach to the
$\eta NN$ system as is presented in \cite{FA1}. Three-body theory being a
natural theoretical frame for treating correctly the $\eta NN$ interaction
provides reasonable quantitative explanation of the energy dependence of the
total cross section measured for $pp\to\eta pp$ in the region of the c.m.\
excess energy $Q=40$ MeV. The obtained results are however subject to
correction because of Coulomb forces.

\subsection{The reaction $np\to\eta d$ near threshold}
\addtocontents{toc}{\hspace{2cm}(H. Garcilazo)\par}

{H. Garcilazo}

{\em Escuela Superior de F\'\i sica y Matem\'aticas

Instituto Polit\'ecnico Nacional, Edificio 9,
07738 M\'exico D.F., Mexico}

The reaction $np\to\eta d$ has been measured recently in the region near 
threshold [1]. Since the cross section shows a strong
enhancement at threshold, as compared with the predictions based on a 
two-body phase space, it has been
speculated that this could be a signal for an $\eta NN$ quasibound state,
predicted some time ago [2]. However, 
the solutions of the three-body equations for
$\eta d$ elastic scattering do not support the existence of a 
$\eta NN$ quasibound state but rather indicate the presence of a quasivirtual
state [3-5].

In Ref. [6] we constructed seven diferent separable-potential
models (labeled 0 to 6) of the coupled $\eta N$-$\pi N$-$\sigma N$ system
in the $S_{11}$ channel 
(the $\sigma$ was taken as a stable particle with $m_\sigma=2m_\pi$)
by fitting the seven $\eta N\to\eta N$
amplitude analyses of Refs. [7-10] and the
$\pi^- p\to\eta n$ cross section in the region of the $S_{11}$ resonance.
The main feature that distinguishes these models among themeselves is
the value predicted for the real part of the $\eta N$ scattering
length which runs from Re $a_{\eta N}=0.42$ fm for model 0 to
Re $a_{\eta N}=1.07$ fm for model 6. For the nucleon-nucleon interaction
in the $^3S_1$ channel we used the rank-1 separable potential [11]
that generates the same deuteron wave function as the Paris potential
(so-called PEST potential). We found that
the quasivirtual state moves closer and closer to the real axis when
the $\eta N$ interaction gets stronger, i.e., when Re $a_{\eta N}$
becomes larger [5].
The pole moves from $q=-0.29-i0.44$ fm to $q=-0.06-i0.16$ fm
when one goes from model 0 to model 6.
The fact that the pole lies in the third quadrant of the
momentum plane means that it lies in the second Riemann sheet of 
the energy plane.

We calculated the $np\to\eta d$ cross section
using the three-body formalism described above.
We found that the shape of the 
$np\to\eta d$ cross section within a wide energy range can only be explained
by an $\eta N$ interaction model corresponding to a small scattering length
(e.g. the J\"ulich model [7]) and this is independent of the meson production
mechanism considered.

\vspace{0.5cm}
{\large \bf References}

\medskip      
\noindent 
[1] H. Cal\'en {\it et al.},   
               Phys. Rev. Lett. {\bf 80}, 2069 (1998).

\noindent 
[2] T. Ueda, Phys. Rev. Lett. {\bf 66}, 297 (1991).
               
\noindent 
[3] S. Wycech and A. M. Green, Phys. Rev. C {\bf 64},
               045206 (2001).

\noindent 
[4] A. Fix and H. Arenh\"ovel, Nucl. Phys. A {\bf 697}, 277 (2002).

\noindent 
[5] H. Garcilazo, Phys. Rev. C {\bf 67}
               067001 (2003).

\noindent 
[6] H. Garcilazo and M. T. Pe\~na, Phys. Rev. C {\bf 66}
               034606 (2002).

\noindent 
[7] A. Sibirtsev {\it et al.},
                Phys. Rev. C {\bf 65},
               044007 (2002).

\noindent 
[8] M. Batini\'c {\it et al.},
                Physica Scripta {\bf 58}, 15
               (1998). 

\noindent 
[9] A. M. Green and S. Wycech, Phys. Rev. C {\bf 55},
               R2167 (1997).

\noindent 
[10] A. M. Green and S. Wycech, Phys. Rev. C {\bf 60},
               035208 (1999).

\noindent 
[11] H. Zankel, W. Plessas, and J. Haidenbauer,  
               Phys. Rev. C {\bf 28}, 538 (1983).

\subsection{Spin Observables 
in
Meson Production in NN Collisions}
\addtocontents{toc}{\hspace{2cm}(Jan Balewski)\par}

Jan Balewski

{\it IUCF, Indiana, USA}


Spin observables are very useful probes of the NN interaction. This talk describes
advantages of using both polarized beam and target to access variety of single
and double spin observables. The experimental technique of kinematically
complete $\stackrel{\rightarrow}p \stackrel{\rightarrow}p \rightarrow p p
\pi^0$ and $ \stackrel{\rightarrow}p \stackrel{\rightarrow}p\rightarrow p
\pi^+$ measurement are given. The significance of higher partial waves for
meson production above threshold is shown on the example of $\phi$-independent
$\Delta\sigma_L$ and $\Delta\sigma_T$. For inclusive $\pi$ measurements the
various $\phi$-distribution are key to disentangle single ($A_y$) and double
($C_{xx}, C_{yy}, C_{xy}$) spin observables.  From 3 independent ratios of
$\phi$-dependent yields $N_{++},N_{+-},N_{-+},N_{--} $, one can extract
uniquely the observables.

Simultaneous measurement of  directions of two outgoing particles 
allows for determination of more exotic single spin observable  $A_z(\Delta\phi)$.

\vspace{0.5cm}
{\large \bf References}

H.O.Meyer et al., {\it Phys. Lett. B} 480, 7, 2000.

H.O.Meyer et al., {\it Phys. Rev. C} {\bf 63}, 064002, 2001.

W.W.Daehnick et al., {\it Phys. Rec. C.} {\bf 65}, 024003, 2002.

\subsection{Double Polarized Observables at ANKE}
\addtocontents{toc}{\hspace{2cm}(F. Rathmann)\par}

F. Rathmann

{\it IKP, Forschungszentrum, J\"{u}lich, Germany}

The study of near--threshold meson production in $pp$ collisions 
involving polarized beams and polarized targets offers
the rare opportunity to gain insight into short--range
features of  the nucleon--nucleon interaction.
The Cooler Synchrotron COSY at  FZ--J\"ulich is a unique environment 
to perform such studies.  Ideally, measurements of polarization observables 
require a cylindrically  symmetrical detector, capable to measure the 
momenta and the  directions of outgoing charged hadrons.
  
In the case of the reaction $\vec{p}\vec{p} \to ppn$, a measurement of some
polarization observables seems feasible using the ANKE magnetic dipole
spectrometer in the Q-range of a few ten MeV. If one assumes a purely
vertically polarized beam ($P_x=P_z=0$) and a purely vertically polarized
target ($Q_x=Q_z=0$), the spin-dependent cross section reads after integration
over the relative momenta $\vec{k}_{NN}$
\begin{eqnarray*}
\sigma/\sigma_0=1 & + & A_y^B \cdot P_y \cdot \cos(\phi)+A_y^T \cdot Q_y \cdot \cos(\phi)\\  
                        & + & \frac{1}{2}(A_{xx}+A_{yy}) \cdot P_y \cdot Q_y  \\
                        & + & \frac{1}{2}(A_{xx}-A_{yy}) \cdot P_y \cdot Q_y \cdot \cos(2\phi)\;,
\end{eqnarray*}
where $\sigma_0$ is the spin-independent cross section and $\phi$ corresponds
to the azimuth of the emitted $\eta$ meson \cite{meyer}.  The azimuthal
acceptance of ANKE near $\phi=180^\circ$ ($\cos(\phi)=-1$, $\cos(2\phi)=1$)
together with identical beam and target particles ($A_y^T=A_y^B=A_y$) leads to
a further simplification. The relative luminosity could either be measured
from inclusive protons scattered near $0^\circ$ or from $pp$ elastic
scattering. Beam and target polarizations could be determined from known
analyzing powers of $pp$ elastic scattering. Thus at $\phi=180^\circ$ a
measurement of $A_y$, $A_{xx}$ would be possible. At any $\phi$, a measurement
of $A_y$, $A_\Sigma=A_{xx}+A_{yy}$ and $A_\Delta=A_{xx}-A_{yy}$ seems
feasible.

\newpage \section{Vector Mesons}

Convenor: M. Hartmann and A. Wirzba

\subsection{Vector meson production - an overview}
\addtocontents{toc}{\hspace{2cm}(Kai-Thomas~Brinkmann)\par}


Kai-Thomas~Brinkmann \\

{\it Institut f\"ur Kern- und Teilchenphysik,
 Technische~Universit\"at~Dresden, D-01062 Dresden, Germany}

Vector meson production in nucleon-nucleon collisions is an important tool in
the study of the nucleon-nucleon interaction at short range. With masses in
the range of 760 MeV for the isovector $\rho$ and 780 MeV for the isoscalar
$\omega$, whose quark content is dominantly $u\bar{u}$\,$\oplus$\,$d\bar{d}$
while the $s\bar{s}$ is contained in the $\phi$ with 1020 MeV, these mesons
are easily produced at COSY.

Early studies of pion-induced $\omega$ production on nucleons
\cite{bin},\cite{key} showed an exotic behavior of the reaction amplitude near
threshold. Recently, this has been attributed to the contributions of
resonances in the production of $\omega$ mesons \cite{pen} and \cite{tit1}
pointed out the importance of resonances and demonstrated that they may
account for differences between $\omega$ and $\phi$ production.

Vector meson production in nucleon-nucleon collisions has gained increasing
interest in recent years. Several experiments at Saturne have addressed
$\omega$ (SPESIII \cite{hib}, DISTO \cite{DIS98}, \cite{DIS00}) as well as
$\phi$ and $\rho$$^0$ production \cite{DIS99}, \cite{DIS02} in proton-proton
collisions. Early theoretical work discussed these cross sections and earlier
measurements at higher excess energies \cite{fla} in the meson exchange
framework \cite{Cas93}, \cite{Sib96}, \cite{Sib00}. Very detailed calculations
have focussed on the near-threshold behavior of the cross section but fail to
reproduce the excitation function at higher energies \cite{Nak98},
\cite{Nak99}, \cite{Nak00}. A recent measurement of TOF at COSY \cite{TOF00}
closed the gap between 30 and 300 MeV in excess energy not covered by the
Saturne experiments. The total cross sections confirm the overall trend and
magnitude of the earlier data. The angular distribution of the $\omega$ mesons
in the overall center-of-momentum system is fairly anisotropic at
$\epsilon$~=~173 MeV in line with the DISTO data at 300 MeV, which indicates a
dominant production through nucleonic currents. This is however at variance
with the DISTO findings in the pp$\rightarrow$pp$\phi$ channel which is
perfectly isotropic at about 90 MeV excess energy. Meanwhile, very detailed
theoretical analyses of the pp$\rightarrow$pp$\omega$ data are available
\cite{Tsu03} which allow predictions to be tested in future COSY experiments.
These calculations also demonstrate the importance of resonance contributions,
but a quantitative analysis will need a much larger body of data. In
particular, additional differential observables such as spectra of invariant
masses in two-body subsystems will provide a valuable test of the model
predictions. Additional measurements at the TOF spectrometer are under way
(see talk of M. Schulte-Wissermann), other COSY experiments study the
production of $\omega$ mesons also in other isospin channels such as
pn$\rightarrow$d$\omega$ (see talks of I. Lehmann and M. Hartmann,
\cite{ANK03}).

Vector meson production in bound systems has only been studied in very few
experiments but exhibits very interesting features. The
pd$\rightarrow$$^3$He$\omega$ reaction was studied in an inclusive experiment
at very backward $\omega$ angles only \cite{Wur95}. The extraction of the
exitation function of the reaction amplitudes assumed isotropic angular
distributions. This is one possible explanation for the steep fall-off of the
amplitudes near threshold, but other explanations such as interactions of the
decay pions with the nucleus or the influence of resonances are not ruled out.
An exploratory measurement at the COSY-TOF spectrometer has demonstrated the
feasibility of experiments covering the full $^3$He angular distribution in
coincidence with the charged pions from $\omega$ decay. These data will
clarify the situation greatly. They can also be compared to exclusive
measurements of $\phi$ production cross sections in pd fusion which were
performed with the MOMO setup at the Big Karl spectrometer at COSY
\cite{Bel02}.

In summary, vector meson production is an inportant tool in the investigation
of nucleon-nucleon interactions. The production of vector mesons with
hadroninc probes on nucleons and nuclei yields information complementary to
other experimental approaches. COSY gives access to many degrees of freedom
which have not been adequately explored to date.



%
%
%

\subsection{Vector Meson Production in $NN$ Collisions}
\addtocontents{toc}{\hspace{2cm}(K. Nakayama)\par}

{K. Tsushima, and K. Nakayama}

{\em Department of Physics and Astronomy, University of Georgia,
Athens, Georgia 30602, USA.}

Using a relativistic effective Lagrangian at the hadronic level,
near-threshold $\omega$ and $\phi$ meson productions in proton proton
($pp$) collisions, $p p \to p p \omega/\phi$, are studied within
the distorted wave Born approximation \cite{TN}. Both initial and final state
$pp$ interactions are included. In addition to total cross section data
\cite{Hibou,DISTO,COSY}, both $\omega$ and $\phi$ angular distribution data
\cite{DISTO,COSY} are used to constrain further the model parameters. For the
$p p \to p p \omega$ reaction we consider two different
possibilities: with and without the inclusion of nucleon resonances.
The nucleon resonances are included in a way to be consistent with the
$\pi^- p \to \omega n$ reaction. It is shown that the inclusion of
nucleon resonances can describe the data better overall than
without their inclusion. However, the SATURNE data in the range of
excess energies $Q < 31$ MeV are still underestimated by
about a factor of two. As for the $p p \to p p \phi$ reaction it is found
that the presently limited available data from DISTO can be reproduced by
four sets of values for the vector and tensor $\phi NN$ coupling constants. 
Further measurements of the energy dependence of the total cross
section near threshold energies should help to constrain
better the $\phi NN$ coupling constant.

\subsection{The production of massive photons in meson-baryon interactions}
\addtocontents{toc}{\hspace{2cm}(Madeleine Soyeur)\par}

Madeleine Soyeur $^a$, Matthias F.M. Lutz $^{b,c}$, Bengt Friman $^{b,c}$

{\it $^a$DAPNIA/SPhN, CEA/Saclay,
     F-91191 Gif-sur-Yvette Cedex, France}

{\it $^b$GSI, Planckstrasse 1, D-64291 Darmstadt, Germany

 $^c$Institut f\"ur Kernphysik, T U Darmstadt, D-64289 Darmstadt, Germany}

We consider the $\pi \, N \rightarrow e^+e^- N$ and
$\gamma\, N \rightarrow e^+e^- N$ reactions
below and in the vicinity of the
$\rho^0$- and $\omega$-meson thresholds. Using the Vector Meson Dominance (VMD)
assumption, the amplitudes describing these processes are related to
the meson-nucleon amplitudes $\pi \, N \rightarrow V \, N$ and
$V \, N \rightarrow V \, N$, where V stands for $\rho^0$- and $\omega$-mesons [1-3].
We restrict our discussion to
$e^+e^-$ pair invariant masses ranging from 0.4 to 0.8 GeV. In these kinematics,
the production of $e^+e^-$ pairs
off proton and neutron targets is closely linked to the coupling of vector mesons to
low-lying baryon resonances.
The
$\pi \, N \rightarrow e^+e^- N$ reaction is determined
by the $\pi\, N \rightarrow \rho^0 N$ and $\pi N
\rightarrow \omega N$ amplitudes. In addition to the photoproduction of vector mesons
materiali\-zing into $e^+e^-$ pairs,
the $\gamma\, N \rightarrow e^+e^- N$ reaction
involves  a large contribution from the
production of $e^+e^-$ pairs through the Bethe-Heitler process.
The dynamics of the $\pi \, N \rightarrow e^+e^- N$ and
$\gamma\, N \rightarrow e^+e^- N$ reactions reflects in
quantum interference patterns. Both reaction cross sections are
sensitive to $\rho^0$-$\omega$ interferences. The quantum interference
between vector meson $e^+e^-$ decays and Bethe-Heitler pair production
plays also an important role in determi\-ning the
$\gamma\, N \rightarrow e^+e^- N$ cross section.
\par\medskip\noindent
We have calculated consistently the $\pi\,N \rightarrow e^+e^- N$
and $\gamma\,N \rightarrow e^+e^- N$
reactions [2,3] in the framework
of a recent relativistic and unitary coupled-channel approach to
meson-nucleon scattering [1].
 The study of the $\pi\,N \rightarrow e^+e^- N$
and $\gamma\,N \rightarrow e^+e^- N$
reactions aims principally at gaining understanding of the structure
of the $\pi\,N \rightarrow \rho^0 N$, $\pi\,N \rightarrow \omega N$,
$\gamma\,N \rightarrow \rho^0 N$ and $\gamma\,N \rightarrow \omega N$
amplitudes arising from the presence of baryon resonances in this energy
range and reflected in $e^+e^-$ pair spectra.
Measurements
of the $\pi \, N \rightarrow e^+e^- N$ and
$\gamma\, N \rightarrow e^+e^- N$ cross sections, planned at GSI and JLab,
will therefore shed light on the role of baryon resonances in these processes
and be sensitive to their couplings to vector meson
nucleon channels.

\vspace{0.5cm}
{\large \bf References}

\par\noindent
1. M.F.M. Lutz, Gy. Wolf and B. Friman, Nucl. Phys. \textbf{A706}, 431 (2002).\par\noindent
2. M.F.M. Lutz, B. Friman and M. Soyeur, Nucl. Phys. \textbf{A713}, 97 (2003).\par\noindent
3. M.F.M. Lutz and M. Soyeur, in preparation.\par\noindent

\subsection{Vector meson production with the DISTO spectrometer}
\addtocontents{toc}{\hspace{2cm}(Ingo Fr\"ohlich)\par}

I. Fr\"ohlich for the DISTO collaboration.

{\it  Physikalisches Institut , 
        Justus-Liebig-Universit\"at,Gie{\ss}en, \\
        Heinrich-Buff-Ring 16,  D-35392 Gie{\ss}en 
    }

Angular distributions for vector meson production have been
proposed~\cite{nak} as a sensitive observable to extract the strength of
nucleonic and mesonic currents responsible in meson exchange models for
the vector meson production near the threshold. Therefore, it is
particularly interesting for various reaction models to compare the
differential cross sections for the production of $\rho^0$, $\omega$ and
$\phi$ mesons. In addition, total and differential cross sections are
important for the understanding of OZI rule violations and the
interpretation of dilepton spectra and the question of in-medium
modifications of hadrons.

With the DISTO spectrometer~\cite{NIM} we measured the production of light
vector mesons in the $pp \rightarrow ppX$
  reaction at 3.67 GeV/c~\cite{rho,JIM3} with the decay $\rho (\omega) \to
\pi^+ \pi^- (\pi^0)$ and $\phi\to K^+ K^-$. Momentum reconstruction has
been done with multi wire proportional chambers and a magnetic dipole
field, particle identification with water \v Cerenkov detectors. A model
independent acceptance correction has been applied. The angular
distribution for $\rho^0$ production exhibits strong forward peaking,
suggesting a dominance of the nucleonic current. The angular distribution
for $\omega$ production at this beam energy is showing strong nucleonic
contribution, too, but also an additional isotropic component signaling
the importance of the mesonic $\pi\rho\rightarrow\omega$
fusion~\cite{nak}. The total cross section of $\sigma_\rho=(23.4\pm 0.8\pm
8)\mu b$ with statistical and systematic errors, respectively, was
determined by normalizing the acceptance corrected $\rho^0$ yield to the
simultaneously measured exclusive $\eta$ yield. For $\omega$ production we
got $\sigma_{\omega}=50\pm 3^{+18}_{-16} \mu b$. For $\phi$ production we
measured $\sigma_{\phi}=0.09\pm 0.007 \pm 0.04 \mu b$. In contrast to the
$\rho^0$ and $\omega$, the polar angular distribution of the $\phi$ meson
is isotropic within the errors, indicating a dominance of mesonic current.
Compared to the model calculation of Sibirtsev~\cite{sib}, our measured
$\phi/\omega$ ration is about a factor of 3 larger.


\subsection{Vector Meson Production at TOF}
\addtocontents{toc}{\hspace{2cm}(M. Schulte--Wissermann)\par}

M. Schulte--Wissermann

{\it Institut f\"ur Kern- und Teilchenphysik,
 Technische~Universit\"at~Dresden, D-01062 Dresden, Germany}

\subsection{Phi Meson Production at ANKE}
\addtocontents{toc}{\hspace{2cm}(M. Hartmann)\par}

M. Hartmann for the ANKE collaboration

{\it IKP, Forschungszentrum, J\"{u}lich, Germany}

As the first experiment with the new detection system for negatively charged
ejectiles at ANKE \cite{ANKE,Nd} $\phi$-meson production has been investigated
in pp collisions \cite{Proposal,pppnPhi,ppPhi} by detecting the $K^+K^-$ decay
mode. First data have been taken at beam energies of 2.83$\,$GeV and
2.7$\,$GeV which correspond to excess energies of 76$\,$MeV and 35$\,$MeV. The
measurements will be continued in the first weeks of 2004 at an excess energy
of 18$\,$MeV ($T_p$=2.65$\,$GeV) (see arrows in the Figure of transparency no.
3) in order to provide the energy dependence in the near threshold region.
The data will cover similar excess energies as the $\omega$-data from SPES-III
\cite{Hibou2} and recent TOF \cite{Samad} experiments.  In addition to the
understanding of the reaction $pp\rightarrow$$pp$$\phi$ near threshold
\cite{ppNakayama2003}, the data will thus provide information on the possible
deviation from the cross-section ratio
$R_\mathrm{OZI}=\sigma_\phi/\sigma_\omega=\tan^2\alpha_\mathrm{v}=4.2{\times}10^{-3}$
given by the the Okubo-Zweig-Iizuka (OZI) rule with the ideal $\phi$-$\omega$
mixing angle $\alpha_\mathrm{v}$ \cite{PDG}.  Such a deviation of the ratio
near threshold could be due to an intrinsic $s{\overline s}$ component in the
nucleon, which would manifest itself in a $\phi$ production cross section
significantly exceeding the limits given by the OZI rule.
  
Subsequent to the measurement $pp{\to}pp\phi$ at $\epsilon$=18$\,$MeV, ANKE
will also collect data for the $\phi$-meson production in $pn$ collision at
beam energy around $T_p$=2.75$\,$GeV \cite{pppnPhi,pnANU}.  The $pn{\to}d\phi$
reaction can be identified by detecting the fast deuteron in coincidence with
the $K^{+}K^{-}$ pairs from the $\phi$ decay using the deuterium cluster-jet
target.  Apart from the unknown cross section \cite{pnGrishina,pnNakayama2000}
this measurement could also be sensitive to the intrinsic strangeness.
Following the arguments of Ref.~\cite{Ellis}, the expected cross section ratio
$\sigma_{pp{\to}pp\phi}/\sigma_{pn{\to}d\phi}$ should be strongly enhanced
according to the negatively polarized strangeness hypothesis. In contrast, no
such enhancement is predicted by meson exchange models
\cite{pnNakayama1999,pnNakayama2000}.

The $\phi$-meson production can also be studied in proton-nucleus reactions at
ANKE by detecting only two particles, one $K^{-}$ candidate in the negative
and one $K^{+}$ candidate in the positive detection system
\cite{ppPhi,Mueller,FZR}.  For the background suppression the clean $K^+$
selection by using the particle range hodoscopes on the positive side is
sufficient.  Rate estimations show the possibility to collect for several
targets up to 10000 $K^+K^-$ correlations ($\approx$$\,$5000 $\phi$-mesons)
per nuclei in few weeks of beam time at ANKE.  Several theoretical
calculations \cite{Lee,Weise,Oset} predict a significant change of $\phi$
properties - the pole mass and decay width - in nuclear matter. If the decay
width increase by one order up to $\Gamma_\phi$=45$\,$MeV, then around 50\% of
the $\phi$-mesons will decay inside of a Cu nuclei for an average momentum of
$p_\phi$=1.2$\,$GeV/c at ANKE.  Naively, one would expect to see a clear
effect in the invariant mass distribution of the $K^+K^-$ mesons. However,
final-state interactions can strongly disturb the measured distributions and
also Coulomb corrections and possible effects of the hadronic kaon potentials
have to be taken into account.  A recent calculation \cite{Mosel} uttered very
skeptical about the visibility of $\phi$ in-medium properties through the
$K^+K^-$ invariant mass distribution. Nevertheless the authors mention that
the reaction can give potentially further evidence for the modification of the
masses of kaons and anti-kaons at finite baryon density.

\subsection{Omega Meson Production in $pn$ Collisions at ANKE}
\addtocontents{toc}{\hspace{2cm}(I. Lehmann)\par}

I. Lehmann, S. Barsov, and  R. Schleichert

{\it IKP, Forschungszentrum, J\"{u}lich, Germany}

The comparison of the cross sections for meson production in
proton-proton and proton-neutron collisions close to threshold,
constrain theoretical models describing the production mechanisms.
For $\eta$ production the observed cross section ratio
$R=\sigma_\mathrm{tot}(pn\to pn\eta)/
\sigma_\mathrm{tot}(pp\to pp\eta)\approx 6.5$
is generally attributed to isovector dominance in model calculations
based on meson exchange.  It is therefore interesting to investigate
whether a similar isospin dependence is found also for the $\omega$,
the next heavier isoscalar meson.  Relatively few experiments were
performed for the $pp\to pp\omega$ reaction, but in proton-neutron
collisions no data whatsoever are available.

The $pn\to d\omega$ reaction was studied in the $pd\to
p_\mathrm{sp}d\omega$ reaction at four proton beam momenta between 2.6
and 2.9$\,$GeV/c at the ANKE spectrometer of COSY-J\"ulich. A
deuterium cluster-jet target was used as an effective neutron target,
detecting the low momentum recoil protons ($p_\mathrm{sp}$), which
have momenta of about $80\,$MeV/c, in a silicon telescope placed close
to the target.  These recoil protons can be treated as ``spectators''
that influence the reaction only through their modification of the
kinematics. By variation of angle and momentum of the spectator
protons, a certain range in excess energy $Q$ is selected
experimentally. This range is used to extract results in $pn$
collisions for the corresponding $Q$ values.  The deuterons emitted at
angles below $8^\circ$ with a momenta around 2$\,$GeV/c were detected
in the forward system of the ANKE spectrometer.  Inclined \v{C}erenkov
counters in combination with two layers of scintillation counters
enabled us to identify these deuterons despite a two orders of
magnitude higher proton background.  Their momenta were reconstructed
using the information from two multi-wire proportional chambers. 

The cross sections extracted for $pn\to d\omega$ at
$Q\approx 26\,$MeV and 60$\,$MeV are significantly smaller than
theoretical predictions.  This suggests that the reaction mechanism
for $\omega$ production differs from the one for the $\eta$, possibly
implying a relatively larger contribution from isoscalar meson
exchange. Measurements with higher precision in both $Q$ and in cross
section are scheduled already for August 2003.  The results are
expected to shed even further light on the basic production
mechanisms.

{\it IKP, Forschungszentrum, J\"{u}lich, Germany}

\newpage \section{Hadrons in Nuclear Medium}

Convenors: M. Nekipelov and A. Sibirtsev

\subsection{Hadrons in the Nuclear Medium}
\addtocontents{toc}{\hspace{2cm}(V. Metag)\par}

V. Metag

{\it  Physikalisches Institut , 
        Justus-Liebig-Universit\"at,Gie{\ss}en, \\
        Heinrich-Buff-Ring 16,  D-35392 Gie{\ss}en 
    }

\subsection{Modification of kaon properties  in nuclear matter}
\addtocontents{toc}{\hspace{2cm}(Th.~Roth)\par}

{Th.~Roth, M.~Buballa, J.~Wambach}

{\em Technische Universit\"at  Darmstadt}

        We investigate the modification of kaons in isospin asymmetric nuclear
        matter.
        Using the leading s-wave couplings of the SU(3) chiral meson-baryon
        Lagrangian we solve the
        coupled channel kaon-nucleon scattering equation selfconsistently.

        The in-medium kaon propagator is calculated for different densities     
        and different proton/neutron ratios. The spectral function of the kaon
        is found to be broadened strongly, its mass shifted downward significantly. 

        However, comparing the effective in-medium mass of the kaon to the
        relevant charge chemical potential of neutron star matter, we
        find  no indication of Kaon condensation.

\subsection{Strangeness Production on Nuclei with Hadronic Probes}
\addtocontents{toc}{\hspace{2cm}(H. Lenske)\par}

H. Lenske

{\it  Physikalisches Institut , 
        Justus-Liebig-Universit\"at,Gie{\ss}en, \\
        Heinrich-Buff-Ring 16,  D-35392 Gie{\ss}en 
    }

\subsection{Investigation of K$^+$-meson production in pA collisions}
\addtocontents{toc}{\hspace{2cm}(M.~Nekipelov)\par}

M.~Nekipelov for the ANKE collaboration

{\em Institut f\"ur Kernphysik, Forschungszentrum J\"ulich GmbH, Germany}

In a series of measurements with the ANKE spectrometer~\cite{anke} at
COSY-J\"ulich, the production of $K^+$-mesons in $pA$ ($A$=C, Cu, Ag and Au)
collisions in a wide range of beam energies, $T_p=1.0\ldots 2.3\,$GeV, has
been investigated.  Double differential cross sections $d^2\sigma/d\Omega{dp}$
for $K^+$ production in $pC$ interactions have been evaluated. For the lowest
incident proton energy it is the first measurement of a complete momentum
spectrum at deep subthreshold energies~\cite{letter}.

The analysis of the target-mass dependence of the cross sections~\cite{phenom}
does not allow to fix the underlying production mechanisms, though reveals
other interesting aspects of the $K^+$ production on nuclei related to the
final-state interactions effects. The strong suppression in the ratios of
$K^+$ production on copper, silver and gold targets to that on carbon observed
for $p_K<200$--$250$~MeV/c may be ascribed to a combination of Coulomb and
nuclear repulsion in the $K^+A$ system.  Our data are consistent with a $K^+A$
nuclear potential of $V_K^0\approx 20\,$MeV at low kaon momenta and normal
nuclear density~\cite{ratios}.


In addition to the direct production on a single target nucleon the $K^+$ mesons can also 
be produced in two-step processes with intermediate pion production, i.e a $pN_1\to\pi X$ 
reaction, followed by $\pi N_2\to K^+ X$ on a second target nucleon. Depending on the beam 
energy, $K^+$-production may be due to both the direct and two-step reaction mechanisms. 
At low beam energies the two-step process is energetically favorable since the intrinsic 
nucleon motion can be utilized twice. A direct test of subthreshold 
K$^+$-production mechanisms comes as a measurement of $K^+d$ and $K^+p$ correlations.

First momentum spectra of protons and deuterons measured in coincidence with the $K^+$-mesons 
have been obtained at $T_p=1.2\,$GeV.  These data supply direct evidence for the two-step reaction 
mechanism, with formation of intermediate pions, leading to kaon production.
It has been shown, that the particular two-step mechanism $pN_1\to d\pi$, $\pi N_2\to K^+\Lambda$ contributes
to about 30\% of the total kaon yield at $T_p=1.2$ GeV \cite{corr}. The data also indicate that a significant fraction
of the $K^+$-mesons is produced on two-nucleon clusters, i.e.\ in the reaction $p(2N)\to (dK^+)\Lambda$.
It is also shown that exploiting criteria of kinematics it is possible to study these mechanisms rather individually.

Another important aspect of $K^+$-production from nuclei is the equality or the difference 
in production on proton and neutron. 
Measurements of correlated $K^+$ mesons and protons allow one to clarify the situation comparing 
resulting spectra with corresponding simulations.
Analysis of the experimental data measured at $T_p\approx2\,$GeV suggests that the production 
on neutron exceeds that on proton, and the ratio of corresponding cross sections is close to 4.

\subsection{Strange Baryon Production at ANKE}
\addtocontents{toc}{\hspace{2cm}(I.Zychor)\par}

{V.Koptev, P.Kravchenko, M.Nekipelov, \underline{I.Zychor}}

{and the ANKE Collaboration}
\ \\
\ \\
 
The production and properties of hyperons have been studied for
more than 50 years, mostly in pion and kaon induced reactions.
Hyperon production in \textit{pp} collisions close to the 
threshold has been studied at SATURNE (Saclay, France) and COSY-J\"ulich.
At COSY beam energies only six hyperons can be produced:
$\Lambda(1116)$, $\Sigma(1192)$, $\Sigma(1385)$, $\Lambda(1405)$,
$\Sigma(1480)$ and $\Lambda(1520)$.
Reasonably complete information on 
$\Lambda(1116)$, $\Sigma(1192)$, $\Sigma(1385)$ and $\Lambda(1520)$
hyperon-production cross sections,
found in the literature, is used for the normalisation of our data.

For the $\Lambda(1405)$, in spite of rather
high statistics achieved (the total world statistics is several 
thousand events),
there are still open questions concerning the nature of
this resonance: is it a singlet \textit{qqq} state in the frame of SU(3)
or a quark-gluon \textit{(uds-q)} hybrid, or a KN bound state? 
Thus more detailed studies of different decay modes of the $\Lambda(1405)$ 
are needed for a better understanding of its nature.

The $\Sigma(1480)$ hyperon is not well established yet.
In the 2002 Review of Particle Physics it is described as a 'bump' with 
unknown quantum numbers. The branching ratios for 
possible decay modes ($N\overline{K}, \Lambda \pi, \Sigma \pi$) 
are known with errors between 60 and 70 \%.  
The total world statistics for this 'bump' is less than two hundred events.

Both~$\Lambda(1405)$ and $\Sigma(1480)$ were produced mainly
from decays of heavier hyperons in kaon and pion induced reactions.
At COSY they are produced directly in $pp{\to}K^+p{Y^*}$ reactions.
It is essential that the ANKE spectrometer permits the simultaneous
observation of different decay modes: $Y^*{\to}{\pi^+}{\Sigma^-}$,
$Y^*{\to}{\pi^-}{\Sigma^+}$ and $Y^*{\to}{K^-}p$.
To realize this project it is necessary to analyse triple
coincidences: positively and negatively charged particles detected 
by the three different parts of the ANKE detector system (side, 
forward and negative). 

The first measurement of \textit{pp} interactions using a cluster-jet 
hydrogen target, carried out in spring 2002, demonstrated the feasibility of these
studies at ANKE.

In the missing mass spectrum in the reaction $pp\to{K}^+pM_X$, resulting from 
detection of the 2-fold K$^+$p coincidences at proton beam energy 
$T_p=2.83\,$GeV, different hyperons ($\Lambda(1116)$, $\Sigma(1192)$, $\Sigma^*(1385)$, 
$\Lambda^*(1405)$, $\Sigma^*(1480)$) and hyperons with additional pions 
($\Lambda\pi$, $\Sigma\pi$, $\Sigma\pi\pi$) were identified.

The 3-fold $K^+p\pi^+$ coincidences were selected
to study heavier hyperons $Y^*$ ($\Sigma^*(1385)$, $\Lambda^*(1405)$,
$\Sigma^*(1480)$). 
A missing mass spectrum in the reaction $pp\to{K}^+p\pi^+M_X$ consists of 
a flat plateau with a peak at approximately 1195~MeV.
The peak corresponds to the decay $Y^*\to\pi^+\Sigma^-$.
If only events with $M_X=(1195\pm20)\,$MeV are selected, then 
the $m_x$ spectrum in the reaction $pp\to{K}^+p\,\,(m_x=\pi^++M_X)$ shows two 
peaks with a width of 45~MeV each.
One of them corresponds to the contribution of $\Sigma(1385)$ and $\Lambda(1405)$
hyperons. The second peak
can be ascribed to the production of the $\Sigma(1480)$. 
In order to prove this a similar analysis of 3-fold $K^+p\pi^-$ 
coincidences is in progress.

\subsection{Hadron Induced two Pion Production in Nuclei }
\addtocontents{toc}{\hspace{2cm}(Rudolf Meier)\par}

\vspace*{6mm}


{Rudolf Meier, Heinz Clement}


{\em Physikalisches Institut, Universit\"at T\"ubingen,
Auf der Morgenstelle 14, 72076 T\"ubingen, Germany}

\vspace*{6mm}

Topic of this talk is the search for medium modifications of the 
sigma meson. The sigma is identified as the $f_0$(600) by the particle
data group [1] with uncertain mass between 400 and 1200 MeV and large 
width of 600 to 1000 MeV. The interpretation of the sigma is controversial,
it has been described as a pion-pion interaction resonance in the I=L=0 
channel, i.e. a $q\bar{q}q\bar{q}$ state, or as an intrinsic $q\bar{q}$
meson [2]. In the latter interpretation, it has been put forward as the 
chiral partner of the pion, which should be severely affected in mass
and width in the nuclear medium due to the partial restoration of
chiral symmetry. The interpretation as a pion-pion-resonance also
leads to property changes as a consequence of medium-modifications 
of the pion-pion-interaction.\\

The sigma channel can be accessed by observing I=L=0 pion pairs.
First information came from the CHAOS experiment at TRIUMF, which measured 
pion-pion invariant mass spectra in the pion-induced pion production 
reactions ($\pi^+,\pi^+\pi^-$) and ($\pi^+,\pi^+\pi^+$) for various nuclei 
(D, $^{12}$C, $^{40}$Ca, $^{208}$Pb). While the I=2 $\pi^+\pi^+$ channel was 
not affected by the choice of target, a clear A-dependent accumulation
of strength at low invariant masses was seen for the $\pi^+\pi^-$ channel [3].   
A similar effect was observed in the ($\pi^-,\pi^0\pi^0$) reaction by
the Crystal Ball collaboration at BNL [4,5]. A calculation by the Valencia
group, including medium effects as modification of the pion-pion-interaction,
failed to describe the size of the enhancement at low invariant masses [4,6].\\

This is contrasted by the situation for the ($\gamma,\pi^0\pi^0$) reaction
in nuclei, which was investigated by the TAPS/A2 collaboration at MAMI.
Also here, modifications of the pion-pion invariant mass spectra were
observed, but the changes are in quite good agreement with the Valencia
group calculations [7,8].\\

 This different behavior 
of the pion- and photon-induced reactions is not
understood. The Valencia group suggests that the failure of the description
for the pion-induced reaction might be due to accidental strong cancellations
of parts of the reaction amplitude in the case of the pion-induced
reaction. This explanation can be checked by measuring the behavior of a
third production mechanism, the proton-induced two pion production.
Investigation of this reaction in nuclei is planned with the TOF detector
at COSY and the WASA detector at TSL.

\medskip

\vspace{0.5cm}
{\large \bf References}

{}[1] K. Hagiwara et al., {\it Review of Particle Physics}, Phys.
      Rev. D {\bf 66} (2002).\\
{}[2] A. F\"a\ss ler et al., Phys.Rev. D68, 014011(2003).\\
{}[3] F. Bonutti et al., Phys. Rev. Lett. {\bf 77}, 603 (1996); 
      Nucl. Phys. {\bf A677}, 213 (2000).\\ 
{}[4] A. Starostin et al., Phys. Rev. Lett. {\bf 85}, 5539  (2000).\\
{}[5] P. Camerini et al., Phys. Rev. C {\bf }, 067601 (2001).\\
{}[6] M. Vicente-Vacas, E. Oset, Phys. Rev. C {\bf 60}, 064621 (1999).\\
{}[7] J. G. Messchendorp et al., Phys. Rev. Lett. {\bf 89}, 222203 (2002).\\
{}[8] L. Roca et al., Phys. Lett. B {\bf 541}, 77 (2002).

\subsection{Double Pion Photoproduction from Nuclei}
\addtocontents{toc}{\hspace{2cm}(S. Schadmand)\par}

 S. Schadmand
 for the TAPS and A2 collaborations

    {\em II. Physikalisches Institut , 
        Justus-Liebig-Universit\"at,Gie{\ss}en, \\
        Heinrich-Buff-Ring 16,  D-35392 Gie{\ss}en 
    }

The interest in modifications of hadron properties in the nuclear medium
is motivated by the origin of hadron masses 
in the context of spontaneous chiral symmetry breaking 
and their modification in a hadronic environment due to chiral 
dynamics and partial restoration of chiral 
symmetry~\cite{Hatsuda:1991ez,Brown:kk}.

The photoabsorption cross section on the proton~\cite{PDG1} for
$E_\gamma<$800~MeV can be explained by the sum of the meson production 
channels:
 single $\pi^+$~\cite{Buechler:jg,MacCormick:1996jz},
 single $\pi^\circ$~\cite{Haerter},
 $\pi^+\pi^-$~\cite{Braghieri:1994rf,Wisskirchen},
 $\pi^+\pi^\circ$~\cite{Langgartner:sg},
 $\pi^\circ\pi^\circ$~\cite{Wolf:qt,Hourany}, and
 $\eta$~\cite{Krusche:nv,Renard:2000iv}. 
In the second resonance region, 
the dominant contribution is the D$_{13}$(1520) resonance
with the strongest coupling to the incident photon.
$\pi\pi$  production reveals sequential resonance decay
via an intermediate $\Delta$ state~\cite{Wolf:qt}.
In $\pi^+\pi^\circ$, a decay branch of 20\% $N^\star\to N\rho$ is 
deduced~\cite{Langgartner:sg}.
The $N^*$ contribution to double pion photoproduction 
by itself is not large. 
The structure stems from an interference with
other terms~\cite{GomezTejedor:1995pe,Nacher:2000eq}.

The comparative study of the nuclear photoabsorption~\cite{Muccifora:yg}
from nuclei (A$>$6) with the elementary photoabsorption show that
the cross sections follow a universal behavior $\sigma\sim 1/A$;
the $\Delta$ resonance is broadened and slightly shifted;
the higher resonance regions appear depleted.
The depletion of nuclear cross sections in the second resonance region
is evidence for modifications of hadron properties in the nuclear medium.
'Trivial' medium effects on baryon resonance parameters like 
Fermi motion, collision broadening, quenching, and Pauli blocking, do not
explain the effect.
An in-medium broadening of the D$_{13}$(1520) resonance could
cause of the depletion on account of
the coupling to the $N\rho$ final state~\cite{ref_mosel}. 
The $\rho$-meson itself is expected to be
appreciably broadened in the nuclear medium~\cite{ref_klingl}.
In ~\cite{Hirata:2001sw}, the disappearance of 
the second resonance structure is related to  a cooperative effect of 
the interference in double pion production processes, Fermi motion, 
collision broadening of the $\Delta$ and $N^*$ resonances, and 
pion distortion in the nuclear medium. 
Systematic studies of quasifree meson production have not yet provided 
a hint for a depletion of the resonance yield.
Single $\pi^\circ$~\cite{Krusche:ku} and $\eta$~\cite{Roebig-Landau:xa,Yamazaki:jz} 
photoproduction cross section from nuclei
show a reduction and change of resonance shapes that can be explained 
by trivial effects like absorption, Fermi smearing and Pauli blocking, 
and collision broadening.
The resonances of the second resonance region
seem to maintain their structure in the nuclear medium.
In a more detailed study, 
invariant $\pi\pi$ mass distributions indicate a modification of 
the pion-pion interaction in the scalar-isoscalar 
state~\cite{Messchendorp:2002au,Roca:2002vd}, 
similar to the dropping mass scenario expected from 
partial chiral symmetry restoration~\cite{Hatsuda:1991ez}.

\subsection{$K^+$ and $K^-$ Mesons in Heavy Ion Collisions
  at SIS Energies}
\addtocontents{toc}{\hspace{2cm}(F. Uhlig)\par}

F. Uhlig

{\em Technische Universit\"at  Darmstadt}

\subsection{Strange Baryons in Matter}
\addtocontents{toc}{\hspace{2cm}(J. Aichelin)\par}

J. Aichelin

{\em SUBATECH, Nantes, France}

\subsection{Baryonic resonances in
$NN$ and $N$--nucleus collisions}
\addtocontents{toc}{\hspace{2cm}(Radhey Shyam)\par}

Radhey Shyam

{\em Institut f\"ur Theoretische Physik, Universit\"at Giessen, D-35394 Giessen,
Germany}

Within an effective Lagrangian model, we present calculations for
cross sections of  kaon [1,2] and dilepton productions [3] in
nucleon-nucleon (NN) collisions for beam energies below
10 GeV. The initial interaction between the two
nucleons is modeled by the exchange of $\pi$, $\rho$, $\omega$ and
$\sigma$ mesons. The particle production in the final channel
proceeds via excitation, propagation and decay into relevant channels
of the nucleon resonances which have large branching ratio for decays into
those channels. In case of $K^+$ production, the relevant resonances are
$N^*(1650)$, $N^*(1710)$, and $N^*(1720)$ while for dileptons they
are $\Delta(1232)$ and $N^*(1520)$. The parameters of the model at
the NN-meson vertices are determined by fitting the NN elastic scattering
$T$ matrix with effective Lagrangians based on the exchange of these
four mesons, while those at the resonance vertices are calculated from
the known decay widths of the resonances. We find that the associated
$K^+\Lambda$ and $\Sigma^0K^+$ productions are dominated by the
contributions coming from $N^*(1650)$ resonance for near threshold
beam energies while $N^*(1710)$ dominates the cross sections at 
higher beam energies. One pion exchange contributions are most
important in the entire beam energy region. The dilepton production
is dominated by the $\Delta(1232)$ contributions for beam energies
in 1-5 GeV range. The effective Lagrangian model calculations are able
to explain the data of the DLS Collaboration on the dilepton
production in proton-proton collisions for beam energies below 1.3 GeV.
However, for incident energies higher than this the inclusion of
contributions from other dilepton sources such as Dalitz decay of
$\pi^0$ and $\eta$ mesons and direct decay of $\rho^0$ and $\omega$
mesons is necessary to describe the data. 
\vskip .1in
We also calculate the exclusive $K^+$ meson production in a
proton-nucleus collision leading to two body final states within a
fully covariant two-nucleon model based on the same effective
Lagrangian and the corresponding coupling constants [4]. The calculated
cross sections show strong sensitivity to the medium effects on
pion propagator and to the final hypernuclear state excited in the
reaction.   
\vskip .1in
This work has been done in collaboration with H. Lenske and U. Mosel 
and is supported by the Forschungszentrum J\"ulich.

\vspace{0.5cm}
{\large{\bf References}} 

\vskip .2in
[1] R. Shyam, Phys. Rev. C {\bf 60}, 055213 (1999).
\vskip .1in
[2] R. Shyam, G. Penner and U. Mosel, Phys. Rev. C {\bf 63}, 022202(R)
(2001).
\vskip .1in
[3] R. Shyam and U. Mosel, Phys. Rev. C {\bf 67}, 065202 (2003).
\vskip .1in
[4] R. Shyam, H. Lenske, and U. Mosel, nucl-th/0308085.

\newpage \section{Accelerator}

Convenors: A. Lehrach and R. Maier

\subsection{COSY Status}
\addtocontents{toc}{\hspace{2cm}(D. Prasuhn)\par}

{D. Prasuhn}

{\em Institut f\"ur Kernphysik, Forschungszentrum J\"ulich GmbH, Germany}

\subsection{Polarization Development}
\addtocontents{toc}{\hspace{2cm}(B. Lorentz)\par}

B.Lorentz, U.Bechstedt, J.Dietrich, R.Eichhorn, R.Gebel, A.Lehrach, R.Maier,
D.Prasuhn,  A.Schnase, H.Schneider, R.Stassen, H.Stockhorst, R.T\"olle

{\em Institut f\"ur Kernphysik, Forschungszentrum J\"ulich GmbH, Germany}

Polarized proton beams at COSY are routinely delivered to internal and
external experiments. The polarized beams from the ion source are
pre-accelerated in the cyclotron JULIC and injected and accelerated in COSY
without significant loss of polarization. Imperfection resonances are
increased in strength by means of vertical orbit distortions, leading to a
complete polarization reversal (spin flip). Intrinsic resonances are overcome
by means of a fast ramping air core quadrupole magnet inducing a rapid change
in tune and therefore preservation of the polarization (tune jump).

The polarization during acceleration is observed utilizing parts of the former
EDDA detector system as polarimeter. Only the fast online polarization
measurements during the COSY acceleration cycle made possible by the EDDA
setup allows an efficient correction of the upto 13 first order depolarizing
resonances encountered.

Polarized deuterons were accelerated for the first time early in 2003. Because
of the lack of depolarizing resonances in the COSY energy range, preservation
of the polarization is less involved for deuterons compared to protons.
However, a polarization measurement technique for vector and tensor
polarization has to be developed. It is intended to use for this task parts of
the former setup of EDDA. So far only a relative measurement of the vector
polarization was possible. However, this was already sufficient to carry out a
first investigation of the polarization reversal of a polarized deuteron beam
by crossing of an RF dipole induced depolarizing resonance [1].

Polarized proton beams with intensities above $10^{10}$ stored protons with a
degree of polarization of upto 0.80 are now routinely delivered to internal
experiments. For external experiments polarized protons can be slowly
extracted from COSY via stochastic extraction without loss of polarization by
keeping the extraction momentum far enough away from depolarizing resonances
and careful adjustment of the betatron tunes during the extraction process. As
an example, $10^8$ protons/second at 1.4 GeV/c with a polarization of 0.80
were delivered to the TOF experiment recently.

A more detailed study of higher order depolarizing resonances in the vicinity
of the strongest first order intrinsic resonance (8-Qy), located close to a
beam momentum of 2.1 GeV/c will be conducted to minimize the remaining small
polarization losses. Furthermore the investigation of spin flipping of
polarized protons and deuterons at fixed energies within an international
collaboration will be continued.

\vspace{0.5cm}
{\large \bf References}

[1] K. Yonehara {\it et al.}, "Spin-Flipping Polarized Deuterons at
  COSY", Proc. {\it CIPANP 2003}, to be published.

\subsection{Intensity Upgrade @ COSY%
}
\addtocontents{toc}{\hspace{2cm}(A. Lehrach)\par}

A. Lehrach, R. Gebel, B. Lorentz, A. Schnase, J. Stein,
 and R. T\"olle for the COSY team

{\em Institut f\"ur Kernphysik, Forschungszentrum J\"ulich GmbH, Germany}

The aim of the Intensity Upgrade @ COSY is to reliably increase the
intensity of the polarized proton beam in COSY typically by a factor of
three compared to what can be delivered to experiments at present.
Therefore, the beam transmission in the whole accelerator chain is
investigated to search for potential improvements. The performance of
the system not only depends on the transmission but also on the beam
quality delivered by the various subsystems. Hence, the measurement of
key beam properties plays a major role in the intensity upgrade program. \\  
Most promising tasks of the intensity upgrade program are improvements
of the source performance and transmission in the source beamline, the
characterization of beam delivered by the cyclotron, securing the long
term availability of cyclotron RF system, fine tuning of the betatron
amplitudes in the injection beamline, further investigation of the
potential of beam debunching for the COSY injection, the maximization
of the COSY acceptance, capture and acceleration efficiency in COSY
and careful revision and optimization of the COSY magnet ramps. \\
In the first beam time an optimized injection setting with unpolarized
beam has been developed. The intensity of the injected beam could be 
increased by a factor of about two, as presented in this talk.
Detailed investigation of all system parameters and beam properties
will allow an adopted set-up of COSY with increased performance. \\ 
The improved capabilities will enable us to further exploit the unique
experimental opportunities of the COSY facility.

\newpage \section{Baryon Resonances}

Convenors: S. Krewald and H.-P. Morsch

\subsection{Baryon Resonances}
\addtocontents{toc}{\hspace{2cm}(H.P. Morsch)\par}

H.P. Morsch 

{\it Institut f\"ur Kernphysik, Forschungszentrum J\"ulich}

Baryon resonances show the dynamical structure of baryons in a direct way and
are therefore uniquely suited to investigate the properties of QCD in the
non-perturbative regime. Presently, there are significant theoretical
efforts to investigate these resonances in different models and approximation
schemes of QCD.
On the other hand, there are large experimental programs with electromagnetic
and hadronic probes to study the detailed properties of the resonances.

Several aspects of N* physics were discussed in the afternoon session;
therefore the present discussion is restricted mainly to the lowest N*
resonances.
A detailed study of the Roper resonance, P$_{11}$(1440), has shown, that this
resonances is excited differently in different reactions: in $\alpha$-p
scattering [1] a N* resonance is observed at a mass of 1400 MeV with a width
of about 200 MeV,
whereas $\pi$-N [2] shows a large resonance with a width of 300-360 MeV.
These observations have led to an interpretation of the Roper resonance
in terms of two structures [3]. Excitation by photons, in which the first 
resonance is not observed, supports this interpretation.

Formerly, N* resonances have been investigated in high energy (p,p'), 
($\pi$,$\pi$') [4] and (e,e') reactions. In proton and pion scattering
N* resonances were observed at masses of the $\Delta$(1232), 1400 MeV, 
1510 MeV, 1690 MeV and 2190 MeV. The last three resonances correspond
to D$_{13}$(1520), F$_{15}$(1680) and G$_{17}$(2190).
With electrons the same resonances were observed, except for the resonance
at 1400 MeV. The cross section of this resonance shows a strong fall-off with 
momentum transfer, characteristic of L=0 excitation. Therefore, this resonance
can be identified as P$_{11}$(1400).    
Remarkably, this structure corresponds in energy and width exactly
to the P$_{11}$ resonance observed strongly in $\alpha$-p scattering [1].  

There are also old exclusive (p,p'x) data available [5], which give 
information
on the decay properties of the observed resonances. In the invariant 
$\pi^+\pi^-$ mass spectrum a prominant peak is observed at the mass of
the P$_{11}$(1400), which indicates strong $2\pi$-N decay of this resonance.

Experiments are in preparation at COSY, in which N* resonances will be studied 
in p-$\alpha$ interactions using a proton beam up to 2.5 GeV and a thin $^4He$
target. The recoil $^4He$-particles will be registered in Silicon microstrip 
$\Delta$E-E detector telescopes, whereas the N* decay products will be
detected in the large acceptance detector TOF. We expect, that in these 
experiments N* production will be strongly increased with respect to the
Saturne experiment and possible up to much larger mass, allowing for more 
detailed investigations.

\vspace{0.5cm}
{\large \bf References}

[1] H.P. Morsch et al. Phys.~Rev.~69 (1992) 1336

[2] R.A. Arndt et al., new results from SAID, 2000, unpublished 

[3] H.P. Morsch and P. Zupranski, Phys.~Rev.~C 61 (1999) 024002 

[4] R.M. Edelstein et al., Phys.~Rev.~D 5 (1972) 1073 

[5] E. Colton et al., Phys.~Rev.~D 3 (1971) 1063

\subsection{Introduction to the Baryon sesssion}
\addtocontents{toc}{\hspace{2cm}(S. Krewald)\par}

{S. Krewald}

{\em Institut f\"ur Kernphysik, Forschungszentrum J\"ulich \\ 
D-52425 J\"ulich, Germany}


Presently, we are living in an exciting period in which experiments present
increasingly strong evidence for a pentaquark\cite{Nakano} predicted a
long time ago by Diakonov, Petrov and Polyakov who assumed the $N(1710,{ \frac {1}{2} }^+)$
to be a member of an antidecuplet. Today's session focusses on another non-$q^3$ candidate,
the so-called Roper resonance, $N(1440,{ \frac {1}{2} }^+)$. Problems of traditional
quark models to reproduce the mass of this resonance have triggered the idea that
the Roper might have an exotic structure
\cite{isgur,capstick}.
 On the other hand,  instanton motivated effective
quark interactions
 or color singlet interactions have been used to reduce the mass of the Roper
assuming a pure three-valence quark structure\cite{loring,glozman}.
Several Lattice groups have reported results concerning excited nucleons and 
found produce Roper masses well above the experimental value, see e.g. \cite{sasaki}.
Recent calculations based on the overlap realisation of a lattice fermion
action possessing an exact chiral symmetry pushed the pion masses down to
$180$ MeV and have obtained a considerable lowering of the lattice Roper mass
which  now  degenerates with the lattice $N(1535)$,however \cite{dong}.

The Juelich theory group\cite{krehl} treats the pion-nucleon
reaction  by coupling
 the reaction channels $\pi N$, $\eta N$, $\rho N$,
$\sigma N$, and $\pi \Delta$ within a meson-exchange model. This model incorporates the most
 relevant reaction channels for one- and   two-pion decays of the resonances. 
The model provides a theory for the physical background due to meson-baryon
interactions. The background may show some structure because the exchange of mesons in the
t-channels can act as a long-range attraction between meson and baryon and thus lead
to a somewhat localized enhancement of strength. In order to reproduce resonances with a
relatively narrow width, such as the $\Delta_{33}$, one has to introduce pole diagrams, however,
which are interpreted as bare triple quark states.
It was found that all
pion-nucleon resonances below 2 GeV require pole diagrams within this formalism,
with one exception: the $ N^*(1440)$.

On the experimental side, there has been considerable progress.
Two-pion production in  proton-proton reactions measured at CELSIUS appear to demand a 
strong two-pion decay of the Roper resonance\cite{brodo,paetz}, as suggested by the
Julich model. One has to note, however,
that the theoretical analysis of the data deserves further scrutiny.

The CRYSTAL BALL collaboration has presented data for the reaction
$ \pi^- p \rightarrow \pi^0 \pi^0 n$ \cite{craig}.  The theoretical model for this
reaction presented  by S. Schneider appears to be compatible with the Crystal ball
data which would be further evidence for a significant role of the Roper.

The alpha particle has been shown to excite the Roper in inclusive reactions\cite{morsch},
exclusive reactions hopefully will be presented in the near future.

At a first glance, photon-induced reactions appear not to be a suitable probe
to investigate the Roper resonance because the photoproduction of two neutral
pions is dominated by the D13(1520)\cite{haert}. This situation may change radically
in the near future because both polarized photons and polarized targets have become
available. As is discussed by R.Beck, these new experimental achievements will
allow to filter out the Roper contribution from the photo-induced cross sections.
Moreover, at Jlab, the CLAS collaboration has started to provide electron-induced
cross sections which allow to map out the momentum dependence of the various reactions.
L.Tiator presents first analyses of the data. Some quark models produce transition amplitudes
for the Roper which vanish at some momentum transfer well below 1 GeV/c\cite{cano}. 
It will be very exciting to see whether such a node in the transition amplitude 
is compatible with the data.

\subsection{Baryon excitations
in a relativistic quark model}
\addtocontents{toc}{\hspace{2cm}(Bernard Metsch)\par}

Bernard Metsch

{\em Helmholtz-Institut f\"ur Strahlen-- und Kernphysik (Theorie),\\ 
Universit\"at Bonn, Nu{\ss}allee 14-16, D53115 Bonn, Germany}

On the basis of the three-particle Bethe-Salpeter equation we
formulated a relativistic quark model for baryons. Because neither the
full quark propagator nor the interaction vertices are reliably known
from QCD, we make two basic phenomenological approximations: Quark
propagators have the free form with an effective constituent mass and
the interactions are assumed to be given by unretarded
potentials, both assumptions are motivated by the apparent success of
non-relativistic quark models. The interaction
potentials include a string-like linearly rising three-body
confinement potential with a suitable spin-dependence and a
spin-flavor dependent $qq$-interaction motivated by instanton effects,
taken into account by the perturbative construction of an effective
three-body kernel. 
This framework was applied to calculate the complete mass spectrum of
light-flavored baryons up to 3 GeV. 
Constituent quark masses and confinement parameters were fixed
to the $\Delta$-Regge trajectory and those of the instanton-induced
interaction (acting on flavor-antisymmetric quark pairs only) to the
ground-state hyperfine splittings, the rest of the spectrum thus being
a genuine prediction. We can describe all the major
features in the experimantal baryon spectra, in particular the
conspicuous low position of the Roper-resonance, and the approximate
parity doublets apparent in the $N$- and $\Lambda$
spectra~\cite{met_1}.
Form factors and decay properties of baryon resonances were calculated
in the present covariant framework in the Mandelstam formalism. 
All model parameters being fixed from the baryon mass spectrum the
calculation of electromagnetic properties constitutes a genuine
prediction: We find a good description of static
electroweak properties, the magnetic and axial nucleon form factors
and the magnetic $N-\Delta$-transition form factor up to moderate (3-4
GeV$^2$) momentum transfers~\cite{met_1}.  
A novel approach to calculate magnetic
moments directly as an expectation value of a local oparator with
Salpeter amplitudes is presented. Results on other transition form
factors and helicity amplitudes are presented.  Strong two-body decay
widths of baryon resonances can also be obtained parameterfree by
calculating in lowest order the contribution of single
quark-loops. First
results on the decay widths of some low-lying baryon resonances are
discussed.

\subsection{Crystal Barrel Results}
\addtocontents{toc}{\hspace{2cm}(V. Crede)\par}

V. Crede

{\em Helmholtz-Institut f\"ur Strahlen-- und Kernphysik,\\ 
Universit\"at Bonn, Nu{\ss}allee 14-16, D53115 Bonn, Germany}

The study of nucleon resonances provides important information 
on many open questions in baryon spectroscopy. The key to any 
progress is the identification of the effective degrees of 
freedom leading to a qualitative understanding of strong QCD. 
The problem of {\it missing resonances} predicted by quark models 
is reviewed on the basis of experimental results of the \mbox{CB-ELSA} 
experiment at the e$^-$~accelerator ELSA in Bonn. Differential and 
total cross sections of $\gamma\,\mbox{p}~\rightarrow~\mbox{p}\,\pi^0$
and $\gamma\,\mbox{p}~\rightarrow~\mbox{p}\,\eta$ have been determined 
for incident photon energies up to $E_{\gamma}=3$~GeV. At low energies,
results of experiments such as GRAAL and CLAS are well reproduced.
New data points have been added to those results for forward angles of 
the meson and at energies above 2~GeV. In the differential cross sections 
of both, $\pi^0$~and~$\eta$~photoproduction, a transition from dominant 
resonance production to a strong peaking in the forward direction can be 
observed around $E_{\gamma}=2$~GeV.
\par\smallskip
Furthermore, resonance production and even cascades of the type\linebreak
\mbox{$\mbox{N}^{**}(\Delta^{**})\rightarrow\mbox{N}^{*}(\Delta^{*})
\rightarrow\mbox{p}\pi^0\pi^0(\mbox{p}\pi^0\eta)$} are observed.
Indications for at least one $\Delta$~resonance around 1900~MeV 
are seen. The latter is particularly interesting if it had negative parity 
because a confirmation of this state would be in contradiction 
with constituent quark models~\cite{Ref-1,Ref-2}.
\par\smallskip
The Crystal Barrel detector is the ideal instrument to study various 
multi-photon final states over the full dynamical range due to its 
almost 4$\pi$ coverage of the solid angle and its large energy 
resolution. It allows to identify highly-excited baryon states by 
observing cascades of high-mass states to the ground state via the 
emission of single pion and eta mesons. The latter could be observed
in the 2001 CB-ELSA data for the reaction 
$\gamma\,\mbox{p}~\rightarrow~\mbox{p}\,\pi^0\,\eta$, 
for instance. It could be shown in partial wave analyses that linearly 
polarised photons are very important in order to avoid ambiguities in
determining the corresponding quantum numbers. In 2002/2003, polarised 
data have been taken off the proton as well as off the neutron with the
Crystal Barrel detector and TAPS in the forward direction as a fast 
trigger forming an ideal photon detector of high granularity (CB-TAPS).

\subsection{Two-Pion Production in Proton-Proton Collisions}%
\addtocontents{toc}{\hspace{2cm}(M. Bashkanov)\par}

{M. Bashkanov, H. Clement,  R. Meier,
  T. Skorodko and G.J. Wagner for the CELSIUS-WASA
 collaboration}%

{\em Physikalisches Institut der Universit\"at T\"ubingen, Auf der Morgenstelle
  14, D-72076 T\"ubingen, Germany}

%
%
In nucleon-nucleon induced two-pion production the correlated two-pion
exchange between the 
interacting nucleons is lifted onto the mass shell. This reaction gives rise
to a number of subsystems, where the different aspects of this process can be
studied. Among these subsystems the $NN\pi$ system is special, since it
facilitates the search for
narrow, in particular $NN$-decoupled dibaryon resonances. First exclusive
measurements at CELSIUS have provided \cite{author1} new and significant
upper limits in the 
range of a few nb for the low-mass region $m_{dibaryon} < 2087$ MeV/c$^2$. For
a review on the present status of dibaryon searches see \cite{author2}

In the $\pi\pi$ subsystem the dynamics in $\sigma$ and $\rho$ channels is of
topical interest. In combination with the $N\pi$ and $N\pi\pi$ subsystems
it gives access to the investigation of nucleon excitations and their decay
properties. Particular emphasis is placed here on the investigation of the
Roper resonance, the second excited state of the nucleon in the non-strange
sector, since its nature and properties  are still very
poorly known. 

Our first exclusive measurements of the $pp \to pp\pi^+\pi^-$
reaction near threshold \cite{author3,author4,author5} reveal this reaction to be
dominated by 
$\sigma$-exchange between the colliding nucleons followed by the Roper
excitation in one of the nucleons with subsequent decay into $N\sigma$ or
$\Delta\pi$ channels. From the observed interference of both decay routes into
the final $N(\pi\pi)_{I=l=0}$ state
their relative amplitudes and branchings have been determined
\cite{author2,author5} in the low-energy tail of the Roper excitation. Though
we observe this low-energy region to be strongly dominated by the $N\sigma$
channel, we also see a small but rapidly growing influence of the $\Delta\pi$
channel as the energy is increased. Due to its $k^2$ dependence the amplitude
of the latter
is even likely to take over at the position of the Roper resonance pole
\cite{author2,author5}. 

These findings are basically in agreement
with theoretical predictions of the Valencia group
\cite{author6,author7}. However, there are also significant deviations
in particular with respect to new measurements of this reaction in the
threshold region with the polarized proton beam at COSY-TOF
\cite{author8}. There the 
preliminary data analysis yields partly non-zero analyzing powers, which
point to perceptible admixtures of $l \neq 0$ partial waves in $pp \to
pp\pi^+\pi^-$, in particular in the $\pi\pi$ system.

With the new WASA $4\pi$ detector at CELSIUS a program has been started to
measure the two-pion production exclusively in all channels over a wide energy
range. First data for the 
$pp\pi^0\pi^0$ channel find the influence of the $N^\ast \to \Delta\pi \to
N\pi\pi$ route to be considerably weaker than expected from the $pp\pi^+\pi^-$
channel assuming isospin invariance, which points to possible $\rho$ channel
admixtures in the latter. At incident energies above 1 GeV we observe drastic
changes in the spectra. In $M_{p\pi}$ spectra we observe a clear peak which
may be associated with $\Delta$ excitation, whereas the $M_{\pi\pi}$ spectra
essentially fall back to phase space. 

%

\subsection{What can we learn about Baryon Resonances from $\pi N \to \pi\pi N$?}
\addtocontents{toc}{\hspace{2cm}(S. Schneider)\par}

S. Schneider, S. Krewald

{\em Institut f\"ur Kernphysik, Forschungszentrum J\"ulich, D-52425
  J\"ulich}

Most of our knowledge of baryon resonances comes from the analysis of $\pi N$
scattering. In order to obtain more detailed information on the decay
properties of resonances, we directly study the decay to the inelastic
channels in the reaction $\pi N \to \pi\pi N$.
We investigate the reaction in the framework of a resonance exchange
model. The intermediate resonances are the $\rho$ and $\sigma$ mesons and the
baryonic resonances $P_{33}(1232)$, $P_{11}(1440)$,
$D_{13}(1520)$, and $S_{11}(1535)$. 
It has been checked that the corresponding
subgraphs of the $\pi N \to \pi\pi N$ model are able to describe $\pi\pi$
scattering and $\pi N$ scattering reasonably well. 
\par
We observe a substantial contribution from the Roper Resonance in the reaction
channels $\pi^-p\to \pi^+\pi^-n$ and $\pi^-p\to
\pi^0\pi^0n$. In the energy range above $T_\pi\approx 270$ MeV we clearly need
the contribution of the Roper Resonance, but in the threshold region it leads
to an overestimation of the cross section data. 
The problem can be mended by switching off the Roper $\to \sigma N$ decay and
readjusting the Roper $\to \pi\Delta$ decay constant. This leads to a stronger
rise of the Roper contribution with energy and even allows for an almost
perfect description of the differential cross sections for $\pi^-p\to
\pi^+\pi^-n$ that are available up to $T_\pi=305$ MeV. 
But switching off the Roper $\to \sigma N$ decay would spoil the
description of the $\pi N$ inelasticities in the $P_{11}$ partial wave. There,
one needs the $\sigma N$ decay in order to reproduce the early onset of the
inelasticities.
\par
These results certainly need further investigation. In particular, it will be
necessary to improve the model by including a 4$\pi$ and a 3$\pi NN$ contact
term, which are needed to respect chiral symmetry.

\subsection{Two $\pi$ Invariant Mass Spectra from $\alpha p\to \alpha' p X$}
\addtocontents{toc}{\hspace{2cm}(A. Prokofiev)\par}

{A. Prokofiev}

{\em High Energy Physics Dept., St. Petersburg Nucl. Phys. Inst.,
188350 Gatchina, Russia}

\subsection{$N^*$ Experiments with Polarized Photons}
\addtocontents{toc}{\hspace{2cm}(R. Beck)\par}

R. Beck

{\em Institut f\"ur Kernphysik,
     Universit\"at Mainz, Mainz, Germany}

\subsection{Electroproduction of Nucleon Resonances with MAID}%
\addtocontents{toc}{\hspace{2cm}(Lothar Tiator)\par}

{Lothar Tiator} 

{\em Institut f\"ur Kernphysik,
     Universit\"at Mainz, Mainz, Germany}

MAID is a unitary isobar model for pion photo- and electroproduction on the nucleon, accessible in
the internet \cite{MAID}. It can be applied for energies from pion threshold up to $W=2$~GeV
covering most of the resonance region. From real photons at $Q^2=0$ up to photon virtualities of
$Q^2=5$~GeV$^2$ it can be used for online calculations of multipoles, amplitudes, cross sections,
polarization observables and sum rules. The model is based on a nonresonant background dominated by
Born and vector meson exchange terms and a resonance part described by Breit-Wigner functions. Both
parts are individually unitarized and fulfill the Watson theorem below the two-pion threshold.
While in MAID2000 only the 8 most prominent nucleon resonances were included, we have recently
extended the resonance sector and now all four-star resonances below $W = 2$~GeV are included.

Using the world data from the GW/SAID database for pion photo- and electroproduction and recent
electroproduction data from Bonn, Bates and JLab we have performed energy dependent single $Q^2$
fits and an energy- and $Q^2$ dependent "superglobal fit" of nucleon resonance excitation.

Our first and preliminary analysis shows quite stable results for the $\Delta$ excitation as well
as for the transverse amplitudes of the $D_{13}(1520)$ and the $S_{11}(1535)$. For most other
resonances the data show large fluctuation that are much bigger than the statistical uncertainties.
Such model-dependent results reflect the poor situation in the data base in some kinematical
regions and the lack of polarization observables that are essential, e.g. for the analysis of the
Roper resonance.

A comparison of our analysis with the results of the hypercentral constituent quark model of the
Genova group (HCQM) \cite{HCQM} shows large differences between the quark model and the
experimental analysis. In the case of the magnetic form factor $G_M^*(Q^2)$  of the $\Delta$
excitation this has long been known from many quark model calculations. But most dramatically this
now appears for the longitudinal $\Delta$ excitation, $G_C^*(Q^2) \sim S_{1/2}(Q^2)$, where the
HCQM gives an almost zero result but the experimental numbers are quite large with relatively small
uncertainties.

In a recently developed dynamical model (DMT) \cite{DMT} for pion-nucleon scattering and pion
electroproduction we have shown that this puzzle can be solved in terms of the pion cloud
contribution. In a dynamical approach the pion-loop integrals give additional contributions at the
resonance position which are neglected in the usually applied K-matrix approach. This pionic
contribution can fully explain the longitudinal $\Delta$ excitation and also solves the old puzzle
of the missing strength in the magnetic $\Delta$ excitation.

In conclusion, we find that microscopic calculations without pionic degrees of freedom (e.g.
constituent quark models) of nucleon resonance excitation cannot be directly compared with
experimental numbers obtained in a K-matrix approach, but have to be compared with a dynamical
model that treats the pionic contribution explicitly.

\newpage \section{Scalar Mesons}

Convenors: M. B\"uscher, V. Kleber and F.P. Sassen

\subsection{Investigation of light scalar mesons $a_0/f_0(980)$ at COSY}
\addtocontents{toc}{\hspace{2cm}(M. B\"uscher)\par}

M. B\"uscher

{\em Institut f\"ur Kernphysik, 
Forschungszentrum J\"ulich GmbH, 52425 J\"ulich,
Germany}

One of the primary goals of hadronic physics is the understanding of
the internal structure of mesons and baryons, their production and
decays, in terms of quarks and gluons. The non-perturbative character
of the underlying theory --- Quantum Chromo Dynamics (QCD) --- hinders
straight forward calculations. QCD can be treated explicitly in the
low momentum-transfer regime using lattice techniques\cite{lattice},
which are, however, not yet in the position to make quantitative
statements about the light scalars. Alternatively, QCD inspired
models, which use effective degrees of freedom, are to be used. The
constituent quark model is one of the most successful in this respect
(see e.g.\ \cite{quarkmodel}). This approach naturally treats the
lightest scalar resonances $a_0/f_0$(980) as conventional $q\bar{q}$
states. However, they have also been identified with $K\bar{K}$
molecules \cite{KK_molecules} or compact $qq$-$\bar{q}\bar{q}$ states
\cite{4q_states}. It has even been suggested that at masses below 1.0
GeV a complete nonet of 4-quark states might exist
\cite{4q_nonet}.

The existing data base is insufficient to conclude on the structure of
the light scalar mesons and additional observables are urgently called
for. In this context the charge-symmetry breaking (CSB) $a_0$-$f_0$
mixing plays an exceptional role since it is sensitive to the overlap
of the two wave functions. It should be stressed that, although
predicted to be large long ago \cite{achasov2}, this mixing has not
unambiguously been identified yet in corresponding experiments.

At COSY an experimental program has been started which aims at
exclusive data on $a_0/f_0$ production close to the $K\bar{K}$
threshold from $pp$, $pn$, $pd$ and $dd$ interactions --- i.e.\
different isospin combinations in the initial state
\cite{cosy-11,momo,a+_proposal,a0f0_proposal,dd_proposal,css2002}.
Data taken at the COSY-11 \cite{cosy-11} and MOMO \cite{momo}
facilities for $pp\to ppK^+K^-$ and $pd\to\ ^3HeK^+K^-$ reactions are
not conclusive about the contribution from the $a_0/f_0$ resonances.
In the first experiment at the ANKE spectrometer the reaction $pp
{\to} dK^+\bar{K^0}$ has been measured exclusively at excess energies
$Q=46$ and 106 MeV above the $K\bar K$ threshold. The data for the
lower energy have already been analyzed and show that most of the
$K\bar K$ pairs are produced in a relative $s$-wave which has been
interpreted in terms of dominant $a_0^+$ production,
$\sigma(pp{\to}da_0^+{\to}dK^+\bar{K}^0) = 83\%\cdot
\sigma(pp{\to}dK^+\bar{K}^0)= 32$~nb \cite{a+_PRL}. Based on these
data (and on model calculations for the different initial isospin
configurations \cite{brat}) it is concluded that the production cross
section for the light scalar resonances in hadronic interactions is
sufficiently large to permit systematic studies at COSY.

If in the future a neutral-particle detector (like WASA) will be
available at COSY, the $a_0/f_0$ can also be detected via their decays
$a_0^\pm\to \pi^\pm\eta$, $a_0^0\to \pi^0\eta$ and $f_0\to
\pi^0\pi^0$, $f_0\to\pi^+\pi^-$.  The  strange decay channel
$a_0/f_0\to K_SK_S$ should be measured in parallel and the results can
be compared with those from ANKE for the charged kaons. Since it is
possible to manipulate the isospin of purely hadronic reactions it is
possible to identify observables that vanish in the absence of
charge-symmetry breaking (CSB) \cite{miller,ANKE_WS}.  Most promising
for the extraction of CSB effects seems to be the reaction $dd\to
\alpha (\pi^0\eta)$. Since the initial deuterons and the $\alpha$
particle in the final state have isospin $I{=}0$ (``isospin filter''),
any observation of $\pi^0\eta$ production in this particular channel
is a direct indication of CSB and can give information about the
$a_0$-$f_0$ mixing \cite{CH}.

The idea behind the proposed experiments is the same as behind recent
measurements of CSB effects in the reactions $np\to d\pi^0$
\cite{opper2} and $dd\to\alpha\pi^0$ \cite{stephenson} which found
broad interest also outside the ``hadron community''. However, the
interpretation of the signal in the case of the scalar mesons is
largely simplified compared to the pion case. Since the $a_0$ and the
$f_0$ are rather narrow overlapping resonances, the $a_0$-$f_0$ mixing
in the final state is enhanced by more than an order of magnitude
compared to CSB in the production operator (i.e.\ ``direct'' CSB
violating $dd\to \alpha a_0$ production) and should give the dominant
contribution to the CSB effect via the reaction chain $dd\to \alpha
f_0(I{=}0) \to \alpha a_0^0(I{=}1) \to \alpha (\pi^0\eta)$ \cite{CH}.

\subsection{Experiments on Light Scalar Mesons}
\addtocontents{toc}{\hspace{2cm}(Eberhard Klempt)\par}

Eberhard Klempt

{\em        Helmholtz-Institut f\"ur Strahlen-- und Kernphysik, \\
        Universit\"at Bonn
        Nu{\ss}allee 14-16, D-53115 Bonn, GERMANY
}

Lattice gauge calculations predict the existence of glueballs$^1$.
In particular a scalar glueball is firmly expected at a
mass of about 1730 MeV. This prediction has led to an intense study
of scalar isoscalar interactions and to the discovery of 
many new meson resonances$^2$. 
The number of scalar states observed seems to exceed
the number of states which can be accommodated in the quark
model even when two states, the $a_0(980)$ and $f_0(980)$, 
are interpreted as \kkb\ bound states and are removed
from the list. However, none of these states has a decay 
pattern which is consistent with that of a pure glueball. 
A reasonable interpretation of the number of states 
and of their decay pattern is found only
when mixing of scalar $q\bar q$ states with the scalar glueball 
is taken into account$^3$. 
\par
However, alternatetive interpretations are conceivable or even
more likely. The pole positions of scalar mesons are strongly
influenced by their couplings to the final--state mesons. If the 
'true' meson masses can be identified with the $K$--matrix poles,
then a very different scalar meson mass spectrum evolves which
is easily mapped on the scalar mass spectrum calculated from
a relativistic Lipman-Schwinger equation using instanton--induced
forces as interaction kernel$^4$. A broad 'background' in the data
can then possibly be identified with a scalar glueball$^5$.
\par
A third interpretation questions the $q\bar q$ nature of 
the $f_0(1370)$. This state is a cornerstone 
in all $q\bar q$-glueball mixing scenarios. If it is interpreted as
$\rho\rho$ system interacting by pion exchange, then 
the remaining scalar states are easily fit into
a nonet classification. If this interpretation should be correct
there would be no room for resonant scalar gluon-gluon interactions,
no room for the scalar glueball$^6$. 
\par

\vspace{0.5cm}
{\large \bf References}

\begin{enumerate}
\vspace*{-3mm}
\item C.~J.~Morningstar and M.~Peardon,
``The glueball spectrum from an anisotropic lattice study,''
Phys.\ Rev.\  {\bf D60} (1999) 034509
\vspace*{-3mm}
\item C.~Amsler, ``Proton antiproton annihilation and meson 
spectroscopy with the Crystal  Barrel,''
Rev.\ Mod.\ Phys.\  {\bf 70} (1998) 1293.
\vspace*{-3mm}
\item C. Amsler and F.E. Close, ''Evidence for a scalar glueball'',
Phys. Lett. {\bf B353} (1995) 385; ''Is $f_0(1500)$ a scalar glueball?''
Phys. Rev. {\bf D53} (1996) 295.
\vspace*{-3mm}
\item M.~Koll, R.~Ricken, D.~Merten, B.~C.~Metsch and H.~R.~Petry,
``A relativistic quark model for mesons with an 
instanton induced  interaction,''
Eur.\ Phys.\ J.\  {\bf A9}, 73 (2000)
\vspace*{-3mm}
\item V.~V.~Anisovich and A.~V.~Sarantsev,
``K-matrix analysis of the (IJ(PC) = 00++)-wave in the mass region 
below  1900-MeV,'' Eur.\ Phys.\ J.\ A {\bf 16}, 229 (2003).
\vspace*{-3mm}
\item E.~Klempt, B.~C.~Metsch, C.~R.~Munz and H.~R.~Petry,
``Scalar mesons in a relativistic quark model with instanton induced forces,''
Phys.\ Lett.\  {\bf B361} (1995) 160
\end{enumerate}

\subsection{Production of the $f_0(980)$ in peripheral
  pion-proton reactions }
\addtocontents{toc}{\hspace{2cm}(F.P. Sassen)\par}

{F.P. Sassen}

{\em Institut f\"ur Kernphysik, 
Forschungszentrum J\"ulich GmbH, 52425 J\"ulich,
Germany}

To describe pion production in the peripheral pion proton reaction
$\pi^- p\rightarrow\pi^0\pi^0 n$ it is sufficient to consider the final
state interaction of the outgoing mesons since due to the interaction
kinematics the nucleon will only interact once with the mesons. This is
why we can use the J\"ulich meson exchange model for meson-meson
interaction to describe the final state interaction. The production
itself is described by the emission of a $\pi$ / $a_1$ from the
nucleon vertex. Because of the high momentum of these particles we use
Regge amplitudes as parametrised in \cite{Achasov:1998pu} to describe the
initial production. We do not account for interference of the two
production mechanisms since they mainly occur in different helicity
amplitudes as the $a_1$ mainly conserves the helicity of the nucleon
and the $\pi$ mainly flips the helicity of the nucleon. To reproduce
the $t$-dependence of the integrated experimental cross section we
need to attach exponential form factors to the nucleon vertex as it is
done in the experimental analysis. We fix the parameters from the
$\frac{d\sigma}{dt}$ data given in \cite{Gunter:2000am}. 
Comparing our predicted
production rates to the BNL data \cite{Gunter:2000am} for
$\pi^-p\rightarrow\pi^0\pi^0n$ in various $t$-bins we find good
agreement. We are further able to reproduce the kaon production data
from \cite{Etkin:1982sg}. We see how the structure at 
$m_{\pi\pi}=1\textrm{ GeV}$ is formed by an interplay  of a
high lying state stemming from the confinement spectrum with a $K\bar{K}$
molecule formed due to the attractive $t$-channel meson exchanges. Sadly the
data is inconclusive when it comes to the properties of this state as
we will discuss now. First let us elaborate in which way results
depend on the momentum of the initial pion beam. First of all we
expect an overall scaling of the data by a factor
$\frac{1}{q_{\mathrm{beam}}^2s_{\mathrm{tot}}}$ which is of no
interest since the data is unnormalised anyway. Then there is the
momentum transfer  $t$ to the nucleon which is either limited by the
cuts applied in the experimental analysis or the kinematic limits
$t_{\mathrm{min}}(m_{\pi\pi},q_{\mathrm{beam}})$ and
$t_{\mathrm{max}}(m_{\pi\pi},q_{\mathrm{beam}})$, which will
produce a different shape by reducing the momentum transfer
range in a $m_{\pi\pi}$ dependent way. We find that for the data considered ($q_{\mathrm{beam}}=18.3,
38.0,100.0\textrm{ GeV}$) this effect only shows up at invariant two
pion masses $m_{\pi\pi} >\approx 1.8\textrm{ GeV}$ 
which is anyhow to high to be dealt with in our model. When we compare
our model and the three data sets of GAMS $38\textrm{ GeV}$
\cite{Alde:1995jj}, GAMS $100\textrm{ GeV}$ \cite{Alde:1998mc} and BNL
$18.3\textrm{ GeV}$ \cite{Gunter:2000am} in
the case of low momentum transfer to the nucleon 
$0.01<-t<0.2\textrm{ GeV}^2$ we find good agreement except for the
GAMS $38\textrm{ GeV}$ data deviating just below 
$m_{\pi\pi}\approx 1\textrm{ GeV}$ and above 
$m_{\pi\pi}\approx 1.3\textrm{ GeV}$. In the high momentum transfer range the
experimental $t$ binnings do not coincide but we can at least join
some bins to guarantee a common lower limit which is where most of the
production should take place anyway. We find that although our model
can reproduce the high momentum transfer $0.3<-t<1.5\textrm{ GeV}^2$
data of BNL it strongly overestimates the $0.3<-t<1.0\textrm{ GeV}^2$
GAMS data. Instead of attributing this to a strong production of high
invariant two pion masses $1.1<m_{\pi\pi}<1.4\textrm{ GeV}$ in
the momentum transfer range $1.0<-t<1.5\textrm{ GeV}^2$ one should
take a revealing look at the momentum transfer
range $0.3<-t<0.4\textrm{ GeV}^2$. Here data from both experiments is
available.  One finds that already at those momentum transfers there is a
discrepancy between the data sets. Adjusting the parameters of the
high lying $f_0^\prime$ confinement pole there is no problem to
describe the full $t$ range of the GAMS data. Since the parameters of
the $f_0^\prime$ influence the physics around 
$m_{\pi\pi}\approx 1\textrm{ GeV}$ a precise measurement of high
resolution close to
threshold $K\bar{K}$ scattering data would be highly appreciated. This
is especially true since the physics of the $f_0^\prime$ differs
considerably between different models where anything from a strong
admixture to
the $K\bar{K}$-molecule \cite{Lohse:1990ew} down to a very tiny
contribution which finetunes the pole position \cite{Krehl:1997rk} is possible.
Another way to extract information on the structure of the $f_0(980)$
and the admixture of the $f_0^\prime$ to it is suggested in
\cite{Krewald:2003lk} where the authors investigate the lifetime of
kaonium. Using a non-relativistic effective field-theory approach
they come to the conclusion, that the life time of kaonium depends
strongly on the scattering length in the $K\bar{K}$-channel which in
turn constrains the structure of the $f_0(980)$ resonance. So a
lifetime measurement of kaonium and a close to threshold measurement
of $K\bar{K}$ production are promising ways to put constrains on the nature
of the $f_0(980)$.

\subsection{ Classification of 
$a_{0}(980)$ and $f_{0}(980)$}
\addtocontents{toc}{\hspace{2cm}(E. van Beveren)\par}

Eef van Beveren and George Rupp

{\it CFT, Universidade de Coimbra, and CFIF, Instituto Superior T\'{e}cnico,
Lisboa, Portugal}

When two scattered mesons are coupled to a resonance in the interaction
region, the scattering cross section for the process shows a peak
near the mass of that resonance. This can be described by a pole in the
scattering matrix $S(E)$, at a complex value of the total invariant mass $E$.
However, when the two scattered mesons are coupled to a confined system,
with an infinite spectrum of resonant states, then, besides an infinity
of complex-energy poles near the masses of the confinement states,
the scattering matrix also contains poles of a different origin.
In general, the latter phenomenon yields no visible signals
in the scattering data, since these poles are far away from the real $E$ axis.
But in some cases their effects can be measured in experiment.
The low-lying scalar mesons $\sigma (600)$, $\kappa (800)$,
$a_{0}(980)$, and $f_{0}(980)$ represent an example of such observable effects
in elastic and inelastic meson-meson scattering \cite{ZPC30p615}.
Other examples are the broad $D(2290)$ and narrow $D_{s}(2317)$ charmed scalar
resonances \cite{PRL91p012003}.

The $S$ matrix for the scattering of meson pairs which in the interaction
region are coupled to an infinite spectrum of confinement states can
be straightforwardly modelled, at least for low energies \cite{HEPPH0201006}.
In the case of elastic scattering one obtains \cite{HEPPH0306155}
\begin{eqnarray*}
\bra{\vec{p}\,} T\ket{{\vec{p}\,}'} & = &
\fnd{\lambda^{2}}{4\pi}\;
\sum_{\ell =0}^{\infty}(2\ell +1)\;
P_{\ell}\left(\hat{p}\cdot{\hat{p}\,}'\right)\;
\fnd{\dissum{n=0}{\infty}
\fnd{{\cal J}^{\ast}_{n\ell}(p){\cal J}_{n\ell}(p')}
{E(p)-E_{n\ell_{c}}}}
{1+i\pi\lambda^{2}\mu p
\left(\fnd{\lambda a}{\mu_{c}}\right)^{2}
\dissum{n=0}{\infty}
\fnd{{\cal J}^{\ast}_{n\ell}(p){\cal H}^{(1)}_{n\ell}(p)}
{E(p)-E_{n\ell_{c}}}} \;\;\;\; .\\ & &
\end{eqnarray*}
For this expression one can study the pole structure
and the related observable signals.
In Ref.~\cite{AIPCP660p353} it is shown why the $a_{0}(980)$ and $f_{0}(980)$
scalar resonances are of exactly the same nature as the $\sigma (600)$ and
$\kappa (800)$.

\subsection{Scalar Mesons and instanton induced Quark Forces}
\addtocontents{toc}{\hspace{2cm}(H.R. Petry)\par}

H.R. Petry

{\em HISKP, Bonn University}

The theory of instantons is reviewed and the canonical instanton induced force
is derived. To demonstrate its effect, this force is studied within a
relativistic quark model. The mass pattern of the pseudoscalars naturally
arises and the $U_A(1)$--problem is solved. For scalar mesons the force
changes sign and produces a low--lying flavor--singlet. It is discussed how
this pattern fits into a dynamical calculation with additional dynamical resonances.

\newpage \section{Mesonic Bound States}

Convenor: A. Gillizer 

\subsection{Mesonic Bound States}
\addtocontents{toc}{\hspace{2cm}(Albrecht Gillitzer)\par}

{Albrecht Gillitzer}

{\it Insitut f\"ur Kernphysik, Forschungszentrum J\"ulich}

The cleanest experimental access to the in-medium properties of hadrons is
opened in the study of bound or quasi-bound states of these hadrons with
nuclei.
The power of this experimental approach has been demonstrated in the study
of deeply bound states~{[1]}.
In this way well-defined spectroscopic information has been obtained for
negative pions in deeply bound 1$s$ states of Pb and Sn nuclei.
These pionic states may be understood as nuclear halo states with significant
overlap of the pion and nuclear density distributions, forming an environment
which is the closest possible approach to the ideal case of ``real'' pions in
nuclear matter at rest.

The strength of the $s$-wave potential is with good accuracy determined by
the 1$s$ binding energy and width.
In comparison to the data on 1$s$ states in light symmetric nuclei, the new
information from 1$s$ states in heavy nuclei allows to separate the isovector
from the isoscalar part.
A significant enhancement of the repulsive isovector strength as compared to
the free pion-nucleon interaction, as determined precisely in the study of
pionic hydrogen, was observed. To the lowest order the isovector $s$-wave
$\pi{}N$ interaction is directly related to the pion decay constant $f_{\pi}$
which is the order parameter of chiral symmetry breaking of QCD.
The deduced strength of the in-medium isovector $\pi{}N$ interaction may be
interpreted as a reduction of the chiral $\overline{q}q$ condensate by 22\%
at the density probed by the $\pi^-$ ($\sim0.6\rho_0$).

Possibilities are open for future experimental studies of meson nuclear bound
states at COSY. Nuclear bound states may exist for $\eta$, $\omega$, and
$K^-$/$\overline{K^0}$ mesons. A study of the $\eta$-$^3$He system at the
TOF detector has been proposed~{[2]} and a first exploratory measurement
was already done. The anomalously high $\eta$ production cross section
observed previously in the reaction $pd\rightarrow{}^3{\rm{}He}\eta$ has been
interpreted as a consequence of the existence of a quasi-bound state in the
$\eta$-$^3$He system.
The strategy of the TOF experiment is based on the expectation that, in case 
such a bound state is formed in $pd$ collisions, the $\eta$ meson will be
predominantly absorbed in the $\eta{}N\rightarrow{}\pi{}N$ channel.
$\eta$ absorption on a neutron will then result in a peculiar final state
consisting of a $p\pi^+$ pair back-to-back in the cm frame and a spectator
di-proton traveling at cm velocity.
The same technique can also be employed in order to search for a quasi-bound
$\omega$-$^3$He state.

Large attractive potentials have been also predicted for antikaons in nuclear
matter. There are indications for a significant attractive in-medium potential
from heavy-ion collisions, but the interpretation is strongly model-dependent.
This issue may be resolvable in the future only in the search for and study of
nuclear bound states of antikaons. The usage of $K^-$ beams seems to be the
most direct way to populate such states, but the availability of good quality
$K^-$ beams in the future is uncertain. The usage of proton beams does not
allow the population of bound $\overline{K}$ states without momentum transfer,
but if the binding is strong enough bound systems might nevertheless be formed.
A possible experiment could be the search for the reaction
$p+ {^4{\rm{}He}}\rightarrow\overline{K^0}\otimes{}^4{\rm{}He}+K^0+p$
with $K^0\rightarrow{}\pi^+\pi^-$ and
$\overline{K^0}\otimes{}^4{\rm{}He}\rightarrow{}\Lambda{}pd\rightarrow{}p\pi^-pd$.
A new magnetic 4$\pi$ detector with charged particle ($K^+$) identification
would allow the even more promising search for a bound $K^-\otimes{}^3{\rm{}He}$
system.

\vspace{0.5cm}
{\large{\bf References}} 

\begin{description}
\item[{[1]}]
 H. Geissel {\it et al.}, Phys. Rev. Lett. 88 (2002) 122301,
 H. Geissel {\it et al.}, Phys. Lett. B 549 (2002) 64,
 K.~Suzuki {\it et al.}, nucl-ex/0211023, subm. to Phys. Rev. Lett.
\item[{[2]}]
 A. Gillitzer {\it et al.}, proposal COSY-102, (2001)
\end{description}

\subsection{The New Pionic-Hydrogen Experiment at PSI: First Results}
\addtocontents{toc}{\hspace{2cm}(D.~Gotta)\par}

  
D.~Gotta for the PIONIC HYDROGEN collaboration

{\em Institut f\"ur Kernphysik, Forschungszentrum J\"ulich, D-52425 J\"ulich}

In pionic hydrogen the hadronic pion--nucleon interaction manifests itself by 
a change of binding energy and natural line width of the atomic s states. Experimentally 
accessible are X--ray transitions to the 1s ground state emitted in the last 
de--excitation step of the atomic cascade. In the framework of $Heavy-Baryon\,\chi PT$,  
1s--level shift $\epsilon_{1s}$ and width $\Gamma_{1s}$ are given unambiguously by the 
isoscalar and isovector scattering lengths $a^{+}$ and $a^{-}$ [1] by Deser--type formulae\,[2]. 
Furthermore, from the isovector scattering length the pion--nucleon coupling constant is 
obtained by means of the Goldberger--Miyazawa--Oehme sum rule\,[3].

To improve on the accuracy for $\epsilon_{1s}$ and $\Gamma_{1s}$ as compared to previous 
measurements\,[4], a thorough study of a possible influence of de--excitation 
processes during the atomic cascade is essential. A first series of measurements has been 
completed by the new pionic--hydrogen experiment at the Paul--Scherrer--Institut (PSI), 
using the new cyclotron trap, a cryogenic target and a Bragg crystal spectrometer 
equipped with spherically bent silicon and quartz crystals and a large--area 
CCD array\,[5]. 

To identify radiative de--excitation of the $\pi H$ system -- when 
bound into complex molecules formed during collisions 
$\pi^{-} p+H_{2}\rightarrow [(pp\pi^{-})p]ee$\,[6] -- the $\pi H(3p-1s)$ transition energy 
was measured in the density range from gaseous $H_{2}$ of 3.5 bar to liquid. X--ray transitions 
from molecular states should show up as low--energy satellites with density dependent 
intensities. No density effect could be established and it is concluded that the decay of 
molecules is dominated by Auger emission. The new (preliminary) value of 
${\it\epsilon_{1s}} = 7.120\pm 0.008 {\,+\,0.009 \atop \,-\,0.008}$~eV for the hadronic shift\,[7,8] 
is in agreement with the result of the previous experiment\,[4]. 

At present, the accuracy for the hadronic broadening (7\%\,[4]) is limited by a 
not precisely known correction to the measured line width originating from the
Doppler broadening due to Coulomb de--excitation. For that reason the 
precisely measured 1s--level shift in pionic deuterium was used together 
with the shift of hydrogen in the determination of the $\pi N$ scattering 
length and the pion--nucleon coupling constant[9]. This procedure, however, requires a 
sophisticated treatment of the 3--body system $\pi D$. In addition, so far it cannot be 
excluded that the radiative decay channel after molecule formation is strongly enhanced in 
deuterium compared to hydrogen.

To study the influence of Coulomb de--excitation, the three $\pi H(2p-1s)$, 
$\pi H(3p-1s)$, and $\pi H(4p-1s)$ transitions were measured. 
An increase of the line width was found for the $2p-1s$ line compared to the 
$3p-1s$ transition, which is attributed to the higher energy release available 
for the acceleration of the $\pi H$ atom. This result is corroborated 
by a reduced line width of the $4p-1s$ line. From the $\pi H(4p-1s)$ transition a 
safe upper limit of 850~meV is determined [8,9]. Data analysis is in progress.

From about 2005 on, Coulomb de--excitation will be studied in detail in the 
absence of strong--interaction effects by measuring K transitions from muonic 
hydrogen. Together with the detailed knowledge of the response function 
by using the ECRIT [10] and a newly developed cascade code [11], which includes
the velocity dependence of the atomic cascade, a sufficiently accurate 
correction for the Doppler broadening in pionic hydrogen should be
achieved to extract the hadronic broadening at the level of about 1\%.

\vspace{0.5cm}
{\large \bf References}

\noindent [1] 
~\,V.~E.~Lyubovitzkij and A.~Rusetski, Phys. Lett. B 494 (2000) 9.

\noindent [2] 
~\,G.~Rasche and W.~S.~Woolcock, Nucl. Phys. A 381 (1982) 405.

\noindent [3] 
~\,M.~L.~Goldberger, H.~Miyazawa and R.~Oehme, Phys. Rev. 99 (1955) 986.

\noindent [4] 
~\,H.--Ch.~Schr\"oder et al., Eur. Phys. J. C 21 (2001) 433.

\noindent [5] 
~\,PSI experiment R--98.01, http://pihydrogen.web.psi.ch.

\noindent [6] 
~\,S.~Jonsell, J.~Wallenius, and P.~Froelich, Phys. Rev.  A 59 (1999) 3440.

\noindent [7] 
~\,M.~Hennebach, thesis, University of Cologne, 2003.

\noindent [8] 
~\,D.~F.~Anagnostopoulos et al.,  proceedings EXA'02, Vienna, Austria, 28--30 November 2002. 

\noindent [9] 
~\,T.~E.~O.~Ericson, B.~Loiseau, and A.~W.~Thomas, Phys. Rev. C 66 (2002) 014005.

\noindent [10] 
D.~F.~Anagnostopoulos et al., Nucl. Instr. Meth. B 205 (2003) 9.
 
\noindent [11] 
T.~Jensen and V.~M.~Markushin, proceedings Menu 2001, $\pi N$ newsletter 16 (2002) 358.

\subsection{Deeply Bound Pionic Atoms and Chiral Restoration in
Nuclei}%
\addtocontents{toc}{\hspace{2cm}(K. Suzuki)\par}

K. Suzuki

{\em Physik-Department, Technische Universit\"at M\"unchen,
D-85748 Garching, Germany,}





A goal of this work is measuring the degree of chiral symmetry
restoration in a nuclear medium through the determination of
the isovector $\pi N$ interaction parameter in the pion-nucleus
potential by studying deeply bound 1s states of $\pi^-$ in
heavy $N>Z$ nuclei \cite{Zphys,Yamazaki:98,PRC,PRL}.
For that we performed a systematic experimental studies of 1s $\pi^-$
states in a series of Sn isotopes, which were produced with the
Sn($d$,$^3$He) reactions.
One of the advantages of using Sn isotopes is that we can produce the
1s $\pi^-$ states as the most dominant quasi-substitutional states,
$(1s)_{\pi^-}(3s)_n^{-1}$, because of the presence of the 3s orbital
near the Fermi surface, as theoretically predicted~\cite{Umemoto}.
Another merit is to make use of isotopes over a wide range of
$(N-Z)/A$ to test the isospin dependence~\cite{Kienle:01}.

We observed spectra, $d^2\sigma/dE/d\Omega$, on mylar-covered
$^{116}$Sn, $^{120}$Sn, $^{124}$Sn targets as function of the $^3$He
kinetic energy~\cite{ksuzuki}.
The overall spectrum shapes for the three Sn targets were found to
be in good agreement with the predicted ones~\cite{Umemoto}.
The spectra were decomposed according to the theoretical
prescription of Ref.~\cite{Umemoto},
from which we could precisely determine the 1s binding energies
($B_{1s}$) and widths ($\varGamma_{1s}$).
The obtained data of binding energies and widths of the 1s $\pi^-$
states in $^{115,119,123}$Sn combined with those of symmetric light
nuclei ($^{16}$O, $^{20}$Ne and $^{28}$Si) yielded
$b_1 = -0.116 \pm 0.007~m_{\pi}^{-1}$~\cite{ksuzuki}.


The magnitude of the observed $|b_1|$ is significantly enhanced over
the free $\pi N$ value~\cite{Schroeder99}, which translates into a
reduction of ${f_\pi ^*}^2$ as~\cite{Weise2000, Kolomeitsev:02}
$$
R  = \frac{b_1^{\rm free}}{b_1} =0.78 \pm 0.05\nonumber
   \approx \frac{b_1^{\rm free}}{b_1^* (\rho_e)}
    \approx \frac{f_{\pi}^* (\rho_e)^2}{f_{\pi}^2}
    \approx 1 - \alpha \rho_e,$$
where we made use of the fact \cite{Geissel:02,YH02}
that the solution with a local-density-dependent parameter,
$b_{1}^* (\rho) = b_1^{\rm free}/(1 - \alpha \rho(r))$,
is equivalent to that using a corresponding constant parameter
$b_1 = b_1^{\rm free}/(1-\alpha \rho_e)$
with an effective density $\rho_e\approx 0.6 \rho_0$.

The above value hence implies that the chiral order parameter,
$f_{\pi}^* (\rho)^2$, would be reduced by a factor of $\approx 0.64$
at the normal nuclear density $\rho = \rho_0$.
If a theoretical value, $m_{\pi}^* \approx m_{\pi} + 3~{\rm MeV}$
(averaged over $\pi^+$ and $\pi^-$ \cite{Meissner2002}), is inserted
into an in-medium Gell-Mann-Oakes-Renner relation
\cite{Hatsuda94, Thorsson},
$\langle \bar{q}q \rangle _{\rho_0}/\langle \bar{q}q \rangle _{0}$
will be $(m_{\pi}^*/m_{\pi})^2 \times (1 - \alpha \rho_0)
\approx 0.67$,
which is in good agreement with the theoretical value of 0.65
\cite{Drukarev}, as an clear evidence for the partial restoration of
chiral symmetry in a nuclear medium.

\subsection{Deeply bound pionic states in $\rm d{^{136}Xe}{\to} {^{135}Xe_{\pi}}{^3He}$ at CELSIUS}
\addtocontents{toc}{\hspace{2cm}(P.-E.Tegn\'er)\par}

Chr.Bargholtz$^{a}$, 
B.Chernyshev$^{b}$, 
L.Ger\'en$^{a}$, 
M.Gornov$^{b}$$\dag$, 
V.Grebenev$^{b}$, 
Y.Gurov$^{b}$, 
B.H\"oistad$^{c}$,
K. Lindberg$^{a}$, 
V.Sandukovsky$^{d}$, 
R. Shafigullin$^{b}$, 
P.-E.Tegn\'er$^{a}$\footnote{Corresponding author},
K. Wilhelmsen Rolander$^{a}$

  {\em $^{a}$Department of Physics, Stockholm University, SE-106 91 Stockholm, Sweden\\
    $^{b}$ Department of Elementary Particle Physics, Moscow Engineering Physics Institute, Kashirskoe sh. 31, RU-115409 Moscow, Russia\\
    $^{c}$Department of Radiation Sciences, Uppsala University, Box 535, SE-751 21 Uppsala, Sweden\\
    $^{d}$Laboratory of Nuclear Problems, Joint Institute for Nuclear Research, Joliot-Curie Street, RU-141980 Dubna, Russia\\
    $\dag$ deceased}

In experiments at the CELSIUS accelerator and storage ring\cite{Ekstrom} we
are studying the possibility to produce and to investigate in detail deeply
bound pionic states of xenon in order to obtain information on pion properties
in the nuclear medium.  The existence of relatively narrow deeply bound pionic
states in heavy nuclei has been predicted theoretically \cite{Toki} and the
observation of a peak corresponding to the $2p$ state of the $\rm {^{207}Pb}$
pionic atom \cite{Yamazaki} has been reported. Recently the production of the
$1s$ state in $\rm {^{205}Pb}$ was also observed \cite{Geissel}. These
experiments confirmed that, in the case of a lead target, the probability to
populate atomic $2p$ states is large compared to that for populating the $1s$
state, due to the lack of $s$-state neutrons in the outer nuclear shell. For
xenon the outer neutron shell contains $s$-state neutrons and a relative
increase in the population probability of the $1s$ pionic-atom state is
expected. The closed shell nucleus $\rm {^{136}Xe}$ is suggested, by Umemoto
et al. \cite{Umemoto2}, to be a particularily good candidate as a target for
the observation of deeply bound $1s$ states in $\rm (d,^3He)$ reactions.

Here we report results of an investigation of the production of pionic atoms
of xenon in $\rm (d,{^{3}He})$ reactions using natural xenon as a target
\cite{Andersson} and a status report on the experiment using an isotopically
pure gas of $\rm {^{136}Xe}$. The production of the pionic $1s$ state in xenon
is observed in both experiments.

\subsection{Chiral Dynamics of Deeply Bound Pionic Atoms}
\addtocontents{toc}{\hspace{2cm}(N. Kaiser)\par}

N. Kaiser

{\em Physik Department T39, Technische Universit\"{a}t M\"{u}nchen,
    D-85747 Garching, Germany}

\medskip

We present a systematic calculation, based on two-loop chiral perturbation
theory, of the pion selfenergy in isospin-asymmetric nuclear matter [1]. A
proper treatment of the explicit energy dependence of the off-shell $\pi^-$
selfenergy together with (electromagnetic) gauge invariance of the
Klein-Gordon equation turns out to be crucial for description of the deeply
bound pionic atom states [2]. The minimal replacement $\omega\to\omega-V_c(r)$,
with $V_c(r)$ the (attractive) Coulomb potential, in the $\omega$-dependent
$\pi^-$ selfenergy generates effectively a large part of the "missing" s-wave
repulsion. Accurate data for the binding energies and widths of the 1s and 2p
states in $^{205}$Pb, $^{207}$Pb and several Sn-isotopes are well reproduced.
The connection with the in-medium change of the pion decay constant $f_\pi$
is clarified. At leading order the energy dependence effects can be interpreted
in an equivalent energy independent optical potential in terms of a reduced
in-medium pion decay constant $f_\pi^*(\rho)$.

\medskip

\vspace{0.5cm}
{\large{\bf References}} 

[1] N. Kaiser and W. Weise, Phys. Lett. B512 (2001) 283.

[2] E.E. Kolomeitsev, N. Kaiser and W. Weise, Phys. Rev. Lett. 90 (2003)
092501.

\subsection{Search for Eta--Nucleus Bound States at Big Karl, COSY }
\addtocontents{toc}{\hspace{2cm}(B.J.Roy)\par}

B.J.Roy for the GEM collaboration

{\em IKP, Forschungszentrum, J\"{u}lich, Germany}

The idea that the eta meson can form a bound system with nucleus was first
put forward long ago by Peng\cite{peng} and also by Liu\cite{liu} and was
based on the observation of Bhalerao \& Liu\cite{bhal} that the low energy
eta-nucleon interaction is attractive. Since then a number of calculations have
been performed to predict binding energy and width of eta-nucleus bound
system. Such a quasi-bound system (known as eta-mesic nucleus) if exists, can
be a very useful tool to investigate eta-nucleus interaction. At the
experimental side, a few attempts were made to look for the existence of
such systems but have failed, so far, to reach any conclusive direct evidence. At
COSY, we have a dedicated experimental programme on the investigation of
eta-nucleus interaction. In one part of the programme, the eta-nucleus
quasi-bound states will be studied employing the magnetic spectrometer
BigKarl and a large acceptance plastic scintillator detector ENSTAR. The
reaction proposed to be studied is p+A$\rightarrow$$^3$He+(A-2)$_{\eta }$ at
recoil free kinematics. $^{3}$He will be detected by BigKarl while the decay
products of ''eta-mesic nucleus'', namely, protons and pions will be
registered at ENSTAR. The coincidence measurement between BigKarl and ENSTAR
is expected to enhance sensitivity of the measurement. The detector ENSTAR
\cite{brep1, brep2} which has been built at BARC, Mumbai, is in its final
stages of fabrication at Juelich and will be tested in fully assembled conditon
at COSY in March 2004.
The data taking measurement will then follow.

The other part of the programme focuses eta-production in a nucleus (not
necessarily through the bound state formation) e.g., p+$^{6}$Li$\rightarrow $%
$^{7}$Be+$\eta $. In this case, the heavy recoil nucleus $^{7}$Be will be
detected at the focal plane of BigKarl and eta-production events can be
identified by missing mass method. In order to improve position resolution
necessary for such a measurement, a multi-wire avalanche counter(MWAC) which
functions at relatively low pressure will be used. Such a MWAC has been
constructed and testing of the chamber is in progress.

\subsection{The $\eta-3N$ system} 
\addtocontents{toc}{\hspace{2cm}(A.\ Fix)\par}

{A.\ Fix and H.\ Arenh\"ovel}

{\em Institut f\"ur Kernphysik, Johannes Gutenberg-Universit\"at Mainz,
D-55099 Mainz, Germany}

Low-energy interaction and photoproduction of $\eta$ mesons on three-body
nuclei is considered \cite{FA2}. The model containing 4-body scattering
approach to the 
$\eta-3N$ system does not predict existence of the $\eta 3N$ bound
states. Strong rise of the $\eta$ yield observed in experiments just above 
threshold is a consequence has to be ascribed to the virtual $\eta 3N$ state.
Only with $\Re ea_{\eta N}$ more than 1 the bound state may be generated. The
energy dependence of the cross section, which is determined by the low-energy
parameters of the $\eta ^3$He elastic scattering agrees quite well with that
observed in the $pd$ collision. On the other hand the magnitude as well as
the form of the experimental cross section for $^3$He$(\gamma,\eta)^3$He
presented in \cite{Pfeiff02} is not explained.

\subsection{Photoproduction of $\eta$ mesic $^3$He}
\addtocontents{toc}{\hspace{2cm}(Marco Pfeiffer)\par}

Marco Pfeiffer

{\em II. Physikalisches Institut, Uni Giessen}

In recent years, a lot of effort has gone into the investigation
of possible quasi-bound states of $\eta$ mesons and nuclei - so-called $\eta$
mesic nuclei. Determinations of the $\eta$-N scattering length \cite{bhalerao}
suggest a strong attraction of the $\eta$ meson and nuclei that might
result in quasi-bound states even for light nuclei \cite{ueda} like $^3$He.
However, up to now there is no solid experimental proof for
such states neither for light nor for heavier nuclear systems, although
recent investigations of $^{11}$B seem to show indications for a bound state
\cite{sokols}.
In the present experiment, the possible existence of an $\eta$ mesic nucleus
is investigated via the $\eta$ photoproduction from $^3$He. The experiment
has been carried out at the MAMI accelerator facility in Mainz
\cite{mami,ahrens94s} with real
photons up to an energy of 820 MeV using the TAPS detector
system \cite{rainerieee}. Two decay
channels of $\eta$ mesic $^3$He have been considered and compared to each
other: the decay into the coherent $\eta$ channel and the competing decay
into $\pi^0$pX. In the latter case, the relative opening angle of the
$\pi^0$-proton pair is expected to be near 180$^\circ$ for the decay of an
$\eta$ mesic state. The excitation functions for both decay channels show a
structure near $\sqrt{s}$=1485 MeV indicating the existence of a bound
$\eta$-$^3$He state. This conjecture is supported by the measured angular
distributions while theoretical calculations
\cite{kamalov,tiator,shev1s,shev2s,fix_co2} for the coherent $\eta$ cross
section fail to reproduce the measured distribution.

\newpage \section{$pn$ induced Reactions}

Convenors: R. Schleichert and C. Hanhart

\subsection{Nucleon-Nucleon (NN) Elastic Scattering}
\addtocontents{toc}{\hspace{2cm}(Heiko Rohdjess)\par}

Heiko Rohdjess 

{\it Helmholtz-Institut f\"ur Strahlen- und Kernphysik der
Universit\"at Bonn}\\

The large amount of recent data by the NN programs at
PSI \cite{PSI}, 
SATURNE II \cite{SAT}, 
the PINTEX experiment at IUCF \cite{IUCF}, 
and EDDA at COSY \cite{EDDA} 
has helped to improve and extend phase-shift parameterizations
\cite{PSA}  
of the world NN database and has set tight limits on the coupling of
dibarionic resonances to the isovector NN channel \cite{rohdjess}.
Up to about 1 GeV precise phase shifts and
scattering amplitudes have been obtained unambiguously, while 
at higher energies PSA solutions tend to disagree 
\cite{PSA} and various sets of
amplitudes fit the data \cite{EDDA}.
Above 1.1 GeV np data is scarce, especially
double-polarization observables and phase shifts are
practically unknown.
However, phase shift parameters of certain partial waves 
are crucial in describing the initial state interaction and thus the
scale of the total cross section
in near-threshold meson production \cite{Hanhart:1998rn}.
Here, COSY could further contribute to the pp data base by measuring
certain triple-spin observables which would resolve the remaining
ambiguities even with moderate experimental precision.
For quasi-free scattering of protons on vector-polarized deuteron
targets -- where the proton in the deuteron acts as a spectator -- seems to
be the most promising way to access np spin observables at COSY in the
near future \cite{Rathmann}. 
A detector for low-energy spectator protons has already
been developed \cite{Barsov:2003zc}.
Quasi-free data in meson production and pp
and np elastic scattering at CELSIUS and SATURNE \cite{QF} indicate that 
the spectator model may be valid at COSY energies over a sizeable
angular range. 

On the theoretical side chiral perturbation theory
 has 
has been pushed to fourth order in the chiral extension and a
description of the NN database up to 300~MeV (e.g. \cite{Entem:2003ft})
almost comparable to high precision potentials has been
achieved. Above the pion-production threshold at 300~MeV 
meson-exchange models, where the inelasticity is introduced by coupling
to the N$\Delta$- and $\Delta\Delta$-channels, give a fair description
of the data up to 1~GeV \cite{Elster}. However, these models fail
badly in the 1-3~GeV range, which may not be surprising due to the
large number of other resonances not included in the
model. Only recently theoretical work has started, taking up the
challenge of describing NN scattering in the resonance region.

\subsection{Introduction}
\addtocontents{toc}{\hspace{2cm}(R. Schleichert)\par}

R. Schleichert

{\em Institut f\"ur Kernphysik, Forschungszentrum J\"ulich,
D-52425 J\"ulich, Germany}

\subsection{PN versus PP induced production reactions}
\addtocontents{toc}{\hspace{2cm}(V. Baru)\par}

V. Baru

{\em Institut f\"ur Kernphysik, Forschungszentrum J\"ulich,
D-52425 J\"ulich, Germany}

Meson production in nucleon-nucleon collisions is a rather attractive
research field in hadron physics at intermediate energies. One of the main
questions to be answered in the investigation of the
reactions $NN\to NNM$ is the dominant mechanism for the
production process. 
For example, the understanding of the
mechanism of meson production in NN-collisions could help to
pin down the special features of baryonic resonances involved in the
production process. Also these reactions may be used as a tool for
the investigation of the NN and MN interactions, in particular as a
test for existing NN and MN-models. 

Note that so far the most efforts both of the theory and
experiment are focused on the study of the isotriplet channel,
i.e. on the reactions such as $pp\to NNM$. We would like to stress,
however, that the study of the channel with I=0 is not less important
since pn-reactions contain 
information about the production process 
complementary to the pp-channel. 
A good example illustrating the potential of the pn-channel is
the cross section ratio $R\displaystyle=\frac{\sigma_{pn\to pnM}}{\sigma_{pp\to ppM}}$.
Consider this ratio in the Born approximation and, for example, for
the production of isoscalar particles.  When isovector
meson exchanges dominate in the production process the cross section
ratio equals to 5, whereas when the isoscalar exchanges are
the most relevant ones the ratio is 1. Thus, the ratio contains
information about the production process which can not be extracted if
only one of two NN channels is investigated. 

However, one  has to
keep in mind that the ratio depends also on effects of the NN
interaction in the initial and final states. Note also that NN FSI and
ISI effects are different in different isospin channels. 
The inclusion of the NN FSI to the ratio is not a problem since NN FSI
effects are known rather well in the near threshold region for
different produced mesons. 
The influence of short ranged NN ISI effects is known much worse or not known at all
for mesons heavier than $\eta$ produced in the pn-channel.
This problem is due to the lack of the experimental data related to
the measurements on the neutron target at high energies, i.e. for pn-scattering.
Therefore data on pn phase shifts and inelasticities, which are the
direct input for the theoretical models, are available only below
$T_{lab}=1300$ MeV ($\eta$ threshold corresponds to  $T_{lab}=1250$ MeV).
Thus, in order to allow a quantitative investigation of the production
of mesons heavier than $\eta$ one has to improve our
knowledge about ISI effects. Therefore new experimental data
on pn elastic scattering at higher energies would be very desirable.

In the summary we would like to emphasize that only combined analysis of
pp and pn channels might allow to draw conclusions about the
mechanisms of production for different mesons. 
In papers \cite{Han,Bar} this was done for $\pi$ and $\eta$ meson
production respectively. 
It is also worth noting that
in a microscopic model calculation such as \cite{Han,Bar}  the cross section
ratio results from a rather delicate interplay between several ingredients, in particular the
actual production mechanism, effects from the FSI as well as from
the ISI and interference effects. 
This means that one should be very cautious with drawing conclusions on the
cross section ratio or other experimental observables from simpler model analyses 
when only one or two of those ingredients are considered explicitly.

\subsection{$\eta$ -- meson production and interaction 
in the 3--N system}
\addtocontents{toc}{\hspace{2cm}(J. Smyrski)\par}

J. Smyrski for the COSY-11 collaboration

{\em Institute of Physics, Jagellonian University, Cracow, Poland}

Recent data on the $pp \rightarrow pp \eta$ reaction measured
by the COSY-11 collaboration and earlier data from experiments at SATURNE and at CELSIUS
turned out to be very useful for studies of the $\eta$ production mechanism 
in nucleon--nucleon collisions [1,2,3].
In order to study the $\eta$ production in the three--nucleon system,
which is much less explored compared to the two--nucleon case,
the COSY-11 collaboration performed measurements of the $pd \rightarrow $$^3He \eta$ reaction
for five excess energies in the range from $Q = $5 to 40~MeV.
The preliminary results for the total cross section 
are consistent with predictions of the two--step
reaction model [4]. Further tests of this model should be possible with data 
for the $dp \rightarrow {^{3}He} \eta$ reaction which were taken for four excess energies:
$Q = $0.7, 4.0, 6.9, 10~MeV and at present are being analyzed.
For simultaneously measured reaction $dp \rightarrow {^{3}He} X$, 
a clear peak is visible in the missing mass spectra at the mass of the $\pi^{0}$--meson. 
This confirms usability
of the COSY-11 detection system for measurements of the $dp\rightarrow {^3He} \pi^{0}$ 
reaction at the $\eta$ threshold and thus opens the possibility of studying
the $\pi^{0} - \eta$ mixing [5] and 
of determining the sign of the ${^3He} - \eta$ scattering length [6].

The COSY-11 system supplemented with a neutron detector and a spectator proton detector
was also successfully applied for 
measurements of the quasi--free $p n \rightarrow p n \eta$ reaction 
with a proton beam scattered on neutrons in a deuteron target.
In the next step, it is planned to study the gluon content in the $\eta'$--meson wave function
by comparison of the ratio 
$R_{\eta'}=\sigma(pn\rightarrow pn\eta') / \sigma(pp\rightarrow pp\eta')$
with $R_{\eta}=\sigma(pn\rightarrow pn\eta) / \sigma(pp\rightarrow pp\eta)$ as proposed by
Moskal [7].

\vspace{0.5cm}
{\large \bf References}

\begin{enumerate}

\item[[1]] J. Smyrski et al., Phys. Lett {\bf B474}(2000)182.
\item[[2]] P. Moskal et al., Nucl. Phys. {\bf A721}(2003)657c.
\item[[3]] P. Moskal et al., submitted to Phys. Rev. {\bf C}.
\item[[4]] G. F\"aldt and C. Wilkin, Nucl. Phys. {\bf A587}(1995)769.
\item[[5]] A. Magiera and H. Machner, Nucl. Phys. {\bf A674}(2000)515.
\item[[6]] V. Baru at al., nucl-th/0303061.
\item[[7]] P. Moskal, COSY Proposal No. 100, IKP, FZ-J\"ulich, Germany, April 2001. 
\end{enumerate}

\subsection{$\vec p\vec n \to pn$ at ANKE}
\addtocontents{toc}{\hspace{2cm}(F. Rathmann)\par}

F. Rathmann

{\em Institut f\"ur Kernphysik, Forschungszentrum J\"ulich,
D-52425 J\"ulich, Germany}

While for the $pp$ system the situation is fair, the nucleon-nucleon database  on $pn$ scattering up to 3 GeV is only scarcely populated. Above about 1 GeV the
$pn$ data base is almost empty. The phase shift analyses
need to be improved/extended towards higher energies. It should be noted that in order to gain further insight
into the $NN$ interaction, {\it both} isospin channels need to be studied. But for the $pn$ system even the data are missing.
A recent theoretical assessment of the situation has clearly shown, that the theoretical understanding of the $NN$  system above 1 GeV is still unsatisfactory \cite{machleidt}.

\begin{table}[h]
\begin{center}
\begin{tabular}{c|r|r|c}
    & $I$ & $I_3$ & status \\\hline
pp  & 1   & 1     & ok\\\hline
np  & 1   & 0     & !\\\hline
nn  & 1   & -1    & ?\\\hline
np  & 0   &  1    & !\\\hline
\end{tabular}
\end{center}
\end{table}

Although the ANKE dipole spectrometer is not ideally suited for polarization experiments because of a lack of azimuthal symmetry,
some polarization observables, i.e. analyzing power $A_y$ and one of the four spin correlation coefficients ($A_{yy}$) could also be measured using the ANKE dipole spectrometer with a vertically polarized polarized hydrogen gas target, bombarded by the polarized proton beam from COSY. In particular at small cm scattering angles below $\theta=30^\circ$, accessible with ANKE, the $pn$ data base is almost empty at all energies \cite{rentmeester}.

\newpage \section{Instrumentation}

Convenor: F. Rathmann

\subsection{Photon Detection at COSY}
\addtocontents{toc}{\hspace{2cm}(V. Hejny)\par}

V. Hejny

{\em Institut f\"ur Kernphysik, Forschungszentrum J\"ulich}

{\bf Motivation}

Photon detection is a very important tool to obtain information in
nuclear physics. Nearly all of the produced neutral mesons
($\pi^\circ, \eta, \omega, \eta', \mathrm{f_0}, \mathrm{a_0}$) have
decay branches resulting in multiple photon final states. Large photon
detectors have been built at many accelerators and are being operated
with great success. However, at COSY-J\"ulich, such a device is
missing up to now. All existing detectors are designed for charged
particle detection and, thus, neutral particles are reconstructed
via missing mass analysis.

{\bf What can be done?}

Although direct identification of neutral particles is important, a
stand-alone photon detector will be a minor improvement only. In order
to get the full power in event reconstruction it has to be combined
with a number of other detector modules. The requirements for
these modules strongly depend on the physics reactions, a fully
universal detector will be hard to realize. Taking into account these
boundary conditions, one can think of three
different ways to get such a detector system: (i) designing a complete
new detector, optimized for use at COSY, (ii) taking over an available
4$\pi$ detector (like it has been done with Crystal Barrel at ELSA or
Crystal Ball at MAMI) or (iii) adding photon detection to an existing
installation at COSY.

{\bf Photon Detection at ANKE}

One accepted proposal for (iii) (COSY PAC \#21 [1]) is the
extension of ANKE [2] with a large acceptance electromagnetic 
calorimeter. This combination allows coincidence measurements of
charged particles at $0^\circ$ and neutral mesons decaying into
photons, and, thus, is ideally suited for reactions close to
threshold, or when one charged particle can be missed. Since the
time this proposal has been submitted, various R\&D studies were
done in order to meet the requirements given by the ANKE setup,
namely the available space, the stray magnetic field and the
compatibility with other installations at ANKE. The result is
a detector based on PbWO$_4$ crystals read out by shielded
fine-mesh photomultipliers (e.g. Hamamatsu R5505). The next
steps would be the final overall design and starting the
construction. 

{\bf Conclusions}

There is common agreement that photon detection has to be
exploited at COSY. There is an existing proposal for a dedicated
photon detector at ANKE, and, furthermore, realization can
start soon. However, there might be the possibility for a
more general solution. Recent discussions have indicated
that the CELSIUS ring (Uppsala) may terminate operation within
the next few years. At the moment negatiations are in progress
to study the possibility to have the WASA detector avaible for
operation at COSY.

\vspace{0.5cm}
{\large \bf References}

[1] COSY Proposal \#83: A photon detector for COSY, November 2000. \\{}
[2] S. Barsov et al., Nucl. Inst. and Meth. A462 (2001) 364.

\subsection{Silicon Microstrips}
\addtocontents{toc}{\hspace{2cm}(R. Schleichert)\par}

{R. Schleichert}

{\em Institut f\"ur Kernphysik, Forschungszentrum J\"ulich}

\subsection{A Large Tracking Detector with Self-Supporting Straw Tubes}
\addtocontents{toc}{\hspace{2cm}(P. Wintz)\par}

{P. Wintz for the COSY-TOF Collaboration}

{\em
  Institute for Nuclear Physics, Research Centre Juelich, 
           52425 Juelich, Germany

  }


  A novel technique to stretch the anode wire inside drift tubes (straws)
  simply by the gas over-pressure will be presented. 
  Combined with a dedicated gluing method of the single straws
  to close-packed double-layers, large but self-supporting
  detector planes can be built. No heavy end or support structures are 
  needed reducing
  the overall detector weight to an absolute minimum, allowing a
  clean and background-free, highly transparent tracking.  

  The detector will consist of 
  more than 3000 straws filling up a cylindrical tracking volume of
  $1m$ diameter and $30cm$ length, close behind the target in the
  COSY-TOF barrel. 
  Each straw tube consists of a $30\mu m$ thick mylar film
  with a length of $1m$ and $10mm$ diameter, the tubes inside being
  aluminised to be used as cathode. In the tube centre a $20\mu m$ thick W/Re
  wire is used as anode. Cylindrical end plugs made from ABS close the
  tubes at both ends.

  The projected spatial resolution of
  $200\mu m$ for a 3-dimensional track reconstruction allows to
  resolve the target interaction point with sub-mm precision.  
  The chosen granularity (straw diameter) of $10mm$  
  provides an almost continuous tracking with up to $30$ hits 
  for a precise reconstruction
  of complex track patterns like the $\Lambda$ decay ($c\tau \simeq
  8cm$) and polarisation by its $p$ and $\pi^-$ decay tracks.
  The detector 
  with a total mass of less than $15kg$ will be operated in vacuum, 
  but will have an added wall thickness of 3mm mylar, only.

  Detector design, production experience and first
  results will be discussed.

\def\Journal#1#2#3#4{{#1} {\bf #2}, #3 (#4)}

\def\NCA{Nuovo Cimento}
\def\NIM{Nucl. Instrum. Methods}
\def\NIMA{{Nucl. Instrum. Methods} A}
\def\NPB{{Nucl. Phys.} B}
\def\PLB{{Phys. Lett.}  B}
\def\PRL{Phys. Rev. Lett.}
\def\PRD{{Phys. Rev.} D}
\def\ZPC{{Z. Phys.} C}
\def\PPNP{Prog. Part. Nucl. Phys.}
\def\EPJ{{Eur. Phys. J.} A}

\subsection{The Crystal Barrel Detector}
\addtocontents{toc}{\hspace{2cm}(V. Crede)\par}

V. Crede

{\em Helmholtz-Institut f\"ur Strahlen-- und Kernphysik,\\ 
Universit\"at Bonn, Nu{\ss}allee 14-16, D53115 Bonn, Germany}

From 1989-1996, the Crystal-Barrel spectrometer~\cite{Ref-11} was used at the
Low-Energy Antiproton Ring (LEAR), CERN, to study the products of
$\overline{\mbox{p}}\mbox{p}$ and $\overline{\mbox{p}}\mbox{d}$ annihilations.
One goal was to investigate the annihilation dynamics in the non-perturbative
regime of QCD. On the other hand, spectroscopy of light mesons was succesfully
carried out and new particles were discovered, the scalar meson~f$_0(1500)$,
for instance.\par\smallskip In 1997, the Crystal-Barrel calorimeter was moved
to Bonn. Since that time, it has formed the basis of a new experimental setup
in order to carry out baryon spectroscopy in photoproduction experiments at
the Electron Stretcher Accelerator (ELSA). The general idea of the experiment
is to use the barrel in combination with suitable forward detectors. In a
first series of experiments, Time-Of-Flight walls were used in order to
measure meson production directly at threshold. At present, the TAPS detector
is operational in current experiments.  The latter has fast trigger
capabilities to cope with high photon rates and provides high granularity in
the forward direction. The total setup covers almost 4$\pi$~solid angle, thus
is an ideal configuration in order to measure multi-photon final states over
the full dynamical range. In 2002/2003, data with linearly polarized photons
have been taken off the proton as well as off the neutron with the
Crystal-Barrel detector and TAPS in the forward direction.
\par\smallskip
The Crystal-Barrel is a modular electromagnetic calorimeter consisting of
1380~elements. The CsI(Tl) crystals, 30~cm long corresponding to 16.1
radiation lengths, are each viewed by a single photodiode mounted on the edge
of a wavelength shifter placed on the rear end of the crystal. This
configuration improves the light collection between the output of the CsI
crystal and the photodiode. Each crystal covers $6^\circ$ in $\theta$ (polar
angle) and $6^\circ$ in $\phi$ (azimuthal angle) except for the two sets of
three layers closest to the beam axis where the angular acceptance is
increased to $12^\circ$ in $\phi$. The range of polar angles covered by the
complete calorimeter is $12^\circ \leq \theta \leq 168^\circ$. In order to
make best use of TAPS, the Crystal-Barrel was opened downstream~($\pm
30^\circ$), i.e.~3 forward crystal rings were removed.
\par\smallskip
In Bonn, the Crystal-Barrel readout electronics is based on a 128-channel-ADC
Fastbus system. Readout software was developed which was optimized for high
data rates, thus to allow for large amount of data in cooperation with the
data aquisition. In addition, the calorimeter is part of the second-level
trigger.  Based on a discriminator signal from each crystal (threshold
15~MeV), a cellular logic, a fast cluster encoder (FACE), is able to count
clusters and, therefore, to trigger on chosen multiplicities of real photons
within $\approx 4~\mu$sec.

\subsection{Pellet Target$^*$}
\addtocontents{toc}{\hspace{2cm}(P.~Fedorets)\par}

$\underline{P.~Fedorets}^a$, V.~Balanutsa$^a$, W.~Borgs$^c$, M.~B\"uscher$^c$, A.~Bukharov$^b$,
V.~Chernetsky$^a$, V.~Chernyshev$^a$, M.~Chumakov$^a$, A.~Gerasimov$^a$, V.~Goryachev$^a$,
L.~Gusev$^a$, S.~Podchasky$^a$

he~ANKE pellet target is being developed to study reactions
with meson production and cross sections less then 0.1 $\mu$b in $pp$, $pn$,
$pd$ and $dd$
collisions~\cite{plan}. It is expected to reach
luminosities higher than $L=10^{32} $cm$^{-2}$s$^{-1}$ with standard
COSY beam conditions (proton beam intensity of a~few times $10^{10}$).

The~target consists of the~following parts~\cite{anke97}:
a~cryostat for the~production of solid hydrogen (or deuterium) pellets;
a~dumping cryostat for the~collection of the pellets;
a~vacuum system;
gas supply systems (hydrogen, nitrogen and helium);
temperature and pressure control systems.

The~upper part of the~cryostat contains a~system of heat exchangers,
which provide a~stable
stream of liquid $H_2$ at given values of temperature and pressure.
The~cooling materials are liquid $N_2$ and evaporated liquid $He$.  
There are  three stages of $H_2$ cooling. First, $H_2$
is cooled by liquid $N_2$ down to 100 K. Then the~$H_2$ is cooled
down to 22 K in an~coaxial-tube heat exchanger by $He$ gas
coming from the~condenser. The~final cooling and transformation
into the~liquid $H_2$ is performed in the~condenser
with the~help of cold $He$ gas.

The~liquid hydrogen jet is produced with a~60 $\mu$m nozzle inside
the~triple point chamber (TPC). The~temperature of the~condenser
channel and the~TPC must be close to the~triple point value
($T_{tr}$=14 K, $p_{tr}\sim 100$ mbar) with an accuracy of not less than 0.1 K.
The~liquid hydrogen jet is broken into microdroplets of about 70 $\mu$m
diameter by acoustic excitation. The~piezo-electric generator with a~resonance
frequency of 3 kHz is mounted coaxially at the~side of the~nozzle.   

Behind the~TPC the~hydrogen droplets pass into two vacuum chambers, in which
the~pressure
is reduced from 100 mbar (triple point value) down to
$10^{-7}$ mbar (COSY ring pressure). During flight through the~vacuum
the~droplets freeze and a~continuous flow of frozen hydrogen 
pellets is formed.

In the~moment all main parts of the~pellet target are produced and
assembled. The~target has been installed at a~test place in the~COSY hall
and operates in tests since May 2001.
The~necessary values of temperatures and pressures can be kept stable for
a long time.
The~production of the~stable liquid hydrogen jet and splitting
this jet into droplets has been achieved.

The~next main step is to pass droplets into the~vacuum chambers through
sluices of 0.6 mm diameter. For this purpose the~adjusting systems
for the~sluices are under construction. First pellet production
is foreseen for test in autumn 2003. 

$^a$ Institute for Theoretical and Experimental Physics, Moscow, Russia\\
$^b$ Moscow Power Energy Institute, Moscow, Russia\\
$^c$ Institut f\"ur Kernphysik, Forschungszentrum J\"ulich, Germany\\
$^*$ Supported by grants RFFI02-02-06518, RFFI02-02-16349, RFFI99-02-04034,
DFG-443RUS-113, WTZ-RUS-684-99, ISTC-1966\\

\subsection{Cluster Target}
\addtocontents{toc}{\hspace{2cm}(N. Lang)\par}

{N. Lang}

{\it Universit\"at M\"unster, Germany}

\subsection{Polarized Internal Target and Lamb-Shift Polarimeter at ANKE}
\addtocontents{toc}{\hspace{2cm}(R. Engels)\par}

\underline{R. Engels$^2$}, R. Emmerich$^1$, J. Ley$^1$, H.Paetz
 gen. Schieck$^1$,
 M. Mikirtichyants$^2$,\\
F. Rathmann$^2$, H. Seyfarth$^2$, T. Ullrich$^2$ and A. Vassiliev$^3$

{\small\it
(1) Institut f\"ur Kernphysik, Universit\"at zu K\"oln, Z\"ulpicher
Str. 77, 50937 K\"oln, Germany\\
(2) Institut f\"ur Kernphysik, Forschungszentrum J\"ulich, 52425 J\"ulich,
Germany\\
(3) High Energy Physics Dept., St. Petersburg Nucl. Phys. Inst.,
188350 Gatchina, Russia\\
}

The following components for the polarized internal gas target at
ANKE have been built and tested:
\begin{enumerate}
\item The polarized atomic beam is produced by an ABS [1]. The intensity
 of this source is about $7.6\times 10^{16}$ atoms/s for hydrogen with
 a polarization of $-0.97\le p_z \le +0.91$. Atomic beams with atoms
 in any single hyperfinestate can be produced. The beam profile
 ($\sigma =3.6 $ mm) is rather small.\\
 With this high intense beam a target density up to $1\times 10^{14}$
 atoms/cm$^2$ can be expected in a storage cell.
\item With a Lamb-shift polarimeter [2] it is possible to measure the
 occupation numbers of the single hyperfine states in an atomic beam of
 hydrogen
 or deuterium. For the full atomic beam of the ABS it takes only 2 s
 to measure the polarization with an error of less than $1\%$.
 When only a small fraction ($10^{-4}$) of the beam intensity is extracted
 from the storage cell, the background dominates the measured signal.
 With a new getter pump in the ionizer of the Lamb-shift polarimeter this
 background will be decreased substantially and the polarimeter can be
 used to measure the polarization of atoms effusing from the storage cell.
\end{enumerate}

\vspace{0.5cm}
{\large \bf References}

{\begin{description}
\item{[1]}
M. Mikirtytchiants et al., Proceedings of the 9th International Workshop
on Polarized Sources and Targets (PST01), Nashville, edited by V. P. Derenchuk
and B. Przewoski (World Scientific, Singapore, 2002), p. 47.\\
\item{[2]}
R. Engels, R. Emmerich, J. Ley, M. Mikirtytchiants, H. Paetz gen. Schieck,
F. Rathmann, H.Seyfarth, G. Tenckhoff and A. Vassiliev; accepted by
Rev. Sci. Instr., to be published Nov. 2003\\
\end{description}}

\IfFileExists{\jobname.bbl}{}
 {\typeout{}
  \typeout{******************************************}
  \typeout{** Please run "bibtex \jobname" to optain}
  \typeout{** the bibliography and then re-run LaTeX}
  \typeout{** twice to fix the references!}
  \typeout{******************************************}
  \typeout{}
 }

\subsection{Solid Polarized Target}
\addtocontents{toc}{\hspace{2cm}(H. Dutz)\par}

{H. Dutz}

{\em Helmholtz-Institut f\"ur Strahlen-- und Kernphysik,\\ 
Universit\"at Bonn, Nu{\ss}allee 14-16, D53115 Bonn, Germany}

\newpage \section{HESR}

Convenors: R. Maier and H. Str\"oher

\subsection{The New Research Facilities at GSI}
\addtocontents{toc}{\hspace{2cm}(H. Gutbrod)\par}

H. Gutbrod

{\it GSI-Darmstadt, Germany}

\subsection{HESR-Design}
\addtocontents{toc}{\hspace{2cm}(B. Franzke)\par}

B. Franzke, K. Beckert, O. Boine-Frankenheim, P. Beller, A. Dolinskii, F.
Nolden, M. Steck,

{\it GSI-Darmstadt, Germany}

A major objective of the new beam facility proposed by GSI [1] is a
High Energy Storage Ring (HESR, [2]) for high luminosity internal
target experiments with stored and cooled antiprotons in the momentum
range 0.8-14.5 GeV/c (see figure). The concept includes suitable
accelerator/storage ring designs as well as adequate techniques for
efficient production, fast stochastic cooling and accumulation of
antiproton beams. It must fit properly into the complete project:
optimal efficiency and extensive time/instrument sharing with heavy
ion acceleration and rare isotope beam production. A new linac has to
be developed delivering 70 MeV protons with pulse currents of 70 mA,
0.1 ms pulse length and 5 Hz repetition frequency for injection to the
existing SIS18. The latter shall l serve as energy booster for a new,
fast cycled 100Tm synchrotron accelerating $2.8\times 10^{13}$ protons per
cycle to 29 GeV. A net accumulation rate of about $1\times 10^8$ antiprotons
every 5 s should allow for experiments at correspondingly high
luminosity of up to $2\times 10^{32} cm^{-2} s^{-1}$.

\begin{figure}
\vspace{7cm}
\includegraphics{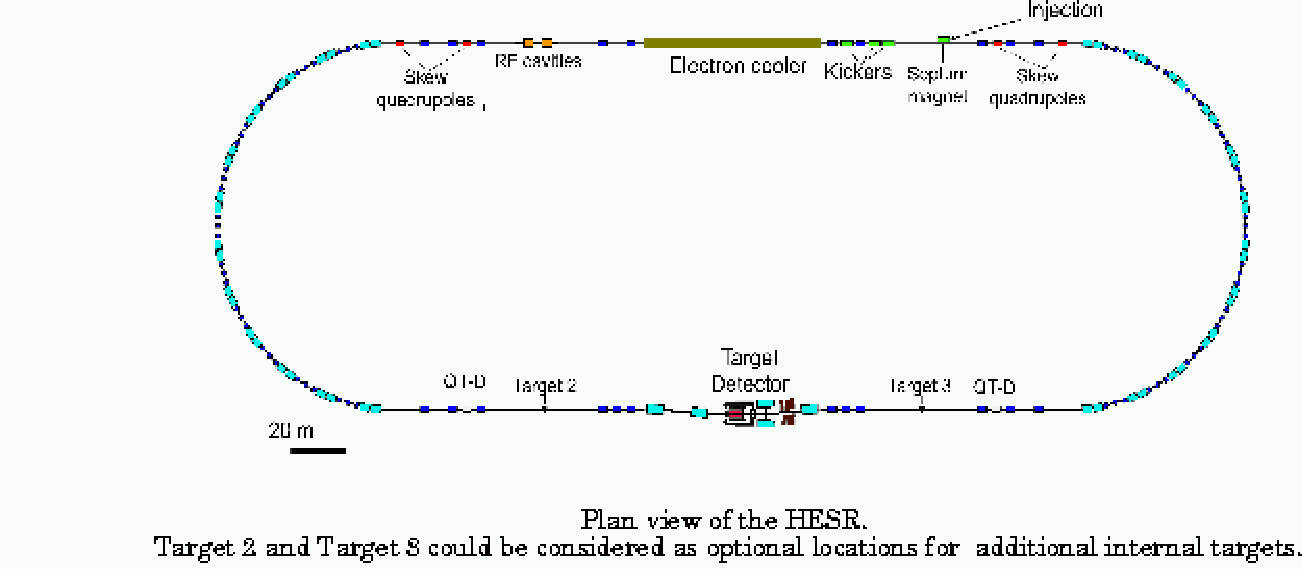}
\end{figure}

The lattice design of the HESR provides a race-track shaped ring with
super-conducting arcs and two approximately 120 m long straight
sections. One of straight sections is required for the installation of
an electron cooling device of about 30 m length capable provide cooled
antiproton beams over the full energy range [3]. Stochastic cooling
might be considered as an complementary installation or as a fall-back
option. Internal target experiments with one or two target positions
and large detectors are foreseen at the other long straight
section. Recent results of numerical calculations of equilibrium beam
properties taking into account intra-beam scattering, internal target
effects and electron cooling give hope that beam heating might be
compensated by means of the proposed EC-device [4]. Investigations of
collective beam effects due to space charge and beam-wall coupling are
underway.

\vspace{0.5cm}
{\large{\bf References}} 

[1]An Internal Accelerator Facility for Beams of Ions and Antiprotons,
Conceptual Design Report, GSI-Darmstadt, November 2001,
http://www-new.gsi.de/zukunftsprojekt/veroeffentlichungen\_e.html.

[2]B. Franzke, K. Beckert, P. Beller, O. Boine-Frankenheim,
A. Dolinskii, F. Nolden, M. Steck, Conceptual Design of a Facility for
Internal Target Experiments with Antiprotons up to 15 GeV/c, Proc. of
the 8th Europ. Particle Accelerator Conf., Paris, 2002, p. 575.

[3]V. Parkhomchuk et al., Technical Feasibility of Fast Electron
Cooling of Antiproton Beams in the Energy Range 0.8-14.5 GeV, Internal
BINP-GSI-Report, Oct. 2003, to be published.

[4]A. Dolinskii et al., Beam Simulations for HESR, Internal
GSI-Report, Aug. 2003, to be published.

\subsection{Spectroscopy in the Charm domain}
\addtocontents{toc}{\hspace{2cm}(H. Koch)\par}

H. Koch

{\em Institut f\"ur Experimental Physik, Ruhr Universit\"at Bochum, Germany} 

Experiments with antiprotons at LEAR and Fermilab have started a new
era in hadron spectroscopy. Candidates for bound states with gluonic
degrees of freedom were found and the spectroscopy in the charmonium
region has reached a new level of precision. It is planned to extend
measurements of this kind at GSI/Darmstadt. Antiprotons with energies
up to 15 GeV will interact with a Hydrogen cluster target in a storage
ring (HESR) with high luminosity. The charged and neutral reaction
products will be registered in a 4p-detector (PANDA). The talk gives
an overview on the physics program envisaged for HESR/PANDA as far as
experiments on H2-targets are concerned. The experiments with nuclear
targets are discussed in [1].  
Fig 1 gives on overview on the
QCD-systems, which can be studied.

\begin{figure}
\vspace{8.5cm}
\includegraphics{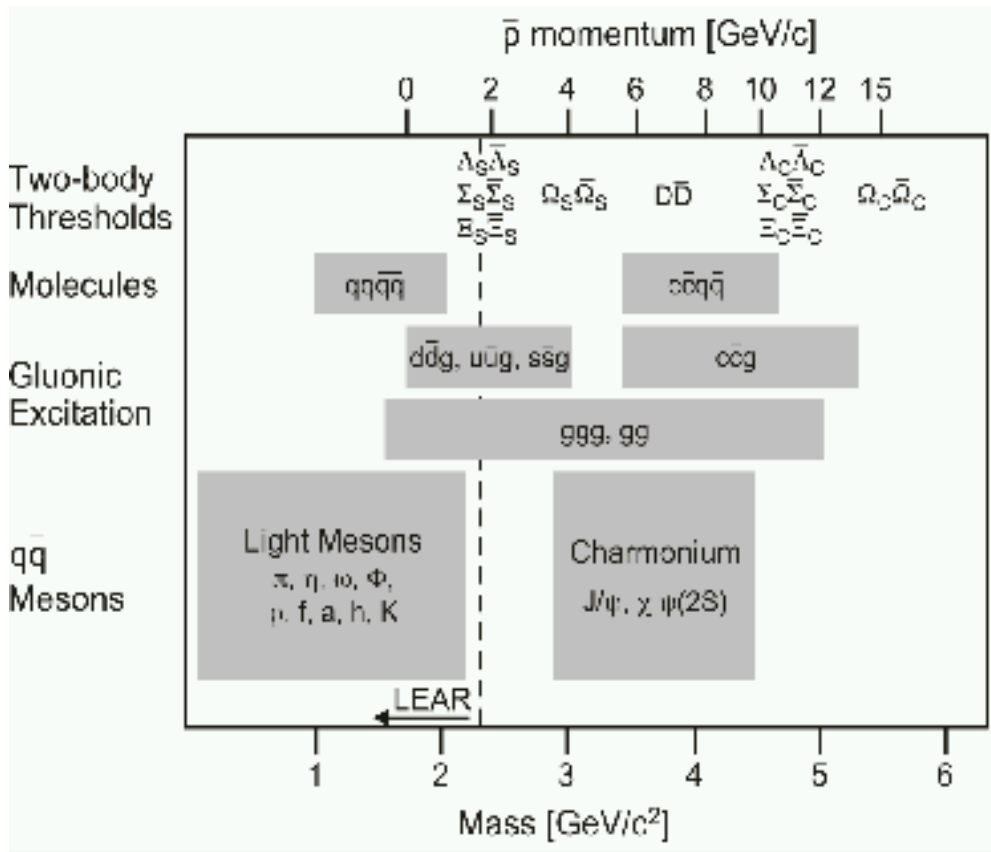}
\caption{Overview on QCD systems}
\end{figure}
Production rates for different kinds of particles are given in Table 1.

\begin{table}
\begin{center}
\begin{tabular}{|c|c|c|}
\hline
Final State
& cross section
& \#reconstr. Events/y \\
\hline
Meson resonance + anything
& 100 $\mu$b
& $10^{10}$ \\
& 50 $\mu$b
& $10^{10}$ \\
& 2 $\mu$b
& $10^8 \ (10^5)$ \\
& 250 nb
& $10^7$ \\
& 630 nb
& $10^9$ \\
& 3.7 nb
& $10^7$ \\
& 20 nb
& $10^7$ \\
& 0.1 nb
& $10^5$ \\
\hline
\end{tabular} 
\end{center}
\caption{
{\it Production cross sections and event rates per year for selected final states}}
\end{table}

Some physics highlights are discussed more in detail Precision measurements in
the Charmonium system.  In contrast to e+e--experiments all states are
accessible in formation processes allowing very good mass resolutions Search
for Charmed Hybrids These states are expected in the mass range above 3.9
GeV/c2. Some of them will have spin exotic quantum numbers, particularly the
ground state (JPC = 1-+) Search for Heavier Glueballs These states are
expected in the energy domain up to 5 GeV/c2. Some of them may be narrow
because of quantum number conservation. They would be searched for in exotic
channels like ff or fh.  Spectroscopy of systems with Open Charm Recently, a
narrow -state has been seen by BaBar (later confined by Belle and Cleo) with
an unexpected mass, making it a good candidate for a charm-exotic state, e.g.
a DK-molecule. More states of this kind are expected to exist and can be seen
in -induced reactions.  Further options of the program are Spectroscopy of
Charmed Baryons Direct CP-violation in the Charm sector Reversed Deeply
Virtual Compton Scattering For more details of the physics program see
references [2,3], for more details on the PANDA-Detector see [4].

\vspace{0.5cm}
{\large{\bf References}}

[1]J. Ritman, Charm in Medium, This Workshop

[2]Conceptual Design Report for an International Accelerator Facility for Beams of Ions and Antiprotons, GSI, November 2001

[3]H. Koch, in Proceedings of the LEAP03-Conference, Yokohama, March 3-7, 2003, to be published in NIM B

[4]PANDA-Collaboration, 
$www-new.gsi.de/zukunftsprojekt/experimente/hesr-panda/index_e.html$
%

\subsection{Charm in the Medium}
\addtocontents{toc}{\hspace{2cm}(J. Ritman)\par}

J. Ritman

    {\em Physikalisches Institut , 
        Justus-Liebig-Universit\"at,Gie{\ss}en, \\
        Heinrich-Buff-Ring 16,  D-35392 Gie{\ss}en 
    }

\newpage \section{Hyperon--Nucleon Interaction}

Convenors: J. Haidenbauer and W. Eyrich

\subsection{Prediction and Discovery of a Pentaquark Baryon}
\addtocontents{toc}{\hspace{2cm}(M. Polyakov)\par}

M. Polyakov

{\it Universit\"at Bochum, Germany}

\subsection{Hyperon-nucleon interaction}
\addtocontents{toc}{\hspace{2cm}(J. Haidenbauer)\par}

J. Haidenbauer

{\em Institut f\"ur Kernphysik, Forschungszentrum J\"ulich,
D-52425 J\"ulich, Germany}

The hyperon-nucleon ($YN$) interaction is an ideal testing ground for
studying the importance of SU(3) flavor symmetry breaking in hadronic
systems.
Existing meson exchange models of the $YN$ force usually assume SU(3)
flavor symmetry for the hadronic coupling constants, and in some cases
\cite{Holz,Reu} even the SU(6) symmetry of the quark model.
The symmetry requirements provide relations between couplings of mesons of
a given multiplet to the baryon current, which greatly reduce the number
of free model parameters.
Specifically, coupling constants at the strange vertices are connected to
nucleon-nucleon-meson coupling constants, which in turn are constrained by
the wealth of empirical information on $NN$ scattering.

Indeed SU(3) symmetry was invoked for the construction of basically all
$YN$ models that one can find in the literature.
One should note, however, that the various models differ dramatically in
their treatment of the scalar-isoscalar meson sector, which describes the
baryon-baryon interaction at intermediate ranges.
For example, in the Nijmegen models \cite{NijIII,NijIV} this interaction
is generated by the exchange of a genuine scalar meson SU(3) nonet.
The T\"ubingen model \cite{Tueb}, on the other hand, which is essentially
a constituent quark model supplemented by $\pi$ and $\sigma$ exchange at
intermediate and short ranges, treats the $\sigma$ meson as an SU(3)
singlet.

In the $YN$ models of the J\"ulich group the $\sigma$ is viewed as arising
from correlated $\pi\pi$ exchange. In practice, however, in the old $YN$
models \cite{Holz,Reu}  the coupling constants of the fictitious $\sigma$
meson (with a mass of 550 MeV) at the strange
vertices ($\Lambda\Lambda\sigma$, $\Sigma\Sigma\sigma$) are treated as free
parameters. A new model, which is presently being developed \cite{HaWa},
includes now explicitly correlated $\pi\pi$ exchange. 

Despite those differences, however, 
essentially all $YN$ interaction models can reproduce the existing $YN$
scattering data. This is primarily due to the poor experimental information
about the $YN$ interaction. There are very few data available at low
energies and, moreover, they are basically all from the 1960's and 
of rather low accuracy. But even the few new data points that have
emerged over the last few years, due to efforts at the KEK facility \cite{Kado}, 
are afflicted with large error bars and, as a consequence, do not provide
better constraints for the existing $YN$ models. 

Therefore, we need to look for other reactions/systems which also
involve the $YN$ system and, thus, might allow to obtain further
information on the $YN$ interaction. One such possibility is offered
by the study of hypernuclei. Here some experimental information is
already available and especially the lighter nuclei 
($^3_\Lambda\rm H$, $^4_\Lambda\rm H$, $^4_\Lambda\rm{He}$) are also accessible
to microscopic, i.e. Faddeev-Yakubovsky type calculations \cite{Nogga}. 

However, I believe that scattering experiments where the $YN$ system
is produced in the final state might be the most promising reactions
for learning more about the $YN$ interaction. In particular, only
here one can study the $YN$ system at very low energies simply by
choosing the reaction energy close to the production threshold. 
The $YN$ system can be produced in a variety of reactions. One can 
use electromagnetic probes such as in
$Ž \gamma + d \to K^+ + YN$ \cite{Yama} or 
$Ž e + d \to  e' K^+ + YN$ and also hadronic probes like in
$Ž K^- + d \to \gamma + YN$ \cite{Ben}, 
$Ž K^- + d \to \pi^- + YN$ \cite{Pig},
$Ž \pi^+ + d \to K^+ + YN$ \cite{Davis}, 
$Ž p + p  \to K^+ + YN$, or
$Ž p + n  \to K + YN$.  
The latter two reactions are the ones that can be studied at
the COSY accelerator and there is already a wealth of data
available on the $pp$ channel, cf. corresponding contributions
of the TOF and COSY-11 collaborations. 

Results reported at this workshop by
B. Gibson (for $Ž K^- + d \to \gamma + YN$) and A. Gasparyan \cite{Gas}
(for $p + p \to K^+ + YN$) show that by  
choosing specific kinematics and observables one can isolate the effect of the
final state interaction in the $YN$ system and then extract 
specific features of the $YN$ interaction such as the ($S$ wave)
scattering lengths from the data. 

\def\Nucl{Nucl.\ }
\def\Phys{Phys.\ }
\def\Rev{Rev.\ }
\def\Lett{Lett.\ }
\def\PL{\Phys\Lett}
\def\PLB{\Phys\Lett B}
\def\NP{\Nucl\Phys}
\def\NPA{\Nucl\Phys A}
\def\NPB{\Nucl\Phys B}
\def\NPBS{\Nucl\Phys (Proc.\ Suppl.\ )B}
\def\PR{\Phys\Rev}
\def\PRL{\Phys\Rev\Lett}
\def\PRC{\Phys\Rev C}
\def\PRD{\Phys\Rev D}

\subsection{Study of the hyperon - nucleon interaction at Cosy-11 }
\addtocontents{toc}{\hspace{2cm}(T.~Ro{\.z}ek)\par}

T.~Ro{\.z}ek  \ \ for the COSY-11 Collaboration

{\em Institut f\"ur Kernphysik, Forschungszentrum J\"ulich GmbH, Germany}

The $\Sigma^0$ and  $\Lambda$ hyperon production near the kinematical
threshold was studied by the COSY-11 collaboration in $pp\rightarrow
pK^+\Lambda/\Sigma^0$ reactions. Data points,16 for the $\Lambda$
and 13 for the $\Sigma^0$ channel, were taken in the excees energy range
between 0.68 MeV and 59.3 MeV for $\Lambda$ (2.8 MeV and 59.1 MeV for
$\Sigma^0$) \cite{bal98,sew99b,kow02}. 
\\
The cross section ratio $\sigma (pp\rightarrow pK^+\Lambda ) /\sigma
(pp\rightarrow pK^+\Sigma^0 )$ below excess energie of about 15 MeV
was measured to be about 28 in contrast to a value of
about 2.5 determined for various excess energies higher than $Q = 300$ MeV
\cite{bald88}. The ratio for higher energies is in good agreement with
the $\Lambda/\Sigma^0$ isospin relation, which is 3. 
\\
To explain this unexpected threshold behaviour, various theoretical
scenarios within meson exchange models were proposed. Calculations
have been performed with pion and kaon exchange added coherently
\cite{gas99,gas01} or incoherently \cite{sib99}, including the
excitation of nucleon resonances \cite{shy01,sib00} and heavy meson exchange
($\rho$, $\omega$ and $K^*$) \cite{shy01}. Although the various
descriptions differ even in the dominant basic reaction mechanism, all
more or less reproduce the trend of an increase in the cross section
ratio in the threshold region. The present data are not sufficient to
definitely exclude possible explanations. Further studies e.g. on the other isospin
projections will help to understand the threshold hyperon
production.  
\\
Within the J\"ulich meson exchange model the cross
section ratio $\sigma (\Lambda) / \sigma (\Sigma ^0)$ is reproduced by
a destructive interference of $\pi$ and $K$ exchange
amplitudes. Calculations of the $\Sigma^+$ production  in this model
predict a factor of three higher cross section compared to the
$\Sigma^0$ channel for the destructive and a factor of three lower for
the constructive interference. Recently the $\Sigma^+$ production was
measured at the COSY-11 installation via $pp\rightarrow nK^+\Sigma^+$
at Q = 13 MeV and Q = 60 MeV \cite{pac02} in order to check the cross
section ratio between $\Sigma^+$ and $\Sigma^0$ production. Apart from
the Juelich model this ratio will also differ strongly if the dominant
production mechanism runs via an intermediate $N^*$ excitation or
not. 
\\
In the COSY-11 detection setup, the $\Sigma^+$ is
identified via the missing mass technique, by detecting the remaining
reaction products - $K^+$ and neutron wich will result in an event
identification comparable to the $\Sigma^0$ channel.

\subsection{The $\Lambda  N$   scattering lengths
from production reactions}
\addtocontents{toc}{\hspace{2cm}(A. Gasparyan)\par}

A. Gasparyan

{\em Institute of Theoretical and Experimental Physics,
117259,\\ B.\-Cheremushkinskaya 25, Moscow, Russia,\\
and  Institut f\"ur Kernphysik, Forschungszentrum J\"ulich}

Our  knowledge about the low-energy $\Lambda  N$ scattering
from direct experiments  is based on the scarce data set
measured in 1960's. This leads to  a very large uncertainty in the
$\Lambda  N$ scattering lengths. Thus different $YN$ models,
which describe the above mentioned experimental data
equally well yield the values for the scattering lengths that
may differ by a factor of 4 \cite{Stoks1999,Haidenbauer2000}.

The way to improve the present situation is to use
an indirect information on the $\Lambda  N$ scattering from
production reactions.
In this work we develop a method for analyzing the
production reactions with  a large momentum transfer
such as $pp\to pK^+\Lambda$. It is based on the dispersion
relation technic and uses the well-known fact that
in the properly chosen kinematical region the
energy dependence of the amplitude is given mostly by
the final state $\Lambda N$ interaction. Our
main result can be derived by means of the methods described
in \cite{Muskhelishvili1953,Frazer1959,Geshkenbein1969} and
can be represented by a simple formula  for the
singlet (triplet) scattering length expressed via
an integral of the differential invariant $\Lambda N$ mass spectra
with fixed spin in the $\Lambda N$ system. The novelty
of our approach consists  in that we take under control all
possible theoretical errors appearing
due to various simplifying assumptions.

In order to separate the triplet and singlet parts
of the $\Lambda N$ mass spectra we consider the possibility
to use polarization observables. It turns out that
looking e.g. at the analyzing power or at the
transverse double polarization observables enables 
to exclude the spin singlet contribution for
certain angles of the emitted kaon. It is very important
that such observables can be measured  at COSY.
To extract the pure singlet component of the
mass spectrum one needs to perform a more
complicated experiment involving longitudinally
polarized beam and target.

\subsection{K$^-$ + d $\to$ n + $\Lambda + \gamma$}
\addtocontents{toc}{\hspace{2cm}(B.\ F.\ Gibson)\par}

B.\ F.\ Gibson

{\em Theoretical Division, Los Alamos National Laboratory\\ 
             Los Alamos, New Mexico 87545 USA}

Strangeness provides a third dimension in nuclear physics, one that
rises above the two-dimensional (neutron-proton) isospin plane. 
Using the strangeness (flavor) degree of freedom one can test our
knowledge (and models) which has been developed over 3/4 of a century
from investigations based upon conventional, non-strange nuclear
structure and reactions.  Do our models extrapolate beyond the
isospin plane or are they merely exquisite interpolation tools
within that realm?

Charge-symmetry breaking (CSB) is a topic of current interest, as we
explore the effect of the u-d quark mass difference at the nonperturbative
QCD scales of nuclear physics.  CSB produces a $^3$He/$^3$H binding energy
difference of about 100 keV.  In comparison CSB in $\Lambda$ hypernuclei
is a factor of 3 larger than that observed in the nonstrange sector, as
is evidenced by the $^4_\Lambda$He/$^4_\Lambda$H binding energy difference
of some 360 keV.  It seems unlikely that the u-d mass difference mechanism
dominates CSB in the strangeness -1 sector of $\Lambda$ hypernuclei.

There exist limited p$\Lambda$ scattering data.  COSY, through the
pp $\to$ p$\Lambda$K$^+$ reaction, can add important low-energy data. 
However, there are absolutely no $n\Lambda$ scattering data.  The stopped
K$^-$ reaction K$^-$d $\to$ n$\Lambda \gamma$~\cite{Gib73} offers a means
to extract information about n$\Lambda$ scattering.  Its advantage lies
in the fact that there are only two strongly interacting particles in the
final state of interest.  A similar analysis of
$\pi^-$d $\to$ nn$\gamma$~\cite{Gib75} was successfully utilized to
extract the neutron-neutron scattering length.  A feasibility study for
the stopped K$^-$ experiment was made by Gall {\it et al.}~\cite{Gal90},
and a theoretical analysis was published in that same time
frame~\cite{Wor90}.  A $\chi^2$ study of the uncertainties in values of
the spin-singlet and spin-triplet n$\Lambda$ scattering lengths extracted
from the shape of the photon spectrum has since been performed~\cite{Gib00}. 
This analysis includes a separate investigation of the use of polarization
to cleanly separate the scattering lengths for the two spin states.

The work of B.F.G.\ was supported by the U.S. Department of Energy under
contract W-7405-ENG-36.

\subsection{Hyperonproduction at COSY-TOF$^*$}
\addtocontents{toc}{\hspace{2cm}(Wolfgang Eyrich)\par}

Wolfgang Eyrich

{\em University of Erlangen-Nuremberg, Germany}

The associated strangeness production in elementary nucleon-nucleon-induced
reactions is stu\-died exclusively at the external COSY beam using the time of
flight spectrometer COSY-TOF[1]. The complete reconstruction of all charged
particle tracks allows the extraction of total and differential cross sections
and Dalitz plots as well. A special start detector system, which consists of
several layers of highly granular detector components, was developed and
optimized for precise track reconstruction, both of primary and secondary
decay tracks. The design of the COSY apparatus provides the opportunity to
cover the full phase space of the reactions from threshold up to the COSY
energy limit. The main goal in the investigations of the reaction channels $NN
\rightarrow KYN$ is to gain an insight into the dynamics of the
$s\overline{s}$-production, which may also be connected with the questions of
the strange content of the nucleon. Theoretical access to the reaction
mechanisms is gained within the meson exchange model including resonance
contributions, final and initial state interaction and other effects based on
coupled channel mechanisms.
\par \noindent 
Especially the reaction channel $pp \rightarrow K^+\Lambda p$ was investigated
recently in detail in high statistic runs and delivered precise results which
show strong $N^*$ and FSI contribution. Alongside the $\Lambda$-production the
production of $\Sigma$-Hyperons is a further point of interest within the
associated strangeness production. Both $\Sigma ^0$ and $\Sigma ^+$ production
channels have been measured with respect to total cross sections and angular
distributions. The verification of observables of different reaction channels
at equal excess energies provides a further tool to test model calculations.
Moreover the reaction channels $pp \rightarrow K^0\Sigma^+ p$ and $pp
\rightarrow K^+\Sigma^+ n$ are of high interest due to a possible exotic
pentaquark resonance ($Z ^+$)[2] which might contribute to the production
mechanisms.
\par \noindent 
This talk will focus on the recent results of the TOF experiment on $\Lambda$-
and $\Sigma$-production, especially discussing the strong energy dependent
influence of the 1650 and 1710 $N ^*$ resonances in the $\Lambda$-production
channel and comparing them with the actual theoretical calculations [3],[4]
within the framework of the meson exchange models. Moreover the status of the
search for the exotic $Z^+$ penta-quark state is presented. An outlook
focussing on polarization observables which will be obtained from a very
recent run using a polarized COSY beam will be given. Additionally a short
discussion on further reaction channels in pn collisions measured in a test
run on a deuterium target will finish the talk.  \thispagestyle{empty}

\vspace{0.5cm}
{\large{\bf References}} 

\par \noindent 
[1] COSY-TOF Collaboration, Bilger R., et al., Strangeness production in the reaction $pp \rightarrow K^+\Lambda p$ near threshold, Phys. Lett. B{\bf{420}}(1998), 217
\par \noindent 
[2] Polyakov, M.V., et al., On the search for a narrow penta-quark
$Z^+$-baryon in NN interactions, Eur. Phys. J. A{\bf{9}}(2000), 115
\par \noindent 
[3] Shyam R., $pp \rightarrow p K^+ \Lambda$ reaction in an effectiv Lagrangian model, Phys. Rev. C{\bf{60}}(1999), 055213
\par \noindent 
[4] Sibirtsev A., et al., The Role of $P_{11}(1710)$ in the $NN
\rightarrow N\Sigma K$ reaction, Nucl. Phys. A{\bf{646}}(1999), 427
and references given therein
\par \noindent 
$^*$ supported by BMBF and Forschungszentrum Juelich

\subsection{Baryons Coupled to Strangeness}
\addtocontents{toc}{\hspace{2cm}(A. Sibirtsev)\par}

A. Sibirtsev

{\it Institut f\"ur Kernphysik, Forschungszentrum J\"ulich}

Strangeness production in $pN$  collisions provides an 
effective tool for investigation the $\Lambda{K}$, $\Sigma{K}$,
$KN$, $\Lambda{N}$ and $\Sigma{N}$ subsystems and allows to
measure the properties of known baryonic resonances as well as to
search for new exotic states. 

While the coupling of $S_{11}(1650)$,
$D_{15}(1675)$, $P_{11}(1710)$ and $P_{13}(1720)$ to $\Lambda{K}$
channel is poorly known~\cite{PDG2}, their coupling to $\Sigma{N}$ 
channel is absolutely unknown experimentally. It is expected~\cite{PDG2} that
$P_{33}(1600)$, $F_{15}(1680)$, $D_{33}(1700)$ and $F_{17}(1990)$ baryons
couple to strangeness, however until now there are no 
established data. Recently SPHINX Collaboration~\cite{Vavilov} 
measured $\Sigma^0K^+$ production in $pN$ collisions and detected 
new baryon with mass $M{=}1995{\pm}18$~MeV and width $\Gamma{=}90{\pm}32$ MeV.
The new $X(2000)$ state predominantly decays to strange channels and
was considered as an exotic baryon with hidden strangeness. These 
measurements are accessible at COSY.

The study of $KN$ subsystem produced in $pN{\to}KNY$ reaction
allows to determine the pentaquark properties, and not only its mass
and width. The $K$-meson angular distribution in Jackson frame is an 
effective tool to measure the $\Theta^+$ parity, taking into account 
that the pentaquark production cross section in $pN$ collision is 
large~\cite{Polyakov}.
Moreover the $pn{\to}K^0pK^-p$ or $pn{\to}K^+nK^-p$ reactions are even better
suited to $\Theta^+$ measurements, since these reactions are dominated
by $K$-meson exchange, which results in large pentaquark production
cross section.  The required proton beam energy is $\simeq$2.83~GeV,
that is close to the COSY limit, but one can explore the neutron
momentum gained from the deuteron target.

The $\Lambda{N}$ and $\Sigma{N}$ subsystems allow to search for strange 
dibaryons, the subject being still under controversial 
discussions~\cite{Barmin}. The precise measurements with high 
intensity beam available at COSY may result either in discovery of 
strange dibaryon or lead to new upper limit, which is expected to be
substantially lower than established presently. Furthermore,
the systematic studies of the $\Lambda{N}$ and $\Sigma{N}$ subsystems
provide access to an evaluation of hyperon-nucleon interaction and 
allows direct measurement of the $\Sigma{\to}\Lambda$ transition 
around $\Sigma{N}$ production threshold.

\subsection{High Resolution Search for Strangeness -1 Dibaryons}
\addtocontents{toc}{\hspace{2cm}(Frank Hinterberger)\par}

Frank Hinterberger for the HIRES Collaboration

{\em Helmholtz-Institut f. Strahlen- und Kernphysik der Universit\"at Bonn}

A high resolution search for the lowest strangeness -1 
dibaryons D$_{\rm s}$ and D$_{\rm t}$ via p+p$\rightarrow $K$^+$+D 
using COSY and BIG KARL has been started. 
Quark model calculations \cite{fhint:aer84,fhint:aer85} 
predict such states about 55 and 95~MeV above the $\Lambda$p threshold 
with invariant masses of 2109 and 2149~MeV and
total widths of less than 100 and 800~keV, respectively.
The HIRES experiment at COSY refers to a previous
SATURNE II experiment \cite{fhint:sie94}  where  
a small and narrow peak at about 2097~MeV was observed  
in the missing mass spectrum
of the reaction p+p$\rightarrow $K$^+$+X 
at proton
energies of 2.3 and 2.7~GeV.
The new experiment intends  improved statistical
accuracy and higher missing mass resolution.
By this, the  sensitivity of the search can be improved by an order of
magnitude.
Using the spectrometer BIG KARL the outgoing kaons are detected at 0$^{\circ}$
which is the optimal scattering angle for
a reaction with angular momentum transfer $\Delta l=0$.

The first HIRES run was in April 2003. 
With a proton beam momentum of 2.730~GeV/c the mean 
momentum of the outgoing
kaons was 0.960~GeV/c. The 
momentum acceptance of BIG KARL
coresponds to a missing mass range  
of  2.08~-~2.11~GeV.
The most important 
experimental achievement was the particle identification of the
kaons against the huge background of pions and protons.
We used two new threshold Cerenkov detectors in the focal
plane of the spectrometer in order to discriminate
pions from kaons. The protons were separated using TOF-information.
So, we were able to identify the kaons as a well separated peak 
in the online TOF spectra gated with the Cerenkov signal.
Another important point was the preparation of the proton
beam by the COSY team. 
A careful fine tuning of the socalled stochastic beam extraction
was done with the aim of a very high beam momentum resolution 
(about $3\cdot 10^{-4}$). The expected 
missing mass resolution is 350~keV.
We achieved rather high beam
intensities of about $1.0\cdot 10^9$ protons/s
during extraction corresponding to a time averaged
beam intensity  of about $0.7\cdot 10^9$ protons/s.
The online missing mass spectrum comprises only
a (small) part of the full statistics and 
it is premature to decide the
existence or nonexistence of a sharp
resonance near 2100~MeV.
The off-line analysis is underway.

\newpage \section{Few Nucleon Systems}

Convenors: F. Rathmann and N.N. Nikolaev

\subsection{Few Nucleon Systems
 at COSY
 and short-range  NN interaction
}  
\addtocontents{toc}{\hspace{2cm}(Yu.N. Uzikov)\par}

{  Yu.N. Uzikov  }

{\it  JINR, Dubna, Moscow region, Russia 141980 }

  Short-range structure of the lightest nuclei is related to 
 fundamental problems of hadron physics and  can be probed in 
 processes with high transferred momentum $Q^2$. 
 Pure meson-nucleon theory with consistent inclusion of
  important relativistic contributions
 works well in case of deuteron elastic formfactors at
 $Q \leq 1$ GeV/c (i.e., at internal momenta in the deuteron $q\leq 0.5 $GeV/c),
 but gives less clear result at higher $Q^2$ and fails for high energy 
 deuteron photodisintegration$^{1}$.
 It is assumed$^1$, that these data probably indicate for new physics with explicite
 quark-gluon degrees of freedom in the deuteron at high  $Q^2$. 
 Hadronic processes 
 at special conditions  can give  a new  independent
 information here. 

   The reaction $pd\to pp(^1S_0)n$ in 
 kinematics of backward elastic $pd$ scattering at beam energies 
 1--2 GeV  provides a new testing ground for the short-range
 $NN$ and $pd$ dynamics$^2$. In contrast to the
 elastic $pd\to dp$ process, advantages of the breakup reaction
 are caused by the isospin state  I=1 of the final diproton
 as compared to the  I=0 for  the deuteron.   
 The contributions  from  intermediate
 states with  nucleon resonances ($\Delta, N^*$), which are
 theoretically not well  under  control,
 are essentially suppressed in the $pd\to pp(^1S_0)n$ reaction due to
 isospin invariance. Another important qualitative  feature of
 this reaction is
 connected to the fact that at low excitation energy of the two protons,
 $E_{pp}<3$ MeV, the spin-singlet s-wave state $^1S_0$ dominates
 in the final diproton, whereas the deuteron wave function $\psi(q)$
  at high $q$ has a large  contribution from the d-wave.

 Measurements of the spin-averaged cross section
 of the reaction $pd\to pp(^1S_0)n$ in kinematics of backward
 elastic pd-scattering have been recently performed at  COSY$^3$ 
 and preliminary data on $A_y^p$ have been obtained.
 The analysis within the
 known  ONE+SS+$\Delta$ model$^4$ of the $pd\to dp$ process shows$^5$ 
 that a reasonable
 agreement with the data can be achieved when a  rather soft
 NN interaction potential at short NN distances
 ($r_{NN}< 1$ fm) is used for the $^3S_1-^3D_1$ and
 $^1S_0$ states. Therefore, the modern high accuracy CD Bonn  NN potential
 is much more preferable within 
 the ONE+SS+$\Delta$ model in contrast to the RSC or Paris potentials.  
 In order to  test  this picture,  spin observables (analyzing powers
 $A_y$ and $T_{20}$, spin correlations $C_{y,y}$, $C_{z,z}$ 
 and $C_{xy,z}$)
 are calculated here and planned to be measured at COSY.
 The main question to experiment is whether  the
 $T_{20}(\theta_{cm}=180^\circ)$ changes
 its sign at beam energy $T_p>1$ GeV as it follows from  this model,
 and others spin observables
 exhibit  remarkable features$^2$ caused by the node 
 in the half-off-shell
 pp($^1S_0)$ scattering t(q)-matrix at $q\approx 0.4 GeV/c$.
  Role of initial and final state interaction effects
 and relativistic
 P-wave  components of the deuteron and diproton  are briefly  discussed.

 Using  more heavy target allows us to probe more 
 high internal momenta $q$ in  nuclei at the same beam energy.
 So, according to Ref.$^6$, 
 the cross section  of the p$^3$He$\to ^3$Hep process
 probes  the $^3$He wave function at high internal momenta q=0.6-1.0 GeV/c 
 (at T$_p>$1 GeV) in the NN($^1S_0$) pairs in  $^3$He and acts as a filter
 for  the d$^*(^1S_0)$+N configuration$^7$ in $^3$He.

{\large \bf References}

 1. R. Gilman, F. Gross, J. Phys.G: Part.Nucl. {\bf  28} (2002) R37\\
%
 2. Yu.N.~Uzikov, JETP Lett. {\bf 75} (2002) 5; 
J. Phys.G: Part.Nucl. {\bf  28} (2002) B13.\\
3. V.I. Komarov et al., Phys. Lett. {\bf B 553} (2003) 179.\\
4. L.A. Kondratuyk, F.M. Lev., L.V. Schevchenko.Yad.Fiz., {\bf 33} (1981) 1208\\
5. J. Haidenbauer, Yu.N. Uzikov. Phys.Lett. {\bf B 562} (2003) 227.\\
6. Yu.N. Uzikov,  Nucl.Phys. {\bf  A644} (1998) 321;
   Phys. Rev. {\bf C 58} (1998) 36.\\
7. Yu.N. Uzikov, J. Haidenbauer,  Phys. Rev. {\bf C 68} (2003) 014001.

\subsection{$pd$ and $dd$ interactions}
\addtocontents{toc}{\hspace{2cm}(N.N. Nikolaev)\par}

N.N. Nikolaev

{\it Institut f\"ur Kernphysik, Forschungszentrum J\"ulich}

There are several fundamental reasons for the interest in the deuteron as
a beam and target. First, as a loosely bound $np$ state it is the
best approximation to the free neutron target. Second, the $pd$ and $dd$
interactions are indispensable as the testing ground for a few body
interaction theories. Third, the high momentum transfer $pd$ interactions
at COSY and $ed$ interactions at CEBAF are the complementary probes of  
the short distance properties of the deuteron. Fourth, as a spin-1 target
with the substantial quadrupole spatial deformation the deuteron is a
testing ground for theoretical ideas on the spin-orbit coupling in
relativistic bound states which is of much relevance to the understanding
of the spin properties of vector mesons produced in deep inelastic
scattering at HERA. Fifth, the tensor polarization effects in $pD$ and 
$DD$ interactions are unique to the spin-1 deuteron, here the data from
COSY will be complementary to the tensor polarization effects studied
by the HERMES collaboration at DESY \cite{HermesTensor}. 
Some of the above aspects of the physics with the polarized deuterons at
COSY have been discussed at this Workshop by Yu. Uzikov, R. Schleichert, 
V. Baru, F. Rathmann, A. Kacharawa, H.P. gen. Schieck, Ch. Elster and
A. Kobushkin, see these Proceedings \cite{Othertalks}, the relativistic 
lightcone treatment of spin-orbit coupling in the vector mesons is found in
\cite{IgorPhD}, for the evaluation of tensor polarization effects in DIS 
off deuterons see \cite{SNtensor}. Here I only want to comment on the use
of the polarized deuteron as the polarized neutron target at COSY.

\begin{figure}[h]
   \centering
   \epsfig{file=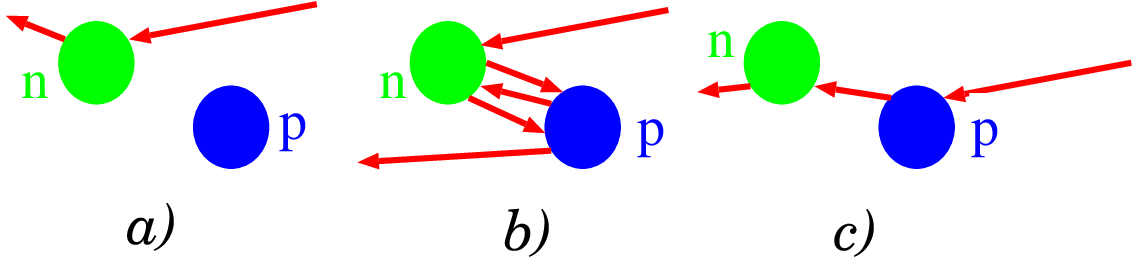,width=140mm}
   \caption{\it The single-scattering/impulse approximation (fig. 1a) and
multiple-rescattering diagrams for the matrix of interaction of
projectile with the deuteron at low (fig. 1b) and high (fig. 1c) 
energies. The elastic scattering is obtained if the scattering matrix
id sandwiched between the initial and final state deuteron wave functions,
the deuteron breakup is obtained if the final state is a two-nucleon
continuum with the appropriate $NN$ final state interaction}
   \label{fig:ivanov-gluondensity}
\end{figure}

The principal issue is whether the $pD$ elastic and breakup data admit
a simple interpretation in terms of the $pp$ and $pn$ scattering
amplitudes or not. To this end, the theoretical description of the $pD$ 
scattering changes drastically form the low energy to intermediate and
to the high energy of COSY. Specifically, at low energies only lowest
partial wave contribute the the projectile-nucleon scattering and the projectile
can rescatter many times back and forth between the proton and the
neutron, see fig. 1b. Here we indicated only the subset of the Faddeev
diagrams, one must allow also for the nucleon-nucleon 
interaction between those rescatterings and needs to solve
the very involved Faddeev equations. A very good summary 
of the relevant issues
was presented at this Workshop by P\"atz gen. Schieck and Elster.
The high energy scattering is
forward peaked and the multitude of multiple rescattering diagrams 
reduces to the double scattering when the two nucleons are aligned
along the nearly straight-line trajectory of the projectile, see fig. 1c.
The emerging formalism has become known as the Glauber theory, its most
important feature is that the double scattering amplitude is calculable 
nearly parameter-free in terms of the projectile-proton and projectile-neutron
scattering amplitudes. With the exception of very small excitation 
energies, the differential cross section of the breakup reactions is
readily related to the free-nucleon cross section with calculable 
double-scattering
corrections. A very instructive demonstration of how the full Faddeev
calculations of the spin observables of the deuteron breakup tend to
the free-nucleon scattering observables as the energy of the beam 
nucleons is increased from 65 MeV to 220 MeV is found in the review by
Gl\"ockle et al \cite{Gloeckle}, the Glauber theory discussion of spin 
phenomena in the proton-deuteron elastic scattering has been reported
by Alberi et al. \cite{Alberi}, who find a good agreement between the
Glauber theory and the experimental data.

The overall conclusion is that the ineterpretation of the experimental
data on the deuteron breakup at ANKE-COSY (for the discussion of the planned 
experiments see the talks of A. Kacharava, R. Schleichert and F. Rathmann
\cite{Othertalks}) in terms of the $np$ elastic and charge exchange 
amplitudes will be fairly straightforward, and the COSY collaboration
can contribute to the databse on polarized and unpolarized $np$ scattering
as much as EDDA-COSY collaboration did for the $pp$ scattering.

\subsection{Lightest nuclei structure at short distances
in exclusive reactions}
\addtocontents{toc}{\hspace{2cm}(A.P.~Kobushkin)\par}

A.P.~Kobushkin$^{a,b}$, E.A.~Strokovsky$^{b,c}$, K.~Hatanaka$^b$ and S.~Ishikawa$^d$\\

{$^a$ \it
Bogolyubov Institute for Theoretical Physics, 03143, Kiev, Ukraine
}\\
{$^b$ \it
RCNP, Osaka University,
10-1 Mihogaoka, Ibaraki,
Osaka,
 567-0047,
Japan}
\\
{$^c$ \it
LPP, Joint Institute for Nuclear Research,
141980,
Dubna,
Moscow region,
Russia}\\
{$^d$\it
Hosei University, Department of Physics, Tokyo, Japan}

\begin{figure}[h!]
\vspace{12cm}
\includegraphics{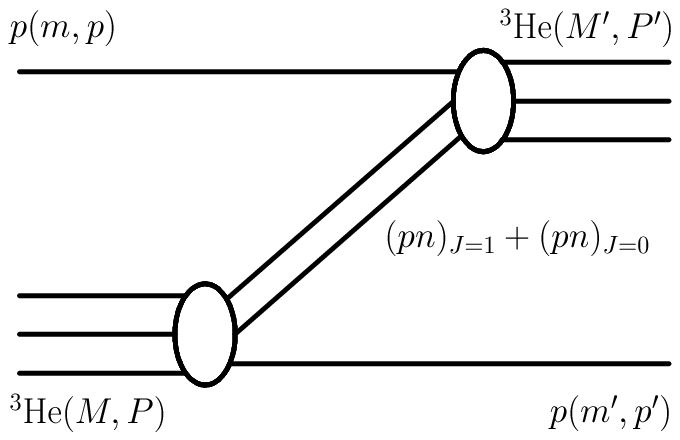}

\includegraphics{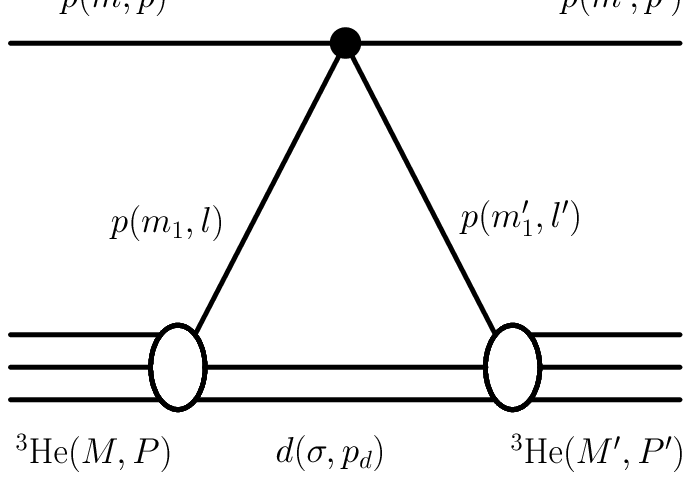}

\includegraphics{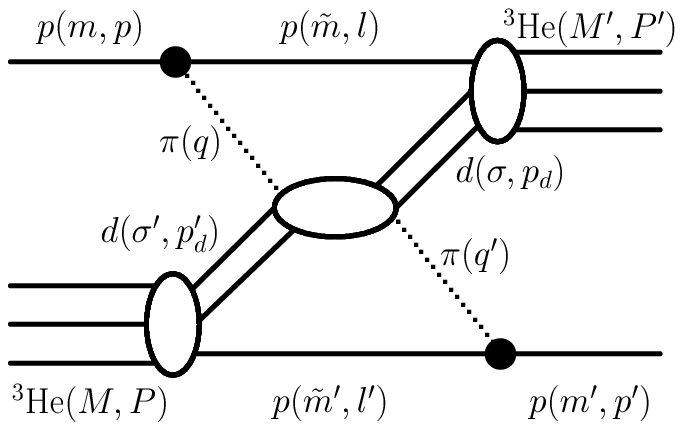}

\includegraphics{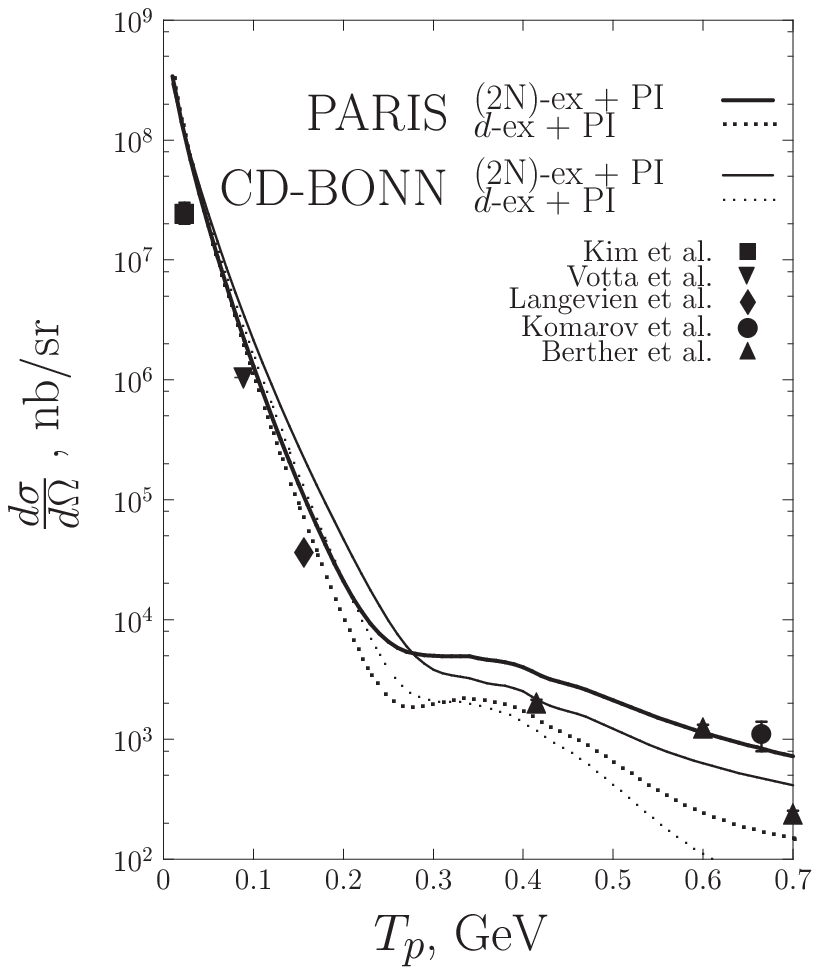}
\includegraphics{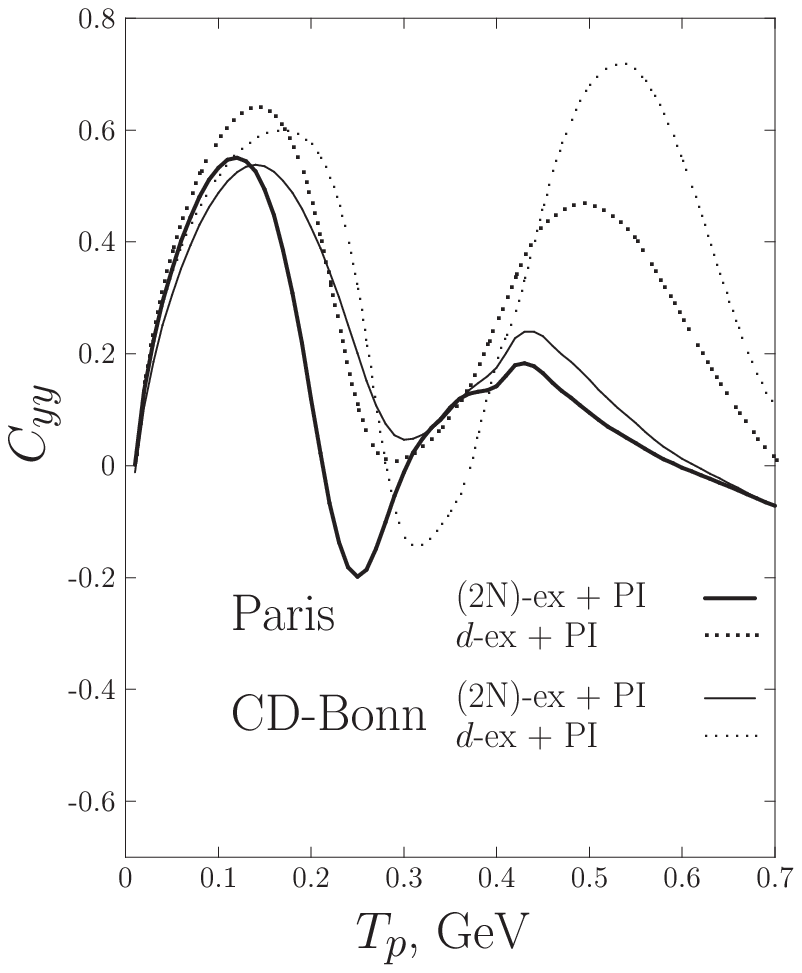}
\end{figure}

Exclusive backward (in the center of mass frame) proton-nucleus and
deuteron-nucleus scattering at intermediate energy involves large momentum
transfer and therefore such reactions can provide an access to the high
momentum components of the wave function lightest nucleus ($d$, $^3\rm{He}$,
$^4\rm{He}$). We discuss some important mechanisms of $p+^3\rm{He}$ elastic
scattering at $\theta=180^0$, two-nucleon exchange (2N), the direct mechanism
(DIR) and rescattering of intermediate pions (PI), and calculate the
differential cross section and polarization transfer $C_{yy}$. We also
consider appropriate mechanisms for $d+^3\rm{He}$ and $d+^4\rm{He}$ elastic
backward scattering and calculate energy dependence of the differential cross
section and the tensor analyzing power $T_{20}$ for this reactions. It is
demonstrated that this observables strongly depend on $d+p$ and $d+d$
components of $^3\rm He$ and $^4\rm He$ wave functions, respectively. It is
also discussed a problem of study $(2N)_{^1S_0} + N$ component of $^3\rm He$
wave function in $n+^3\rm He$ elastic backward scattering.

\subsection{Double Spin-Observables for $\rm pn$ Systems in $\rm pd$ Interactions}
\addtocontents{toc}{\hspace{2cm}(A.~Kacharava)\par}

  A.~Kacharava$^{a}$ and F.~Rathmann$^{b}$ 
   for the ANKE Collaboration. 

  {\it $^{a}$Physikalisches Institut II, Universit\"at
    Erlangen--N\"urnberg,\\
$^{b}$Institut f\"ur Kernphysik, Forschungszentrum J\"ulich}

The nucleon-nucleon ($\rm NN$) interaction represents one of the simplest and most
fundamental systems involving the strong interaction.  An appreciation of the
($\rm NN$) interaction constitutes a necessary step towards a consistent understanding
of the binding of neutrons and protons in nuclei, as well as giving
an insight into reactions between nucleons (or nuclei) and nuclei and, most
importantly, into the makeup of nuclear matter.

The complete description of the $\rm NN$ interaction requires precise data
in order to carry out phase--shift analyses (PSA), from which the scattering
amplitudes can be reconstructed.
For this purpose one generally needs data from experiments where
both the beam and target particles are polarised in the
initial state, as well as polarisation determination of final state
particles. 

Experiments of this kind were carried
out for the $\rm pp$ system to about 3.0~GeV \\({\it R.Arndt et al., Phys. ReV. C62(2000)034005;
EDDA Collaboration: Phys. ReV. Lett. 90(2003)142301}.) 


The challenging physics goals of the proposed research program in frame of the
ANKE collaboration comprise the following objectives:  \\
1. Substantial and systematic enhancement of the world 
database of $\rm pn$ elastic scattering through a series of high-precision experiments,
using polarised protons quasi-elastically scattered off the polarised deuteron target
($\rm \vec{p}\vec{d}$). \\
2. Measurement of the spin structure of the amplitudes of the elementary $\rm np$
charge-exchange (CE) process via deuteron-induced reactions ($\rm \vec{d}\vec{p}$). \\
$[http://www.fz-juelich.de/ikp/anke/doc/Proposals.html]$ \\

{\bf $\rm pn$: Elastic scattering observables} \\
Up to now, the world data base on elastic $\rm pn$ scattering is rather scarce
({\it F.Rathmann, W.T.H. Van Oers, and C.Wilkin, Intermediate Energy Spin Physics,
Progress Report, 11/1998}); above about 1.1 GeV there exist practically no data.
Within the proposed experiment it will become possible to substantially extend
the $\rm pn$ data base into the unchartered territory up to 2.83 GeV incident
proton energy. Polarisation data are the essential precondition for a partial 
wave analysis, because the existing low energy amplitudes can not be
extrapolated to higher energies. Among the different polarisation observables,
that can be measured, spin correlation parameters are most easily accessible.
Their measurement requires a polarised beam incident on a polarised neutron (deuteron)
target. \\

{\bf $\rm np$: Charge-Exchange break-up (CE)} \\
Information can be obtained on the spin--dependent $\rm np$ elastic amplitudes near the
backward direction (the charge--exchange region) by measuring the
charge--exchange breakup of polarised deuterons on an unpolarised
hydrogen target ({\it D.Bugg, C.Wilkin, Nucl. Phys. A467(1987)575}).
The effect can be understood qualitatively as follows. The two
nucleons in the deuteron are in $\rm T=0$, $ ^3{\rm S_{1}}$ or
$^3{\rm D_{1}}$. The spatial and spin states are symmetric so that, by the
generalised Pauli principle, the isospin state is antisymmetric.
In the charge--exchange reaction under special kinematic conditions
(scattering angle $ \theta$ close to zero and momentum transfer $t \sim 0$),
the transition to a spin antisymmetric $ ^1{\rm S_{0}}$ state of two protons 
therefore requires a spin flip.
The overall intensity of the
spin--dependent parts of the elementary $\rm np \to pn$ CE amplitude
can thus be inferred from the probability of the $\rm dp$ CE process.
The experimental programme is divided into two parts:
\begin{enumerate}
\item The first stage will utilise unpolarised and tensor polarised
  deuteron beams incident on an unpolarised hydrogen cluster target.
  The differential cross section gives the overall intensity of
  the spin--dependent parts of the elementary CE process. Tensor
  polarised deuteron beam enables us to separate the absolute values
  of three spin--dependent amplitudes.
\item Using transversely polarised deuterons incident on a polarised
  internal hydrogen gas target and measuring the spin--correlation
  coefficient opens the possibility of obtaining the relative phase
  between amplitudes.
\end{enumerate}

Since we plan to measure the cross section in parallel with polarisation
observables, our experiment will be the first that can provide complete
data necessary to determine the spin-dependent part of the elementary
$\rm np$ process in the energy range above LAMPF (800 MeV) up to
maximum beam energy per nucleon (1150 MeV) achievable at COSY.
ANKE spectrometer is sensitive to the angular range of
$\theta_{c.m.} \approx 0^{\circ}-30^{\circ}$, which is unchartered 
territory for the $\rm pn$ system.

The measurements will be carried out at the internal beam of the Cooler
Synchrotron COSY using the magnetic  spectrometer ANKE, 
including spectator detection system integrated
into the complete experimental setup. Experiment will alow to calibrate 
for the first time the tensor (vector) polarisation of the COSY
deuteron beam.

It should be noted, that on a worldwide scale the above outlined 
experimental program in a near future is possible only at COSY.

\subsection{Nd Interaction at Low Energies: Open Questions}
\addtocontents{toc}{\hspace{2cm}(H. Paetz gen. Schieck)\par}
{H. Paetz gen. Schieck}

{\em Institut f\"ur Kernphysik, Universit\"at zu K\"oln, Cologne, Germany}

Much emphasis in low-energy hadron research in the past has been on the
three-nucleon system, being the simplest system beyond the 
two-nucleon system. The present situation and future of low-energy studies of
the interaction of three nucleons is characterized by 
different trends.

One is the tremendous progress made by few-body theory, first, by numerically
exact calculations using precision meson-exchange nucleon-nucleon potentials
including different three-body forces and, recently, in applying realistic
effective-field theory approaches to low-energy problems.

Both approaches have been highly successful in describing many few-body
observables but still have been incapable of solving long-standing puzzles
such as the A$_y$ puzzle of p-d elastic scattering or the cross-section
discrepancies in deuteron breakup. At low energies progress has been made in
incorporating the Coulomb force into the Faddeev calculations thus giving the
existing precise charged-particle data renewed importance as compared to
neutron data.

Experimentally the number of low-energy working groups has decreased
substantially in recent years, partly because accelerators were phased out or
redirected towards other fields. Though the agreement between low-energy data
for many observables is good, for a number of observables some long-standing
completely unresolved discepancies persist:

\begin{itemize} 
\item d+N elastic cross section (Sagara) anomaly
\item d+N elastic A$_y$ (and iT$_{11}$) puzzle
\item d+N breakup anomaly in the space-star situation
\item d+N breakup anomaly in the quasi-free scattering situation
\end{itemize}

The low-energy discrepancies decrease with energy in contrast to the behavior
of newly discovered medium-energy discrepancies. This suggests that different
three-body forces may be acting in these energy regions. One serious problem
has been that on the one hand the number of n+d observables is smaller and
data quality partly not as good as that of the corresponding p+d data, on the
other hand theory has made progress in incorporating the Coulomb force in the
elastic scattering, but not yet in deuteron breakup.

Therefore besides new and innnovative theoretical efforts there is a need for
more precise and also for new data to track down the origin of these
discrepancies. In the elastic channel new polarization observables should be
measured whereas for the breakup the exploration of much larger regions of the
available phase space should be performed. Inclusion of the Coulomb force in
the theory and exploitation of the higher precision and larger number of
observables in the charged-particle channels would be necessary.

\subsection{Questions and Problems in Few  Body Reactions}
\addtocontents{toc}{\hspace{2cm}(Ch. Elster)\par}

Ch. Elster

{\em Department of Physics and Astronomy, Ohio University, Athens, OH 45701}

\newpage \section{Closing Session}
\subsection{Summary I}
\addtocontents{toc}{\hspace{2cm}(A.W. Thomas)\par}
 
A.W. Thomas

{\it 
Special Research Centre for the Subatomic Structure of Matter \\
The University of Adelaide  \\
Adelaide, SA 5005, Australia}

This was an exciting and stimulating meeting, both in terms of the physics
opportunities available to be explored at COSY and eventually GSI and
because the meeting was in large part organised by a very spirited group
of young scientists who see the possibilities for an exciting 
scientific future at these facilities. I would like to extend my
appreciation to these young people for doing a fine job.

One cannot look to the future without first appreciating the
achievements of the past and COSY certainly has a string of achievements
of which any laboratory would be proud. For example, 
together with its partners at
IUCF and the Svedberg Lab in Uppsala, COSY has done a wonderful job of
mapping the systematics of meson production near threshold in few
nucleon systems. These precise measurements have stimulated a great deal
of theoretical effort, using methods from $\chi PT$ to
somewhat older but effective meson exchange models. The studies of
$\eta$ and $\omega$ production are especially interesting for the
information they yield as to whether predictions of possible
meson-nucleus bound states are correct \cite{Tsushima:1998qw}.

There is currently enormous interest in the problem of baryon
spectroscopy. COSY has already produced important data on the $\Sigma$
and $\Lambda$ hyperons in the mass region around 1.4 GeV and in the
future it will be extremely important to exploit the capabilities of
COSY as a probe which is complementary to Jefferson Lab. Only a
concerted effort involving theory and experiment working together from
all possible angles will allow us to resolve the many puzzles we face in
this field -- puzzles such as ``missing states'', exotics and so on.
Here, as in most other aspects of the COSY program the lab can be proud
of the support it has received from the theoretical physics group, under
the leadership of Josef Speth. Their work on the reaction mechanisms
associated with the formation of various baryon resonances has led to
new insights into the nature of these states \cite{Krehl:1999km}. 
It will be especially
important, given the recent advances in our ability to calculate excited
states in lattice QCD \cite{Leinweber:2002bw}, 
to make use of the best of both approaches in
analyzing new experimental data.

The recent discovery of the $\theta^+$ 
\cite{Nakano:2003qx}, 
a strangeness +1 baryon whose
minimal quark content is $uudd\bar{s}$, caused enormous excitement at
the meeting. The remarkable precision with which it had been predicted
was discussed in detail \cite{Diakonov:1997mm}, along with
the alternative explanations that had already appeared. It is quite
clear that there is a very exciting program to be carried out at COSY 
as well as JLab and other facilities, to determine 
the properties of this state,
especially its spin and parity, and to search for other exotic baryons
which have been predicted. The existence of such exotic states opens a
completely new chapter in the development of strong interaction physics
and we all look forward to the discoveries of the next few years with
great anticipation.

While on the topic of exotic states we note that theory group has also 
made important progress in the analysis 
of possible ``molecular states'' in the coupled $\pi \pi - K \bar{K}$
system \cite{Sassen:2002qv}. Such states are also of tremendous interest
as we struggle to solve non-perturbative QCD. This work is of direct
relevance to recent discoveries of unexpected charmed mesons in $e^+ -
e^-$ annihilation \cite{Aubert:2003fg}, as well as possible exotic 
mesons reported at BNL \cite{Ivanov:2001rv}.

Of course, a number of new results from other laboratories were reported
at the conference. We mention particularly the new results on charge
symmetry violation from TRIUMF and IUCF. The former involved the first
report of a non-zero forward-backward asymmetry in the reaction $n+p
\rightarrow d + \pi^0$, while the latter involved a very clean signal
made in the last weeks of operation of the Cooler. There is considerable
theoretical interest in both results.

Finally, we should look to the future and apart from the examples
already cited there were many other important research directions to be
explored. Data on the behaviour of hadrons in dense matter obtained from
heavy ion collisions requires comparable data from proton and deuteron
induced reactions where density related effects are not expected to be
so large. This is particularly needed for $K^+$ and $K^-$ mesons in
matter \cite{Tsushima:2000hs}. One can also make and 
explore the properties of hypernuclei,
produce the $a_0$ meson using various initial states (and therefore
together with various nuclear final states), excite the Roper and other
resonances with new projectiles and thus look for new insight into the
structure of these resonances. 

On top of the extensive program involving COSY for the next decade,
FZ-J\"ulich also has the opportunity to play a major role in
building and designing experiments for the new GSI facility. This
facility, along with JHF in Japan and JLab in the USA, will provide the
key platforms for unravelling the secrets of hadron physics over the
next 20 years. It is a tremendous opportunity to be   
be a major partner in such a significant project from the very beginning
and the enthousiasm shown at the workshop clearly demonstrated the
willingness of the community to accept the challenge.

\subsection{Summary II}
\addtocontents{toc}{\hspace{2cm}(H. Str\"oher)\par}
 
H. Str\"oher

{\it IKP, Forschungszentrum, J\"{u}lich, Germany}

From the experimental point of view, the aim of the workshop "Hadron Physics at COSY" was two-fold:

\begin{itemize} 
\item[a)]to obtain an overview of what has been
achieved so far in hadron physics with different probes, and
\item[b)]to identify possible directions and - better - specific
experiments which should be given a high priority in future
measurements.  \end{itemize} 
Due to time constraints and breadth of
the field "hadron physics", it could not be expected to achieve a full
overview, but certainly many important issues were covered and very
recent (and exciting) results were presented.

As far as new experiments and future directions are concerned, the
meeting was a very good starting point, which in the meantime seems to
result in more specific developments like "WASA at COSY".

\end{document}